\documentclass[11pt]{article}%
\usepackage{amsfonts}
\usepackage{amsmath}
\usepackage{hyperref}
\usepackage{amssymb}
\usepackage{amsfonts,setspace}
\usepackage{caption}
\usepackage[caption=false]{subfig}
\usepackage{epsfig,latexsym,graphicx}
\usepackage{appendix}
\usepackage{graphicx}%
\usepackage{soul,xcolor}
\usepackage{enumitem}
\providecommand{\U}[1]{\protect\rule{.1in}{.1in}}
\begin{document}
	\setstcolor{red}
	\title{\bf Robust inference for linear regression models \\ with  possibly skewed error distribution}
		\author{Amarnath Nandy, Ayanendranath Basu and Abhik Ghosh\footnote{Corresponding author}\\
		Interdisciplinary Statistical Research Unit\\
		Indian Statistical Institute, Kolkata, India  
		\\
		{\it amarnath\_r@isical.ac.in, ayanbasu@isical.ac.in, abhik.ghosh@isical.ac.in}}
	\date{}
	\maketitle
	\begin{abstract}
		Traditional methods for linear regression generally assume that the underlying error distribution, equivalently the distribution of the responses, is normal. Yet, sometimes real life response data may exhibit a skewed pattern, and assuming normality would not give reliable results in such cases. This is often observed in cases of some biomedical, behavioral, socio-economic and other variables. In this paper, we propose to use the class of skew normal (SN) distributions, which also includes the ordinary normal distribution as its special case, as the model for the errors in a linear regression setup and perform subsequent statistical inference using the popular and robust minimum density power divergence approach to get stable insights in the presence of possible data contamination (e.g., outliers). We provide the asymptotic distribution of the proposed estimator of the regression parameters and also propose robust Wald-type tests of significance for these parameters. We provide an influence function analysis of these estimators and test statistics, and also provide level and power influence functions. Numerical verification including simulation studies and real data analysis is provided to substantiate the theory developed.
\end{abstract}


\section{Introduction}
\label{SEC:Intro}
In real life problems, we are often interested in investigating how the different sources of variation (covariates or explanatory variables) affect the desired output (response variable). The linear regression model (LRM) is one of the most popular tools for doing so, where we assume that the covariates (or transformed covariates) have linear relationships with the response. In traditional linear regression, our canonical assumption is that the errors are normally distributed. But in real life, for example in case of real data related to health and behavioral sciences and many related variables (e.g., medical expenditure), skewed patterns may be observed \cite{Boos:1987}. In such cases, it may be necessary to assume a skewed model distribution for the random error component in the LRM; 
a few previous attempts have used gamma or generalized gamma distributions for this purpose (see \cite{Agostinelli:2020}, \cite{Malehi:2015}, \cite{Petrinco:2012}) which are always skewed. 
However, observed skewness patterns in response distribution may arise due to data contamination (where the true underlying distribution is symmetric)  or they can be the inherent characteristics of the response variables. 
So, we propose to use the class of skew normal distributions -- which also includes the ordinary normal distribution as a special case -- as the model for the errors in a linear regression model. 
Recently, several authors have used this skew-normal distribution in regression models where response variables are skewed in nature but all these attempts have used the classical non-robust likelihood based inference procedures (see \cite{Alhamide/etc:2015}, \cite{Guedes:2014}, \cite{Hui-qiong:2014}, \cite{Kheramandi:2015}, \cite{Lyu:2018}).

Yet, it is well known that likelihood based inference, for all its efficiency properties, may perform very poorly in terms of robustness against data contamination and other types of anomalies which are not uncommon in modern complex datasets. There can be additional (model-specific) outliers even when the underlying distribution is skewed so that the classical likelihood based inference would again fail to provide correct inference.  It is generally difficult, sometimes impossible, to identify and remove such anomalies in an objective manner. So, a robust procedure that automatically takes care of the effects of such data contamination to produce stable inference would be an important and useful alternative in case of practical applications \cite{Deb:2017, Rousseeuw:1987, Swallow:1996, Zaman/etc:2001}. Some recent attempts to model the outliers in such situations have been made using mixture error distributions; see, \cite{Beath:2018, Dogru:2017, Liu:2014, Zeller:2016}.

In this paper, we have presented a robust method of fitting the linear regression model under skew normal errors based on the minimization of the density power divergence (Basu et al. (1998) \cite{Basu/etc:1998}). The corresponding minimum density power divergence estimator (MDPDE for short) is indexed by a single tuning parameter $\alpha$, which provides a trade-off between the efficiency and the robustness of the method. Recently, Nandy et al. \cite{Nandy:2021} have implemented this method for the skew normal distribution in case of homogeneous data. In our paper, we have extended this idea for the regression set up, where the data are independent but not homogeneous. We have found robust estimators of the regression coefficients as well as the parameters of the error distribution. In addition, we have provided the asymptotic distribution of the estimators and constructed a robust Wald-type test for testing the significance of the regression coefficients. We have also presented an influence function analysis of this method and demonstrated the boundedness of the same for positive $\alpha$.

The rest of the paper is organized as follows: in Section 2, we have defined our MDPDE for the skew regression model and provided the formula for its asymptotic variance as also an influence function analysis. In Section 3, we have proposed a robust Wald-type test based on the MDPDE and described its usefulness. We have provided the results obtained from different simulation studies in Section 4. We have also implemented our method on two different real life datasets and we have summarized the results in Section 5. These examples are instructive and help to understand how some of the MDPDEs provide stable inference under the presence of anomalous observations. We have also discussed a method for obtaining the optimal value of the tuning parameter in Section 6. Finally, Section 7 provides some concluding remarks.

\section{The model and the proposed estimator}
\label{SEC:2}
\subsection{The linear regression model with skew-normal errors}
\label{SEC:Skew_Norm_LRM}
We first give a brief description of the  linear regression model with skew-normal (SN) errors. The skew normal (SN) distribution (see \cite{Azzalini:1985}, \cite{Azzalini:1986}, \cite{Azzalini:2005}) is defined in terms of three parameters, namely the location parameter $\mu \in \mathbb{R}$, 
the scale parameter $\sigma \in \mathbb{R}^{+}$ and the skewness parameter $\gamma \in \mathbb{R}$, and is denoted by $SN(\mu,\sigma,\gamma)$. 
In particular, if $\mu=0$ and $\sigma=1$, it is referred to as the standard SN distribution and is denoted by $SN(\gamma)$. 
The probability density function (pdf) and the cumulative distribution function (cdf) of 
the $SN(\mu,\sigma,\gamma)$ distribution are given, respectively, as
\begin{eqnarray}
	f(x,\boldsymbol{\eta}) &=& \frac{2}{\sigma} \phi\left(\frac{x-\mu}{\sigma}\right)\Phi\left(\gamma\frac{x-\mu}{\sigma}\right),
	~~~~ x\in \mathbb{R}, 
	\label{EQ:SN_pdf}
	\\
	F(x,\boldsymbol{\eta})&=&\Phi\left(\frac{x-\mu}{\sigma}\right)-2T\left(\frac{x-\mu}{\sigma},\gamma\right),~~~~ x\in \mathbb{R},
	\label{EQ:SN_cdf}
\end{eqnarray}
where $\boldsymbol{\eta}=\big(\mu,\sigma,\gamma\big)^T$ is the vector of unknown parameters, 
$\phi$ and $\Phi$ are the pdf and cdf of the standard normal distribution, respectively, 
and $T(h,a)$ is the Owen's function defined as 
$$
T(h,a)=\frac{1}{2\pi}\int_{0}^{a} \frac{e^{-\frac{h^{2}}{2}(1+x^{2})}}{1+x^{2}}dx,
\hspace{5mm}h,a\in \mathbb{R}.
$$ 
The mean, variance and skewness ($\gamma_{1}$) of a random variable $X$ having the $SN(\mu,\sigma,\gamma)$ distribution  are  given by
$E(X)=\mu+\sigma\delta\sqrt{\frac{2}{\pi}}$, 
$Var(X)=\sigma^{2}\left(1-\frac{2\delta^{2}}{\pi}\right)$ and 
$\gamma_{1}=\frac{1}{2}(4-\pi)\gamma^3\left(\frac{\pi}{2}+(\frac{\pi}{2}-1)\gamma^{2}\right)^{-3/2}$,
with $\delta={\gamma}{(1+\gamma^{2})^{-1/2}}$.
Clearly, the SN distribution is positively and negatively skewed according to the sign of the parameter $\gamma$. For the particular case $\gamma=0$, the SN distribution $SN(\mu, \sigma, 0)$ has skewness zero and coincides with the ordinary (symmetric) normal distribution, $N(\mu, \sigma^2)$, having mean $\mu$ and variance $\sigma^2$.

Now, let us consider the skew normal linear regression model
\begin{equation}
	y_i=\boldsymbol{x_{i}^{T} \beta} + \epsilon_{i},~i=1,2, \ldots,n
	\label{EQ:lm}
\end{equation}
where $\epsilon_{i} \sim SN(0,\sigma,\gamma)$, independently. Note that here $\boldsymbol{x_{i}^{T} \beta}$ is not the conditional mean of $y_i$, but is the location parameter of the distribution of $y_i$ given $\boldsymbol{x_i}$. Here, $\boldsymbol{x_{i}}=(x_{i1}, x_{i2}, \ldots,x_{ip})$ is the $p$-dimensional covariate vector corresponding to $i$-th observation and $\boldsymbol{\beta}=(\beta_{1},\beta_{2}, \ldots, \beta_{p})$ is the vector of regression coefficients; $\sigma$ and $\gamma$ are the scale and skewness parameters of the error distribution. In our case, $\boldsymbol{\theta}=(\boldsymbol{\beta},\sigma,\gamma)$ and $f_{i}(y,\boldsymbol{\theta})$ is the density of $SN(\mu_{i}=\boldsymbol{x_{i}^{T}\beta},\sigma,\gamma)$ given by
\begin{equation}
	f_{i}(y,\boldsymbol{\theta})=\frac{2}{\sigma} \phi\left(\frac{y-\boldsymbol{x_{i}^{T} \beta}}{\sigma}\right)\Phi\left(\gamma\frac{y-\boldsymbol{x_{i}^{T} \beta}}{\sigma}\right),
	~~~~ y\in \mathbb{R}.
	\label{EQ:SN_LRM}
\end{equation}

\subsection{The minimum density power divergence estimator (MDPDE)}
\label{SEC:MDPDE}
The density power divergence and the corresponding MDPDE for independent and identically distributed (IID) data was first proposed by Basu et al. \cite{Basu/etc:1998}, which was later extended for non-homogeneous data (independent but not identically distributed) by Ghosh and Basu \cite{Ghosh:2013}, where they considered the standard linear regression model. Later, this technique has been applied to GLM and survival data models \cite{Ghosh/Basu:2016,Ghosh et.al.:2016,ghosh,Ghosh et.al.:2019}. The DPD family \cite{Basu/etc:1998} is indexed by a single tuning parameter  $\alpha\geq 0$, controlling the trade-off between robustness and efficiency of the procedure. For two densities $g$ and $f$, both being absolutely continuous with respect to some common dominating measure $\mu$,
the DPD measure between $g$ and $f$ is defined as
\begin{eqnarray}
	d_{\alpha}(g, f)&=& \int \bigg\{{f^{1+\alpha}}-\left(1+\frac{1}{\alpha}\right)gf^{\alpha}+\frac{1}{\alpha} g^{1+\alpha}\bigg\}, 
	\hspace{1cm} \mbox{ if }\alpha>0,
	\label{EQ:dpd}\\
	d_{0}(g, f)&=& \lim\limits_{\alpha\downarrow 0}d_{\alpha}(g, f)
	=\int g\log\frac{g}{f}.
	\label{EQ:dpd0}
\end{eqnarray}
Note that, the DPD at $\alpha=0$ is nothing but the well-known Kullback-Leiber divergence (KLD) associated with the likelihood approach. Let $Y_1, Y_2, \ldots, Y_n$ represent independent and identically distributed observations drawn from the true distribution $G$ (having pdf $g$). Let ${\cal F} = \{F(\cdot, \boldsymbol{\theta}): \boldsymbol{\theta} \in \Theta \subset {\mathbb R}^p\}$ be the family of distributions modeling it, with $f(\cdot,  \boldsymbol{\theta})$ being the pdf of $F(\cdot, \boldsymbol{\theta})$. Minimization of the divergence $d_\alpha(g, f(\cdot, \boldsymbol{\theta}))$ over the parameter space $\Theta$ generates the ``best fitting" target parameter. As $g$ will always be unknown, we need an empirical estimate of this divergence which may be minimized to obtain the parameter estimate. Note that from (\ref{EQ:dpd}) one gets
\begin{eqnarray}
	d_{\alpha}(g, f(.,\boldsymbol{\theta}))
	= \int f^{1+\alpha}(y,\boldsymbol{\theta}) dy-\left(1+\frac{1}{\alpha}\right) \int f^{\alpha}(y,\boldsymbol{\theta}) dG(y) + K,
	\nonumber
\end{eqnarray}
where the last term $K=\frac{1}{\alpha}\int g^{1+\alpha}(y)d(y)$ is independent of the parameter $\boldsymbol{\theta}$
and has no role in the optimization. Now, the second term in the above equation can be estimated just by plugging in the empirical estimate of $G$, namely the empirical cdf $G_n$ obtained from the sample $Y_{1},\ldots,Y_{n}$, and then the MDPDE can be obtained by minimizing the empirical objective function

\begin{eqnarray}
	H_{n,\alpha}(\boldsymbol{\theta})=\int f^{1+\alpha} (y,\boldsymbol{\theta}) dy 
	- \left(1+\frac{1}{\alpha}\right)\frac{1}{n}\sum_{i=1}^{n} f^{\alpha} (Y_{i},\boldsymbol{\theta}).
	\nonumber
\end{eqnarray}
 In the regression case, $Y_1, Y_2, \ldots, Y_n$ are $n$ observations (realizations of the response variable), where $Y_{i} \sim g_{i}$ independently. Here, we want to model $g_i$ by $f_{i}(.,\boldsymbol{\theta})$, where $f_{i}(.,\boldsymbol{\theta})$ is as defined in (\ref{EQ:SN_LRM}). Although the distribution of each $Y_i$ is distinct, they share the common parameter $\boldsymbol{\theta}$. However, here the model density is different for each $Y_{i}$, so we need to calculate the divergence between data and model separately for each point. Subsequently, we take the average of these individual divergences and we minimize it to estimate $\boldsymbol{\theta}$. So, if $d_{\alpha}(g_{i},f_{i}(.,\boldsymbol{\theta}))$ be the density power divergence measure between the data and model for the $i$-th case, then we minimize the objective function \[\frac{1}{n}\sum_{i=1}^{n}d_{\alpha}(g_{i},f_{i}(.,\boldsymbol{\theta}))\]
with respect to $\boldsymbol{\theta} \in \Theta$. In the presence of only one data point $Y_{i}$ from $g_{i}$, the best possible density estimate of $g_{i}$ is the degenerate density which puts the whole mass on $Y_{i}$. Hence, the divergence between $g_i$ and $f_i(\cdot, \boldsymbol{\theta})$ is estimated as
\begin{equation}
	\int f_{i}^{1+\alpha} (y,\boldsymbol{\theta}) dy-\left(1+\frac{1}{\alpha}\right)f_{i}^{\alpha} (Y_{i},\boldsymbol{\theta})+K
	\label{EQ:2.1}
\end{equation}
where $K$ is independent of $\boldsymbol{\theta}$. Dropping the constant, our entire objective function then becomes
\begin{equation}
	H_{n,\alpha}\left(\boldsymbol{\theta}\right)=\frac{1}{n}\sum_{i=1}^{n}\left[\int f_{i}^{1+\alpha} (y,\boldsymbol{\theta}) dy-\left(1+\frac{1}{\alpha}\right)f_{i}^{\alpha}(Y_{i},\boldsymbol{\theta})\right]=\frac{1}{n} \sum_{i=1}^{n}V_{i}(Y_{i},\boldsymbol{\theta}).
	\label{EQ:2.2}
\end{equation}
Note that, as $\alpha \to 0$, the corresponding objective function, which is to be minimized, becomes
 \begin{equation}
 	\frac{1}{n}\sum_{i=1}^{n}\left[-\ln(f_{i}(Y_{i},\boldsymbol{\theta}))\right].
 	\label{EQ:2.3}
 \end{equation}
Thus the MDPDE with $\alpha=0$ coincides with the maximum likelihood estimator (MLE) which is obtained by solving the score equation $\sum\limits_{i=1}^{n}u_{i}(Y_{i},\boldsymbol{\theta})=\boldsymbol{0}$,
 where $u_{i}(Y_{i},\boldsymbol{\theta})= \frac{\partial \ln(f_{i}(Y_{i},\boldsymbol{\theta}))}{\partial \boldsymbol{\theta}}$ is the score function. In general, the estimating equation for obtaining the MDPDE becomes
 \begin{equation}
 	\sum_{i=1}^{n}\left[f_{i}^{\alpha}(Y_i,\boldsymbol{\theta})u_{i}(Y_{i},\boldsymbol{\theta})-\int f_{i}^{1+\alpha}(y,\boldsymbol{\theta})u_{i}(y,\boldsymbol{\theta}) dy\right]= \boldsymbol{0}.
 	\label{EQ:2.6}
 \end{equation}
 Note that this defines an M-estimator with a model dependent $\psi$ function. In our case, the objective function becomes
\begin{eqnarray}
	H_{n,\alpha}(\boldsymbol{\theta}) 
	&=&\frac{1}{n} \sum_{i=1}^{n}\left[\int_{-\infty}^{\infty}\bigg(\frac{2}{\sigma}\bigg)^{1+\alpha}\phi^{1+\alpha}\bigg(\frac{y-\boldsymbol{x_{i}^{T} \beta}}{\sigma}\bigg)\Phi^{1+\alpha}\bigg(\gamma\frac{y-\boldsymbol{x_{i}^{T} \beta}}{\sigma}\bigg)dy \right.
	\nonumber\\
	&& 
	~~~~~~~~~~~~~~ - \left. \left(1+\frac{1}{\alpha}\right)\bigg(\frac{2}{\sigma}\bigg)^{\alpha}\phi^{\alpha}\bigg(\frac{Y_{i}-\boldsymbol{x_{i}^{T} \beta}}{\sigma}\bigg)\Phi^{\alpha}\left(\gamma\frac{Y_{i}-\boldsymbol{x_{i}^{T} \beta}}{\sigma}\right)\right].
	\label{EQ:2.7}
\end{eqnarray}
By standard differentiation, we get the estimating equations of our MDPDE to be 
 \begin{eqnarray}
 	\sum_{i=1}^{n} \boldsymbol{u}_{i}(Y_{i},\boldsymbol{\theta})\bigg(\frac{2}{\sigma}\bigg)^{\alpha}
 	\phi^{\alpha}\bigg(\frac{Y_{i}-\boldsymbol{x_{i}^{T} \beta}}{\sigma}\bigg)\Phi^{\alpha}\bigg(\gamma\frac{Y_{i}-\boldsymbol{x_{i}^{T} \beta}}{\sigma}\bigg)
 	=\sum_{i=1}^{n}\boldsymbol{\xi}_{i,\alpha}(\boldsymbol{\theta}),
 	\label{EQ:2.8}
 \end{eqnarray}
 where $\boldsymbol{u}_{i}(y,\boldsymbol{\theta})$
 is the score function of the SN distribution for the $i$-th case and has the form
 \begin{eqnarray}
 	\fontsize{9pt}{2pt}
 	\boldsymbol{u}_{i}(y,\boldsymbol{\theta})
 	=\left(\left(\frac{y-\boldsymbol{x_{i}^{T}\beta}}{\sigma^{2}}-\frac{\gamma\phi\left(\gamma\frac{y-\boldsymbol{x_{i}^{T}\beta}}{\sigma}\right)}{\sigma\Phi\left(\gamma\frac{y-\boldsymbol{x_{i}^{T}\beta}}{\sigma}\right)}\right)\boldsymbol{x_{i}},\frac{(y-\boldsymbol{x_{i}^{T}\beta})^{2}}{\sigma^{3}}-\gamma\frac{(y-\boldsymbol{x_{i}^{T}\beta})}{\sigma^{2}}\frac{\phi\left(\gamma\frac{y-\boldsymbol{x_{i}^{T}\beta}}{\sigma}\right)}{\Phi\left(\gamma\frac{y-\boldsymbol{x_{i}^{T}\beta}}{\sigma}\right)}-\frac{1}{\sigma},\frac{(y-\boldsymbol{x_{i}^{T}\beta})}{\sigma}\frac{\phi\left(\gamma\frac{y-\boldsymbol{x_{i}^{T}\beta}}{\sigma}\right)}{\Phi\left(\gamma\frac{y-\boldsymbol{x_{i}^{T} \beta}}{\sigma}\right)}\right)^T,
 	\nonumber
 \end{eqnarray}
 and 
 \begin{eqnarray}
 	\boldsymbol{\xi}_{i,\alpha}(\boldsymbol{\theta})
 	=\int \boldsymbol{u}_{i}(y,\boldsymbol{\theta})f_{i}^{1+\alpha}(y,\boldsymbol{\theta})dy
 	=\left(\xi_{i,\alpha}^{(1)}(\boldsymbol{\theta})~~
 	\xi_{i,\alpha}^{(2)}(\boldsymbol{\theta})~~
 	\xi_{i,\alpha}^{(3)}(\boldsymbol{\theta})\right)^T.
 	\label{EQ:2.10}
 \end{eqnarray}
 Clearly, the estimating equation presented in (\ref{EQ:2.8}) is very complicated and the involved integral to be computed numerically. Here, we have used the R function \textsf{optim} to minimize $H_{n,\alpha}(\boldsymbol{\theta})$.
 
 \subsection{The asymptotic distribution and the standard error of the MDPDE}
 \label{SEC:Asymptotic MDPDE}
 The asymptotic distribution of the MDPDE for the non-homogeneous case was developed by Ghosh and Basu (2013) \cite{Ghosh:2013}. They demonstrated that, under some regularity conditions, the minimum DPD estimators are $\sqrt{n}$-consistent and asymptotically normal. We have used the same theory for our present case. At a given $\alpha\geq 0$, if the corresponding MDPDE  based on a random sample of size $n$ 
 is denoted by $\widehat{\boldsymbol{\theta}}_{\alpha,n}$, and the true parameter value is $\boldsymbol{\theta}_0$, 
 we have 
 $$
 \sqrt{n}~ \boldsymbol{\Omega_{n}}^{-\frac{1}{2}} ~\boldsymbol{\Psi_{n}} \left(\widehat{\boldsymbol{\theta}}_{\alpha,n} - \boldsymbol{\theta}_0\right)
 \mathop{\rightarrow}^\mathcal{D} N_{p+2}\left(\boldsymbol{0}_{p+2}, \boldsymbol{I}_{p+2}
 \right),
 $$
 where $\boldsymbol {I}_b$ is the identity matrix of dimension $b$, $\boldsymbol {0}_b$ ia a $b$-dimensional vector with all entries zero,  
 $\boldsymbol{\Psi_{n}}=\frac{1}{n}\sum_{i=1}^{n}\boldsymbol{J}_{\alpha}^{(i)}(\boldsymbol{\theta})$, $\boldsymbol{\Omega_{n}}=\frac{1}{n}\sum_{i=1}^{n}Var[\nabla V_{i}(Y_{i},\boldsymbol{\theta})]=\frac{1}{n}\sum_{i=1}^{n}\boldsymbol{K}_{\alpha}^{(i)}(\boldsymbol{\theta})$ (say), and $\nabla$ represents derivative with respect to $\boldsymbol{\theta}$. Here, for SN distributions with $\boldsymbol{\theta}=(\boldsymbol{\beta}, \sigma, \gamma)^T$, the $(p+2)\times (p+2)$ matrices $\boldsymbol{K}_{\alpha}^{(i)}(\boldsymbol{\theta})$ 
 and $\boldsymbol{J}_{\alpha}^{(i)}(\boldsymbol{\theta})$ are given by

 \begin{eqnarray}
 \boldsymbol{J}_{\alpha}^{(i)}(\boldsymbol{\theta})
 &=&\int \boldsymbol{u}_{i}(y,\boldsymbol{\theta})\boldsymbol{u}_{i}^{T}(y,\boldsymbol{\theta})f_{i}^{1+\alpha}(y,\boldsymbol{\theta})dy
 =\begin{pmatrix}
 \begin{array}{ccc}
 N_{i,\alpha}^{(11)}(\boldsymbol{\theta})  & N_{i,\alpha}^{(12)}(\boldsymbol{\theta}) & N_{i,\alpha}^{(13)}(\boldsymbol{\theta})\\
 N_{i,\alpha}^{(12)}(\boldsymbol{\theta})  & N_{i,\alpha}^{(22)}(\boldsymbol{\theta}) & N_{i,\alpha}^{(23)}(\boldsymbol{\theta})\\
 N_{i,\alpha}^{(13)}(\boldsymbol{\theta})  & N_{i,\alpha}^{(23)}(\boldsymbol{\theta}) & N_{i,\alpha}^{(33)}(\boldsymbol{\theta})
 \end{array}
 \end{pmatrix},
 \label{EQ:J}
 \\
 ~~~~~
 \boldsymbol{K}_{\alpha}^{(i)}(\boldsymbol{\theta})
 &=&\int \boldsymbol{u}_{i}(y,\boldsymbol{\theta})\boldsymbol{u}_{i}^{T}(y,\boldsymbol{\theta})f_{i}^{1+2\alpha}(y,\boldsymbol{\theta})dy
 - \boldsymbol{\xi}_{i,\alpha}(\boldsymbol{\theta})\boldsymbol{\xi}_{i,\alpha}(\boldsymbol{\theta})^{T}
 \nonumber\\
 &=&\begin{pmatrix}
 \begin{array}{ccc}
 N_{i,2\alpha}^{(11)}(\boldsymbol{\theta}) - \xi_{i,\alpha}^{(1)}(\boldsymbol{\theta}) \xi_{i,\alpha}^{(1)}(\boldsymbol{\theta})^{T}
 & N_{i,2\alpha}^{(12)}(\boldsymbol{\theta}) - \xi_{i,\alpha}^{(1)}(\boldsymbol{\theta})\xi_{i,\alpha}^{(2)}(\boldsymbol{\theta})
 & N_{i,2\alpha}^{(13)}(\boldsymbol{\theta}) - \xi_{i,\alpha}^{(1)}(\boldsymbol{\theta})\xi_{i,\alpha}^{(3)}(\boldsymbol{\theta})
 \\
 N_{i,2\alpha}^{(12)}(\boldsymbol{\theta}) -\xi_{i,\alpha}^{(1)}(\boldsymbol{\theta})\xi_{i,\alpha}^{(2)}(\boldsymbol{\theta})
 & N_{i,2\alpha}^{(22)}(\boldsymbol{\theta}) - \xi_{i,\alpha}^{(2)}(\boldsymbol{\theta})^2
 & N_{i,2\alpha}^{(23)}(\boldsymbol{\theta}) - \xi_{i,\alpha}^{(2)}(\boldsymbol{\theta})\xi_{i,\alpha}^{(3)}(\boldsymbol{\theta})
 \\
 N_{i,2\alpha}^{(13)}(\boldsymbol{\theta})  -\xi_{i,\alpha}^{(1)}(\boldsymbol{\theta})\xi_{i,\alpha}^{(3)}(\boldsymbol{\theta})
 & N_{i,2\alpha}^{(23)}(\boldsymbol{\theta}) -\xi_{i,\alpha}^{(2)}(\boldsymbol{\theta})\xi_{i,\alpha}^{(3)}(\boldsymbol{\theta})
 & N_{i,2\alpha}^{(33)}(\boldsymbol{\theta}) - \xi_{i,\alpha}^{(3)}(\boldsymbol{\theta})^2
 \end{array}
 \end{pmatrix},~~~~~~~~
 \label{EQ:K}
 \end{eqnarray}
 with 
 $\boldsymbol{\xi}_{i,\alpha}(\boldsymbol{\theta})$ and $f_{i}(y,\boldsymbol{\theta})$ are as defined in (\ref{EQ:2.10})
 and (\ref{EQ:SN_LRM}), respectively, 
 and 
 \begin{eqnarray}
 N_{i,\alpha}^{(11)}(\boldsymbol{\theta})&=&\int_{-\infty}^{\infty}\Bigg(\bigg(\frac{y-\mu_i}{\sigma^{2}}\bigg)-\frac{\gamma}{\sigma}\frac{\phi\big(\gamma\frac{y-\mu_i}{\sigma}\big)}{\Phi\big(\gamma\frac{y-\mu_i}{\sigma}\big)}\Bigg)^{2}\boldsymbol{x_i}\boldsymbol{x_i}^{T}f_{i}^{1+\alpha}\big(y,\boldsymbol{\theta}\big)dy
 \nonumber\\
 N_{i,\alpha}^{(22)}(\boldsymbol{\theta})&=&\int_{-\infty}^{\infty}\Bigg(\frac{\big(y-\mu_i\big)^{2}}{\sigma^{3}}-\frac{1}{\sigma}-\frac{\gamma\big(y-\mu_i\big)}{\sigma^{2}}\frac{\phi\big(\gamma\frac{y-\mu_i}{\sigma}\big)}{\Phi\big(\gamma\frac{y-\mu_i}{\sigma}\big)}\Bigg)^{2}f_{i}^{1+\alpha}\big(y,\boldsymbol{\theta}\big)dy
 \nonumber\\
 N_{i,\alpha}^{(33)}(\boldsymbol{\theta})&=&\int_{-\infty}^{\infty}\Bigg(\frac{y-\mu_i}{\sigma}\frac{\phi\big(\gamma\frac{y-\mu_i}{\sigma}\big)}{\Phi\big(\gamma\frac{y-\mu_i}{\sigma}\big)}\Bigg)^{2}f_{i}^{1+\alpha}\big(y,\boldsymbol{\theta}\big)dy
 \nonumber\\
 N_{i,\alpha}^{(12)}(\boldsymbol{\theta})&=&\int_{-\infty}^{\infty}\Bigg(\bigg(\frac{y-\mu_i}{\sigma^{2}}\bigg)-\frac{\gamma}{\sigma}\frac{\phi\big(\gamma\frac{y-\mu_i}{\sigma}\big)}{\Phi\big(\gamma\frac{y-\mu_i}{\sigma}\big)}\Bigg)\Bigg(\frac{\big(y-\mu_i\big)^{2}}{\sigma^{3}}-\frac{1}{\sigma}-\frac{\gamma\big(y-\mu_i\big)}{\sigma^{2}}\frac{\phi\big(\gamma\frac{y-\mu_i}{\sigma}\big)}{\Phi\big(\gamma\frac{y-\mu_i}{\sigma}\big)}\Bigg) \boldsymbol{x_i}f_{i}^{1+\alpha}\big(y,\boldsymbol{\theta}\big)dy
 \nonumber\\
 N_{i,\alpha}^{(13)}(\boldsymbol{\theta})&=&\int_{-\infty}^{\infty}\Bigg(\bigg(\frac{y-\mu_i}{\sigma^{2}}\bigg)-\frac{\gamma}{\sigma}\frac{\phi\big(\gamma\frac{y-\mu_i}{\sigma}\big)}{\Phi\big(\gamma\frac{y-\mu_i}{\sigma}\big)}\Bigg)\Bigg(\frac{y-\mu_i}{\sigma}\frac{\phi\big(\gamma\frac{y-\mu_i}{\sigma}\big)}{\Phi\big(\gamma\frac{y-\mu_i}{\sigma}\big)}\Bigg)\boldsymbol{x_i}f_{i}^{1+\alpha}\big(y,\boldsymbol{\theta}\big)dy
 \nonumber\\
 N_{i,\alpha}^{(23)}(\boldsymbol{\theta})&=&\int_{-\infty}^{\infty}\Bigg(\frac{\big(y-\mu_i\big)^{2}}{\sigma^{3}}-\frac{1}{\sigma}-\frac{\gamma\big(y-\mu_i\big)}{\sigma^{2}}\frac{\phi\big(\gamma\frac{y-\mu_i}{\sigma}\big)}{\Phi\big(\gamma\frac{y-\mu_i}{\sigma}\big)}\Bigg)\Bigg(\frac{y-\mu_i}{\sigma}\frac{\phi\big(\gamma\frac{y-\mu_i}{\sigma}\big)}{\Phi\big(\gamma\frac{y-\mu_i}{\sigma}\big)}\Bigg)f_{i}^{1+\alpha}\big(y,\boldsymbol{\theta}\big)dy
 \nonumber
 \end{eqnarray}
where $\mu_{i}=\boldsymbol{x_{i}^{T}\beta}$. Here also we can see that the above integrals do not have closed forms, so we must numerically compute them to obtain the asymptotic covariance matrix at different given values of $\alpha$ and $\boldsymbol{\theta}$. 
 
The above asymptotic variance formula can also help us to obtain the standard errors of the MDPDEs
in any practical application. For the MDPDE $\widehat{\boldsymbol{\theta}}_{\alpha,n}=(\widehat{\boldsymbol{\beta}}_{\alpha,n}, 
~\widehat{\sigma}_{\alpha,n},~ \widehat{\gamma}_{\alpha,n})^T$, obtained on the basis of a sample of size $n$, 
its standard errors are obtained as $\sqrt{\text{diag}(\boldsymbol{\Sigma}_{\alpha,n}(\widehat{\boldsymbol{\theta}}_{\alpha,n}))/n}$,
where $\boldsymbol{\Sigma}_{\alpha,n}(\widehat{\boldsymbol{\theta}}_{\alpha,n})$ is consistent for $\boldsymbol{\Sigma}_{\alpha}(\boldsymbol{\theta}_{0})=\boldsymbol{\Psi_{n}}^{-1}\boldsymbol{\Omega_{n}}\boldsymbol{\Psi_{n}}^{-1}.$  

\subsection{Influence function analysis of the MDPDE}
\label{SEC:MDPDE_IF}

The robustness of an estimator can be theoretically examined through the classical influence function (IF) analysis
\cite{Hampel/etc:1986}. The IF measures the asymptotic (standardized)  bias of the estimator caused by an infinitesimal
contamination at a distant contamination point (say $y$).
Therefore, the boundedness of the IF over the contamination point $y$ restricts the extent of possible bias finitely 
for  the corresponding estimator indicating its robust nature.
On the other hand, an unbounded IF indicates possibly unbounded bias and non-robustness of the estimator. Further, with similar intuition, the supremum of the absolute IF taken over all possible contamination points
naturally indicates the extent of (bias) robustness of the corresponding estimator.

From the theory of the M-estimators \cite{Hampel/etc:1986} or the general MDPDE \cite{Basu/etc:1998,Basu/etc:2011}, 
one can obtain the influence function of the MDPDE functional under the IID set up. Ghosh and Basu \cite{Ghosh:2013} have studied the influence functions for the non-homogeneous set up \cite{Ghosh:2013}. We will present an influence function analysis along these lines for the estimators of our skew normal linear regression model.

Let, $G_{i}$ denotes the true distribution of $Y_{i}$ and $\boldsymbol{T_{\alpha}}(G_{1}, G_{2}, \ldots,G_{n})$ be the MDPDE functional defined as the minimizer of (\ref{EQ:2.2}). Now, we consider the contaminated density $g_{i,\epsilon}= (1-\epsilon)g_{i}+ \epsilon \delta_{y_{i}}$, where $\delta_{y_{i}}$ is the degenerate distribution at the point of contamination $y_{i}$ and let $G_{i, \epsilon}$ be the corresponding distribution function, $i = 1, 2, \ldots, n$. Let $\boldsymbol{\theta}=\boldsymbol{T_{\alpha}}(G_{1}, G_{2}, \ldots, G_{n})$ and let $\boldsymbol{\theta}_{\epsilon}^{i_{0}}= \boldsymbol{T_{\alpha}}(G_{1}, \ldots, G_{i_{0}-1}, G_{i_{0},\epsilon}, \ldots, G_{n})$ be the minimum density power divergence functional with contamination only in the $i_{0}$-th direction, where $G_{i_{0},\epsilon}=(1-\epsilon)G_{i_{0}}+ \epsilon \Delta_{y_{i_{0}}}$ with $\Delta_{y_{i_{0}}}$ being the distribution \textcolor{black}{of a degenerate random variable with point mass at $y_{i_{0}}$ and density} $\delta_{y_{i_{0}}}$. Then, along the lines of  Ghosh and Basu \cite{Ghosh:2013}, we can get the influence function of the functional with contamination only in $i_0$-th direction as follows:
\begin{eqnarray}
	IF_{i_{0}}(y_{i_{0}},T_{\alpha},\boldsymbol{\theta})
	= \boldsymbol{\Psi_{n}}^{-1} \frac{1}{n}
	\left[\boldsymbol{u}_{i_{0}}(y_{i_{0}},\boldsymbol{\theta})f_{i_{0}}^{\alpha}(y_{i_{0}},\boldsymbol{\theta}) -\boldsymbol{\xi}_{i_{0},\alpha}(\boldsymbol{\theta})
	\right],
	\label{EQ:MDPDE_IFi0}
\end{eqnarray}
where $\boldsymbol{\xi}_{i,\alpha}(\boldsymbol{\theta})$ and $\boldsymbol{\Psi_{n}}$ are as defined in Section \ref{SEC:Asymptotic MDPDE}. On the other hand, if we consider contamination in all directions, the form of the influence function turns out to be
\begin{eqnarray}
	IF(y_{1}, \ldots, y_{n},T_{\alpha},\boldsymbol{\theta})
	= \boldsymbol{\Psi_{n}}^{-1} \frac{1}{n} \sum_{i=1}^{n}
	\left[\boldsymbol{u}_{i}(y_{i},\boldsymbol{\theta})f_{i}^{\alpha}(y_{i},\boldsymbol{\theta}) -\boldsymbol{\xi}_{i,\alpha}(\boldsymbol{\theta})
	\right].
	\label{EQ:MDPDE_IFall}
\end{eqnarray}

\subsection{Illustration for a simple set-up}
\label{SEC:Asymptotic_variance_special}

Here we have considered a linear model with a single covariate. Our model is $y_{i}=x_{i}\beta+\epsilon_{i}, ~i=1, 2, \ldots, n$, where $\epsilon_{i} \sim SN(0,\sigma,\gamma)$ and $x$ is the covariate. Thus, $y_{i}$'s are distributed as $SN(\mu_{i}=x_{i}\beta,\sigma,\gamma)$, independently of one another. Using the formulas discussed in Section \ref{SEC:Asymptotic MDPDE}, we can find the asymptotic relative efficiency (ARE) of the proposed MDPDE which are presented in Table \ref{TAB:ARE} for different values of $\boldsymbol{\theta}=(\beta,\sigma,\gamma)^T$ for the SN linear regression model. 
We have considered three different SN distributions for the random error component.
From Table \ref{TAB:ARE}, we observe that ARE of $\beta$ and $\gamma$ are same when we take the error distribution to be $SN(0,1,0)$ and $SN(0,4,0)$, i.e., we take the estimate of $\gamma=0$. This is not an accident and we will show later that the relevant estimated quantities are functionally identical in terms of the estimates of $\sigma$ and $\beta$ when $\gamma=0$.

\begin{table}[h]
	\caption{Asymptotic Relative Efficiency of the MDPDEs of $(\beta,\sigma,\gamma)^T$ for different SN error distributions. The $\alpha=0$ case corresponds to the MLE}
	\centering	
		\begin{tabular}{c c|c c c c c c}
			\hline
			Error & & & &  $\alpha$ & &  &\\
			Distribution & & 0(MLE) &  0.1 & 0.3 & 0.5 & 0.7 & 1\\
			\hline
			$SN(0,1,-2)$ & $\beta$ & 100 & 98.66 & 91.30 & 82.25 & 73.68 & 62.75\\
			& $\sigma$ & 100 & 97.52 & 85.11 & 71.84 & 60.94 & 48.72\\
			& $\gamma$ & 100 & 98.18 & 86.76 & 71.81 & 58.08 & 42.35\\
			\hline
			$SN(0,1,0)$ & $\beta$ & 100 & 98.75 & 92.03 & 83.57 & 75.30 & 64.45\\
			& $\sigma$ & 100 & 97.49 & 85.38 & 72.79 & 63.01 & 53.41\\
			& $\gamma$ & 100 & 98.75 & 92.03 & 83.57 & 75.30 & 64.45\\
			\hline
			$SN(0,4,0)$ & $\beta$ & 100 & 98.75 & 92.03 & 83.57 & 75.30 & 64.45\\
			& $\sigma$ & 100 & 97.49 & 85.38 & 72.79 & 63.01 & 53.41\\
			& $\gamma$ & 100 & 98.75 & 92.03 & 83.57 & 75.30 & 64.45\\
			\hline 
			$SN(0,1,2)$ & $\beta$ & 100 & 98.63 & 91.11 & 81.85 & 73.12 & 62.04\\
			& $\sigma$ & 100 & 97.98 & 86.90 & 74.08 & 63.13 & 50.68 \\
			& $\gamma$ & 100 & 97.90 & 85.13 & 69.21 & 55.36 & 40.15 \\
			\hline
		\end{tabular}
	\label{TAB:ARE}
\end{table}

\noindent
In our example,  the score function becomes
\begin{eqnarray}
	\fontsize{9pt}{2pt}
	\boldsymbol{u}_{i}(y,\boldsymbol{\theta})
	=\left(\left(\frac{y-\mu_i}{\sigma^{2}}-\frac{\gamma\phi\left(\gamma\frac{y-\mu_i}{\sigma}\right)}{\sigma\Phi\left(\gamma\frac{y-\mu_i}{\sigma}\right)}\right)x_{i},\frac{(y-\mu_i)^{2}}{\sigma^{3}}-\gamma\frac{(y-\mu_i)}{\sigma^{2}}\frac{\phi\left(\gamma\frac{y-\mu_i}{\sigma}\right)}{\Phi\left(\gamma\frac{y-\mu_i}{\sigma}\right)}-\frac{1}{\sigma},\frac{(y-\mu_i)}{\sigma}\frac{\phi\left(\gamma\frac{y-\mu_i}{\sigma}\right)}{\Phi\left(\gamma\frac{y-\mu_i}{\sigma}\right)}\right)^T,
	\label{EQ:Score_special}
\end{eqnarray}
where $\mu_{i}=x_{i}\beta$, $\boldsymbol{\theta}=(\beta,\sigma,\gamma)$. When the true value of  $\gamma$ is $0$, the integrals within the matrices $\boldsymbol{\Omega_{n}}$ and $\boldsymbol{\Psi_{n}}$ become simpler and can be calculated easily. In this case, we obtain, after some algebra, the final form of $\boldsymbol{\Psi_{n}}$ and $\boldsymbol{\Omega_{n}}$ as:
\begin{eqnarray}
	\boldsymbol{\Psi}_{n}
	&=\sigma^{-\alpha}\begin{pmatrix}
		\begin{array}{ccc}
			\psi_{11,\alpha}  & 0 & \psi_{13,\alpha}\\
			0& \psi_{22,\alpha} & 0\\
			\psi_{31,\alpha} & 0 & \psi_{33,\alpha}
		\end{array}
	\end{pmatrix},
	\label{EQ:Psi_special}
	\\
	~~~~~
	\boldsymbol{\Omega}_{n}
	&=\sigma^{-2\alpha}\begin{pmatrix}
		\begin{array}{ccc}
			\omega_{11,\alpha}
			& 0
			& \omega_{13,\alpha}
			\\
			0
			& \omega_{22,\alpha}
			& 0
			\\
			\omega_{31,\alpha}
			& 0
			& \omega_{33,\alpha}
		\end{array}
	\end{pmatrix},~~~~~~~~
	\label{EQ:Omega_special}
\end{eqnarray}
where

\begin{eqnarray}
	\psi_{11,\alpha}&=&\phi(\alpha)~\frac{\sum_{i=1}^{n}x_{i}^{2}}{n},   ~~~~~~~~ \psi_{13,\alpha}=\psi_{31,\alpha}=  \phi(\alpha)~\sqrt{\frac{2}{\pi}} \bar{x},
	\nonumber\\
	\psi_{22,\alpha}&=&  \phi(\alpha)~\frac{\alpha^{2}+2}{\alpha+1},   ~~~~~~~~ \psi_{33,\alpha}= \phi(\alpha)~\frac{2}{\pi},~\phi(\alpha)= (2\pi)^{-\frac{\alpha}{2}}~(1+\alpha)^{-\frac{3}{2}},
	\nonumber\\
	\omega_{11,\alpha}&=&  (2\pi)^{-\alpha}~(1+2\alpha)^{-\frac{3}{2}}~\frac{\sum_{i=1}^{n}x_{i}^{2}}{n},
	\nonumber\\
	\omega_{13,\alpha}&=&\omega_{31,\alpha}	=~ (2\pi)^{-2\alpha}(1+2\alpha)^{-\frac{3}{2}}\sqrt{\frac{2}{\pi}}\bar{x},\nonumber\\
	\omega_{22,\alpha}&=& (2\pi)^{-\alpha}\left[(1+2\alpha)^{-\frac{5}{2}}(4\alpha^{2}+2)-\alpha^{2}(1+\alpha)^{-3}\right],
	\nonumber\\
	\omega_{33,\alpha}&=& (2\pi)^{-\alpha}~\frac{2}{\pi}(1+2\alpha)^{-\frac{3}{2}}.
	\nonumber
\end{eqnarray}


Now, after some calculations, the asymptotic variance matrix becomes 
\begin{eqnarray}
	\boldsymbol{\Sigma}_{\alpha}(\boldsymbol{\theta})
	&=\begin{pmatrix}
		\begin{array}{ccc}
			\sigma_{11,\alpha}
			& 0
			& \sigma_{13,\alpha}
			\\
			0
			& \sigma_{22,\alpha}
			& 0
			\\
			\sigma_{31,\alpha}
			& 0
			& \sigma_{33,\alpha}
		\end{array}
	\end{pmatrix},~~~~~~~~
	\label{EQ:Sigma_special}
\end{eqnarray}
where
\begin{eqnarray}
	\sigma_{11,\alpha}&=&\sigma(\alpha,\boldsymbol{x})~\frac{2}{\pi}(1+2\alpha)^{-\frac{3}{2}} \left(\frac{\alpha^{2}+2}{\alpha+1}\right)^{2},
	\nonumber\\
	\sigma_{13,\alpha}&=&\sigma_{31,\alpha}= -\sigma(\alpha,\boldsymbol{x})~\sqrt{\frac{2}{\pi}} \bar{x}(1+2\alpha)^{-\frac{3}{2}}\left(\frac{\alpha^{2}+2}{1+\alpha}\right)^{2},
	\nonumber\\
	\sigma_{22,\alpha}&=& \sigma(\alpha,\boldsymbol{x})~\frac{2}{\pi}s_{xx}\left[(1+2\alpha)^{-\frac{5}{2}}(4\alpha^{2}+2)-\alpha^{2}(1+\alpha)^{-3}\right],
	\nonumber\\
	\sigma_{33,\alpha}&=&\sigma(\alpha,\boldsymbol{x})~\frac{2}{\pi}(1+2\alpha)^{-\frac{3}{2}} \left(\frac{\alpha^{2}+2}{\alpha+1}\right)^{2},
	\nonumber\\
	\sigma(\alpha,\boldsymbol{x})&=& \frac{(2\pi)^{-\alpha}}{s_{xx}\phi^{2}(\alpha)\frac{2}{\pi}\left(\frac{\alpha^{2}+2}{1+\alpha}\right)^{2}},
	\nonumber\\
	\phi(\alpha)&=& (2\pi)^{-\frac{\alpha}{2}}~(1+\alpha)^{-\frac{3}{2}},~s_{xx}=\frac{1}{n}\sum_{i=1}^{n}(x_{i}-\bar{x})^{2},~\bar{x}=\frac{1}{n}\sum_{i=1}^{n}x_{i}.
	\nonumber
\end{eqnarray}
From the expressions above, it may be seen that the forms of the asymptotic variances at the estimated $\beta$ and $\gamma$ are functionally the same whenever $\gamma=0$.
This is clearly an interesting new observation in the context of the asymptotic distributions of the MDPDEs.\\

\begin{figure}[!h]
	\centering
		\subfloat[IF for $\beta$ in $i_{0}=10$th direction]{
			\includegraphics[width=0.5\textwidth]{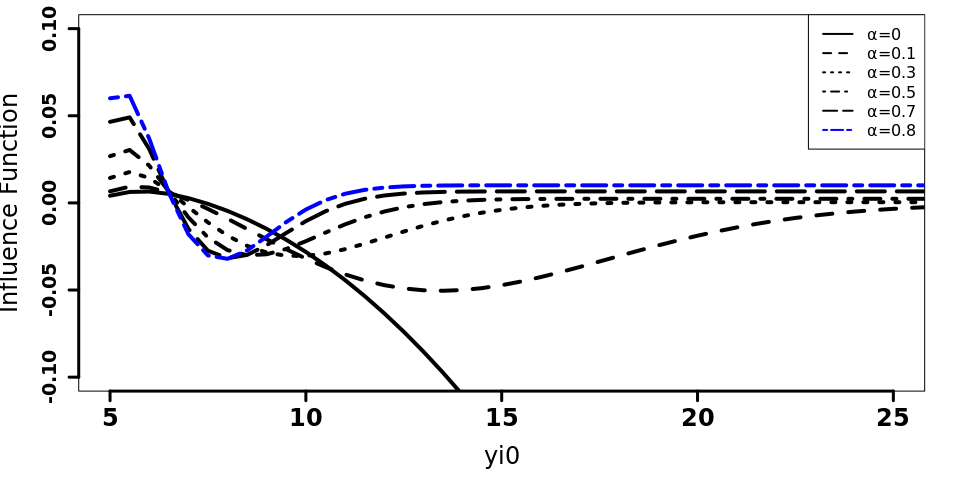}
			\label{FIG:IF_beta_i0_latest1}}
				\subfloat[IF for $\beta$ in all directions]{
			\includegraphics[width=0.5\textwidth]{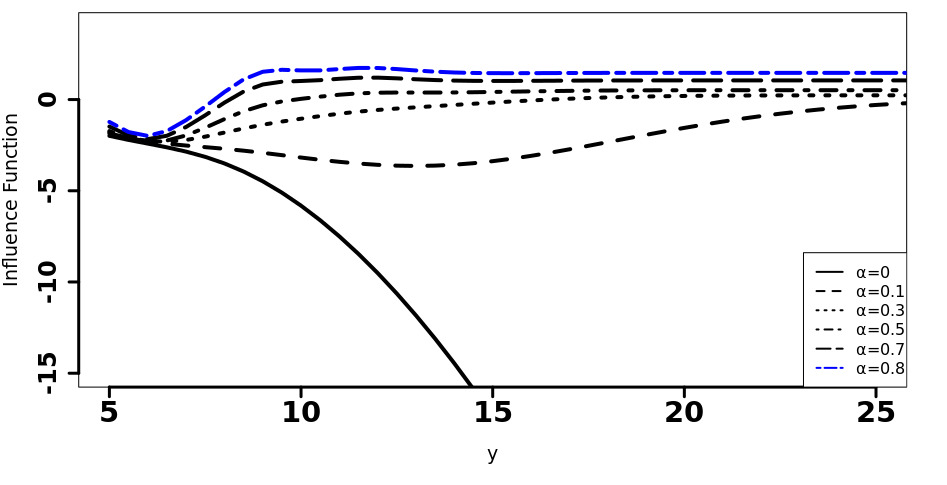}
			\label{FIG:IF_beta_all_latest1}}\\
		\subfloat[IF for $\sigma$ in $i_{0}=10$th direction]{
			\includegraphics[width=0.5\textwidth]{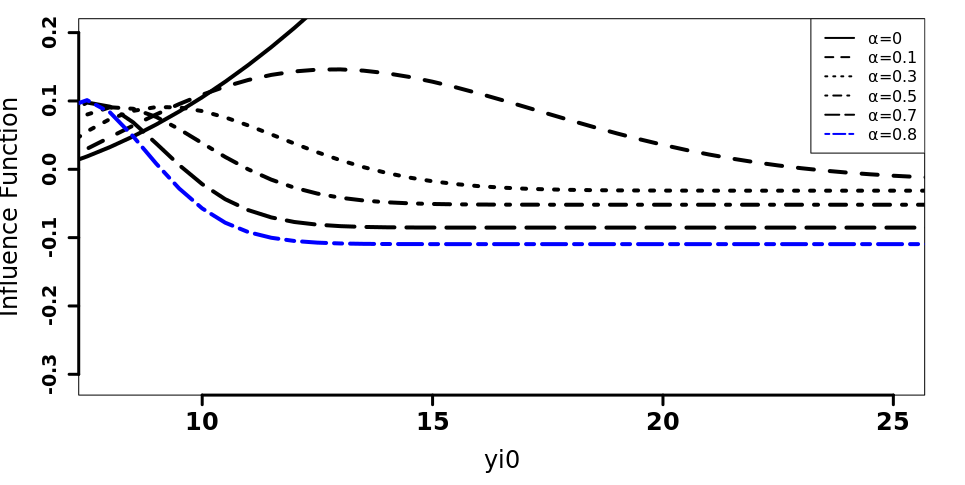}
			\label{FIG:IF_sigma_i0_latest1}}
				\subfloat[IF for $\sigma$ in all directions]{
			\includegraphics[width=0.5\textwidth]{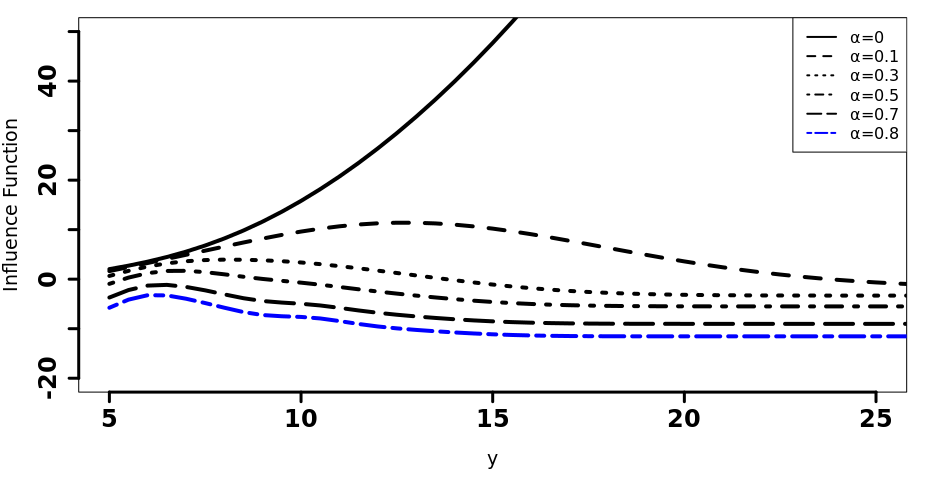}
			\label{FIG:IF_sigma_all_latest1}}\\
		\subfloat[IF for $\gamma$ in $i_{0}=10$th direction]{
			\includegraphics[width=0.5\textwidth]{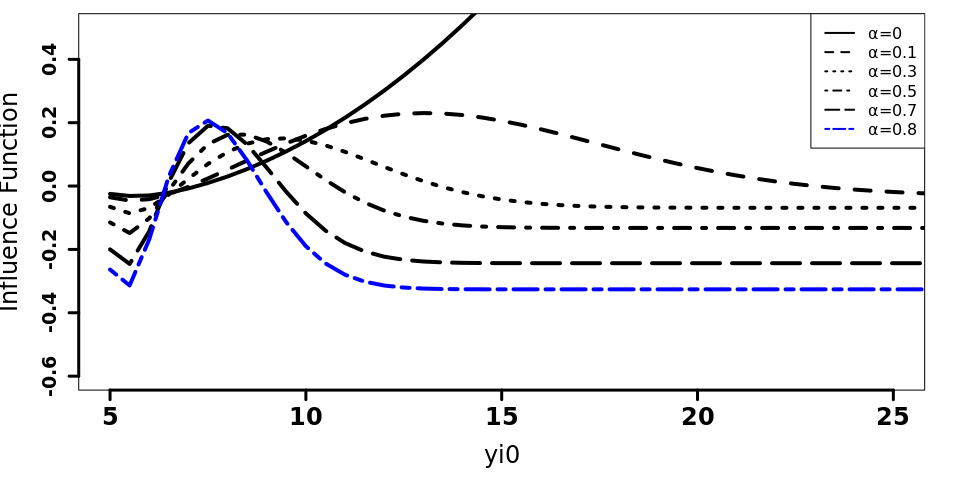}
			\label{FIG:IF_gamma_i0_latest1}}
				\subfloat[IF for $\gamma$ in all directions]{
			\includegraphics[width=0.5\textwidth]{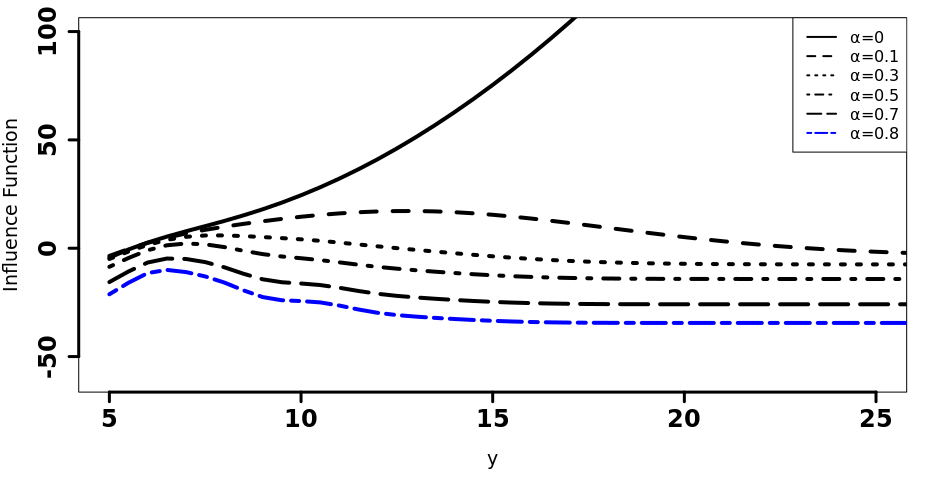}
			\label{FIG:IF_gamma_all_latest1}}
	\caption{Influence functions (IFs) of the MDPDEs of the parameters in $i_{0}=10$th direction (left panel) and in all directions (right panel), with $\boldsymbol{\theta}=(3,2,2)$ for different $\alpha$.}
	\label{FIG:MDPDE_IF_particular_latest1}
	\label{FIG:MDPDE_IF_all_latest1}
\end{figure}

Further, the IFs of the three parameters $\alpha$, $\beta$ and $\gamma$ are plotted in Figures \ref{FIG:IF_beta_i0_latest1} -- \ref{FIG:IF_gamma_all_latest1}. It may be clearly observed that these influence functions are bounded for $\alpha > 0$. For $\alpha = 0$, on the other hand, the function is unbounded for all the three parameters, indicating the robustness deficiencies of the maximum likelihood estimator.

%
%

\section{Robust Wald-type tests for the linear regression model with skew-normal errors}
 \label{SEC:Test}
 
 \subsection{General theory for composite hypotheses}
 \label{SEC:Test_Gen}
 
 We now consider the problem of testing the general composite null hypothesis
 given by%
 \begin{equation}
 	H_{0}:\boldsymbol{\theta}\in\Theta_{0}~\text{against}~H_{1}:\boldsymbol{\theta
 	}\notin\Theta_{0},\label{EQ:Hyp_Comp}%
 \end{equation}
 where $\Theta_{0}$ is a given subset of the parameter space $\Theta\in
 \mathbb{R}^{p+2}$. This null parameter space $\Theta_{0}$ is often
 defined by a set of $r$ restrictions of the form
 \begin{equation}
 	\boldsymbol{m}(\boldsymbol{\theta})=\boldsymbol{0}_{r},\label{2.8}%
 \end{equation}
 where $\boldsymbol{m}:\mathbb{R}^{p+2}\rightarrow\mathbb{R}^{r}$ is a known function with $r\leq
 p+2$. So $\Theta_{0}=\{$$\boldsymbol{\theta}$ $\in\Theta:$ $\boldsymbol{m}%
 ($$\boldsymbol{\theta}$$)=\boldsymbol{0}_{r}\}$. Assume that the $\left(
 p+2\right)  \times r$ matrix
 \begin{equation}
 	\boldsymbol{M}(\boldsymbol{\theta})=\frac{\partial\boldsymbol{m}%
 		^{T}(\boldsymbol{\theta})}{\partial\boldsymbol{\theta}}\label{2.9}%
 \end{equation}
 exists, is continuous in $\boldsymbol{\theta}$, and $\mathrm{rank}\left(
 \boldsymbol{M}(\boldsymbol{\theta})\right)=r$. An important example is the test of significance for the regression model with the null hypothesis $H_{0}:\boldsymbol{\beta}=\boldsymbol{0}%
 _{p}$ indicating no covariate effect; this is a special case of (\ref{EQ:Hyp_Comp}) with  $\boldsymbol{m}(\boldsymbol{\theta})=\boldsymbol{\beta}$ so that $\Theta_{0}=\{ \boldsymbol{\theta
 }$$\in\Theta:$ $\boldsymbol{m}(\boldsymbol{\theta
 })=\boldsymbol{\beta}=\boldsymbol{0}_{p}\}$ and $\boldsymbol{M}=[\boldsymbol{I}_p;O]^{T}$ with $r=p$, where $O$ is the null matrix of order $2 \times p$.
 
Now, the Wald-type test statistic from the MDPDE that utilizes the consistent estimator of their asymptotic variance matrix is given by
 \begin{equation}
 		W_{n}\left(\widehat{\boldsymbol{\theta}}_{\alpha,n}\right)=n\boldsymbol{m}^{T}%
 		\left(\widehat{\boldsymbol{\theta}}_{\alpha,n}\right)\left(  \boldsymbol{M}%
 		^{T}\left(\widehat{\boldsymbol{\theta}}_{\alpha,n}\right)\boldsymbol{\Sigma
 		}_{\alpha,n}\left(\widehat{\boldsymbol{\theta}}_{\alpha,n}\right)\boldsymbol{M}%
 		\left(\widehat{\boldsymbol{\theta}}_{\alpha,n}\right)\right)  ^{-1}\boldsymbol{m}%
 		\left(\widehat{\boldsymbol{\theta}}_{\alpha,n}\right),
 		\label{2.10}%
 	\end{equation}
 	where $\widehat{\boldsymbol{\theta}}_{\alpha,n}$ is the MDPDE of $\boldsymbol{\theta}$ obtained based on a random sample of size $n$ and tuning parameter $\alpha \geq 0$ and utilizes the matrix
 	\[
 	\boldsymbol{\Sigma}_{\alpha,n}\left(\widehat{\boldsymbol{\theta}}_{\alpha,n}\right)=\frac{1}%
 	{n}\boldsymbol{\Psi}_{n}^{-1}\left(  \widehat{\boldsymbol{\theta}%
 	}_{\alpha,n}\right)  \boldsymbol{\Omega}_{n}\left(
 	\widehat{\boldsymbol{\theta}}_{\alpha,n}\right)  \boldsymbol{\Psi}_{n}^{-1}\left(  \widehat{\boldsymbol{\theta}}_{\alpha,n}\right).
 	\]
 	
 From the asymptotic distribution of the MDPDE $\widehat{\boldsymbol{\theta}}_{\alpha,n}$ in Section \ref{SEC:Asymptotic MDPDE}, it may be observed that $W_{n}$ asymptotically follows a (central) chi-squared distribution, $\chi_{r}^{2}$,
 	with $r$ degrees of freedom under the null hypothesis in (\ref{EQ:Hyp_Comp}). Now, the null hypothesis in (\ref{EQ:Hyp_Comp}) is rejected if
 \begin{equation}
 	W_{n}\left(\widehat{\boldsymbol{\theta}}_{\alpha,n}\right)>\chi_{r,\tau}^{2},
 	\label{2.101}%
 \end{equation}
 where $\chi_{r,\tau}^{2}$ \ is the $100(1-\tau)$ percentile of a chi-square distribution with $r$ degrees of freedom.
 
 From the general theory of Basu et al. (2018) \cite{Basu/etc:2017}, it may be observed that the MDPDE based Wald-type test is consistent at any fixed alternative. We will check its behavior in the case of contiguous alternative and try to find an approximate power function of the test. Under the contiguous alternative of the form $H_{1,n}$: $\boldsymbol{\theta} = \boldsymbol{\theta}_{n}$ where $\boldsymbol{\theta}_{n}=\boldsymbol{\theta}_{0}+n^{-1/2}\boldsymbol{d}$  with $\boldsymbol{\theta}_0\in\Theta_{0}$ and $\boldsymbol{d}\in\mathbb{R}^{p+2}\setminus\{\boldsymbol{0}_{p+2}\}$, the statistic $W_{n}(\widehat{\boldsymbol{\theta}}_{\alpha,n})$ asymptotically follows a non-central chi-squared distribution, denoted as $\chi^2_r(\delta)$, 
 having $r$ degrees of freedom and the non-centrality parameter is $\delta = \boldsymbol{d}^{T}\boldsymbol{Q}_\alpha\big(\boldsymbol{\theta}_{0}\big)
 \boldsymbol{d}$,
 with $\boldsymbol{Q}_\alpha\big(\boldsymbol{\theta}_{0}\big) = \boldsymbol{M}\big(\boldsymbol{\theta}_{0}\big)
 \big[\boldsymbol{M}^{T}\big(\boldsymbol{\theta}_{0}\big) \boldsymbol{\Sigma}_{\alpha}\big(\boldsymbol{\theta}_{0}\big)
 \boldsymbol{M}\big(\boldsymbol{\theta}_{0}\big)\big]^{-1} \boldsymbol{M}^{T}\big(\boldsymbol{\theta}_{0}\big)$.
 Based on this result, an approximate expression of the contiguous power function of the test based on $W_{n}\left(\widehat{\boldsymbol{\theta}}_{\alpha,n}\right)$ 
 can be calculated as 
 $$
 \phi_{\alpha}\big(\boldsymbol{\theta}_{n}\big)=1-G_{\chi^2_r(\delta)}\big(\chi_{r,\tau}^{2}\big),
 $$
 where $G_{\chi^2_r(\delta)}$ is the cdf of the $\chi^2_r(\delta)$ distribution.
 
 \subsection{Influence analysis of the robust Wald-type test}
 \label{SEC:Wald_influence}
 The robustness properties of the MDPDE based Wald-type tests were discussed for non homogeneous cases by Basu et al. (2018) \cite{Basu/etc:2017} and Ghosh and Basu (2018) \cite{Ghosh/Basu:2018}, which also hold true for our cases. For completeness, we restate the main results briefly.
 
 At the null distribution with $\boldsymbol{\theta}_0\in\Theta_0$, 
 the first order IF of the Wald-type test statistic becomes identically zero but the second order IFs (either in a single direction or along all directions) have the form
 \begin{eqnarray}
 	IF_{2,i_{0}}\big(y_{i_{0}},W_{n},{\boldsymbol{\theta}_{0}}\big)
 	&=&2 IF_{i_{0}}\big(y_{i_{0}},T_{\alpha},{\boldsymbol{\theta}_{0}}\big)^{T}\boldsymbol{Q}_\alpha\big(\boldsymbol{\theta}_{0}\big)
 	IF_{i_{0}}\big(y_{i_{0}},T_{\alpha},{\boldsymbol{\theta}_{0}}\big),\\
 	IF_{2}\big(y_{1}, \ldots,y_{n},W_{n},{\boldsymbol{\theta}_{0}}\big)
 	&=&2 IF\big(y_{1}, \ldots, y_{n},T_{\alpha},{\boldsymbol{\theta}_{0}}\big)^{T}\boldsymbol{Q}_\alpha\big(\boldsymbol{\theta}_{0}\big)
 	IF\big(y_{1}, \ldots, y_{n},T_{\alpha},{\boldsymbol{\theta}_{0}}\big),
 	\label{EQ:IF2_test_Gen}
 \end{eqnarray}
 where $T_\alpha$ is the MDPD functional. Note that, this (second order) IF of the MDPDE based Wald-type test statistic directly depends on the IF of corresponding MDPD functional. Based on our earlier exploration in Section \ref{SEC:MDPDE_IF}, the influence function of the Wald type test statistic in (\ref{EQ:IF2_test_Gen}) will then be bounded in the contamination point $y$ for any $\alpha>0$ which implies the robustness of the proposed test statistic in (\ref{2.10}) based on the MDPDE.
 
 Again from the general theory of \cite{Basu/etc:2017}, one can observe the robustness of the level and power of the MDPDE based Wald-type tests for any $\alpha>0$ through their bounded level influence function (LIF) and the power influence function (PIF). In particular, the LIF of any order is identically zero and the PIF for testing at the significance level $\tau$ has the form  
 \begin{eqnarray}
 	PIF\big(y_{1}, \ldots, y_{n}, W_{n}, {\boldsymbol{\theta}_{0}}\big) 
 	=C_{r}^{*}\left(\boldsymbol{d}^{T}\boldsymbol{Q}_\alpha\big(\boldsymbol{\theta}_{0}\big)\boldsymbol{d}\right)
 	\boldsymbol{d}^{T}\boldsymbol{Q}_\alpha\big(\boldsymbol{\theta}_{0}\big)IF(y_{1}, \ldots, y_{n}, T_{\alpha}, {\boldsymbol{\theta}_{0}}),
 	\label{EQ:PIF_gen}
 \end{eqnarray} 
 where  
 $C_{r}^{*}\big(s\big)=e^{-\frac{s}{2}}\sum_{v=0}^{\infty}{s^{v-1}}{2^{-v}}(2v-s)P\big(\chi^{2}_{r+2v}>\chi^{2}_{r,\tau}\big)/v!.
 $
 Again the PIF is a linear function of the IF of the MDPDE 
 and hence bounded for all $\alpha>0$ indicating power robustness of the Wald-type test based on the statistic in (\ref{2.10}).

 \subsection{An illustration: testing for the regression coefficient with one covariate}
 \label{SEC:Test_Regression_Coef}
 
 We now discuss, in detail, a particular testing problem regarding the regression coefficient. Let us consider the linear model with a single covariate as $y=x \beta + \epsilon$, with $\epsilon \sim SN(0, \sigma, \gamma)$. Here, we consider the sample size to be $100$ and we generate $x$ from $N(1,1)$. Very often we want to test the significance of the regression coefficients, i.e. in our context we want to test $H_0: \beta=0$. Let us consider a slightly more general problem of testing
 \begin{eqnarray}
 	H_{0} : \beta =\beta_{0} \hspace{2mm} \mbox{against} \hspace{2mm} 
 	H_{1}: \beta \neq \beta_{0},
 	\label{EQ:Hyp_beta}
 \end{eqnarray}
 for a pre-fixed real $\beta_{0}$. The choice $\beta_{0}=0$ gives the test of significance of the covariate.
 Here $\sigma$ and $\gamma$ are unknown nuisance parameters.
 
 In the notation of Section \ref{SEC:Test_Gen}, we have 
 $\Theta_{0}=\big\{\boldsymbol{\theta}=(\beta_{0},\sigma,\gamma)^T : \beta_{0} \in \mathbb{R},~ \sigma \in \mathbb{R}^{+}, ~\gamma \in \mathbb{R} \big \}$,
 $r=1$, $\boldsymbol{m}(\boldsymbol{\theta})=\beta-\beta_{0}$ and $M(\boldsymbol{\theta})= (1,0,0)^{T}$.
 
\begin{figure}[h]
 	   	\centering
 	   	
 	   	\includegraphics[width=0.9\textwidth]{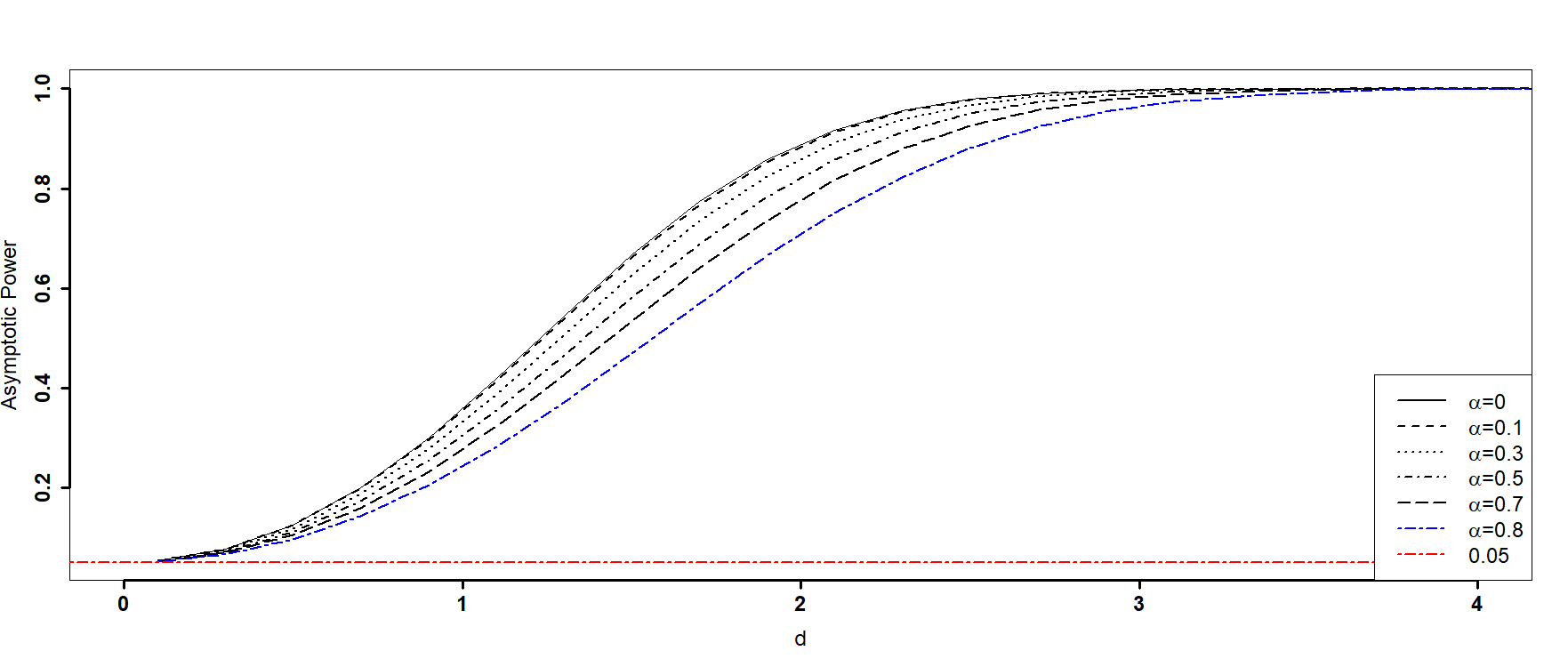}
 	   	\label{FIG:asymp_power}\\
 	   	
 	   \caption{Asymptotic Contiguous Power of the MDPDE based Wald-type test for testing regression coefficient ($\beta_0=3$) at 5\% level of significance with $\boldsymbol{\theta}_{0}=(3,1,2)^T$ and different values of $\alpha$.}
 	   \label{FIG:AsymP_power}
 	\end{figure}

 Denoting the MDPDE as $\widehat{\boldsymbol{\theta}}_{\alpha,n}=\big(\widehat{\beta}_{\alpha,n},~\widehat{\sigma}_{\alpha,n},~\widehat{\gamma}_{\alpha,n}\big)$,
 our MDPDE based Wald-type test statistics (\ref{2.10}) has a simplified form for testing (\ref{EQ:Hyp_beta})
 which is given by 
 \begin{eqnarray}
 	W_{n}\left(\widehat{\boldsymbol{\theta}}_{\alpha,n}\right)= \frac{n\left(\widehat{\beta}_{\alpha,n} - \beta_{0}\right)^{2}}{\Sigma_{\alpha,n}^{(11)}\left(\widehat{\boldsymbol{\theta}}_{\alpha,n}\right)},
 	\label{EQ:TestStat_beta}
 \end{eqnarray}
 where $\Sigma_{\alpha,n}^{(11)}(\boldsymbol{\theta})$ is the $(1,1)$-th element of $\boldsymbol{\Sigma}_{\alpha,n}(\boldsymbol{\theta})$. 
 Then, under the null hypothesis in (\ref{EQ:Hyp_beta}), $W_{n}\left(\widehat{\boldsymbol{\theta}}_{\alpha,n}\right)$ asymptotically follows $\chi_{1}^{2}$ distribution
 and the test can be performed by comparing $W_{n}\left(\widehat{\boldsymbol{\theta}}_{\alpha,n}\right)$ with the corresponding critical values.
 Further, the approximate expression of power function at the contiguous hypothesis of the form 
 $H_{1,n}$: $\beta= \beta_{n}$, where $\beta_{n}=\beta_{0}+ n^{-1/2}d$, with $d\in \mathbb{R}\setminus\{0\},$
 is given by  
 $$
 \phi_{\alpha}\big(\boldsymbol{\theta}_{n}\big)=1-G_{\chi_{1}^{2}(\delta)}\big(\chi_{1,\tau}^{2}\big),
 ~~~~~\mbox{with}~~\delta=\frac{d^{2}}{\Sigma_{\alpha}^{(11)}(\boldsymbol{\theta}_0)}, ~\boldsymbol{\theta}_0\in\Theta_{0}.
 $$
 We have numerically calculated this asymptotic contiguous power for testing $\beta=3$ at 5\% level by using the MDPDE based Wald-type test with different values of $\alpha$, which is graphically presented in Figure \ref{FIG:AsymP_power}
 for $\boldsymbol{\theta}_{0}=(3,1,2)^T$. It is clear that, just like the ARE of the MDPDE, 
 the contiguous power of the MDPDE based test also decreases as $\alpha$ increases but this loss is not quite significant at small $\alpha>0$.

\begin{figure}[h]
	\centering
		\subfloat[$IF2$ for particular direction]{
			\includegraphics[width=0.5\textwidth]{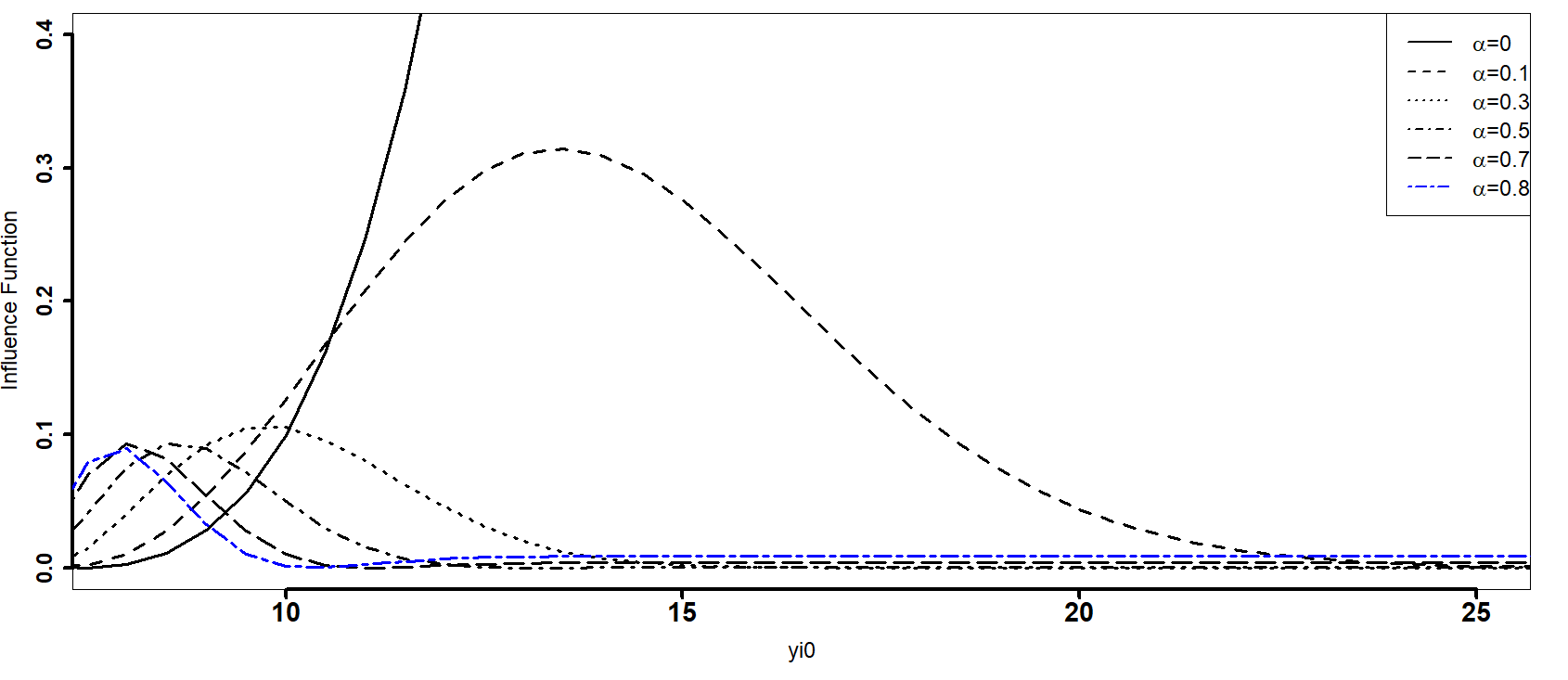}
			\label{FIG:IF_test_stat_i_latest1}}
		\subfloat[$IF2$ for all directions]{
			\includegraphics[width=0.5\textwidth]{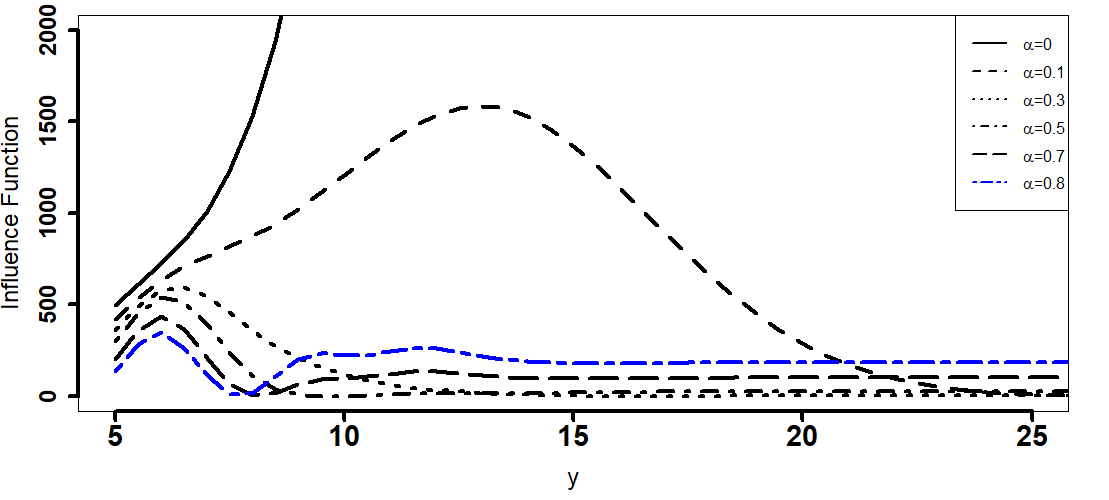}
			\label{FIG:IF_test_stat_all_latest1}}
	\caption{Second order influence functions ($IF2$) of the robust Wald-type test statistic in $i_{0}=10$th and all directions respectively with $\boldsymbol{\theta}=(3,2,2)$ for different $\alpha$.}
	\label{FIG:IF2_test_stat_latest1}
\end{figure}

\begin{figure}[h]
 	\centering
 		
 		\includegraphics[width=0.6\textwidth]{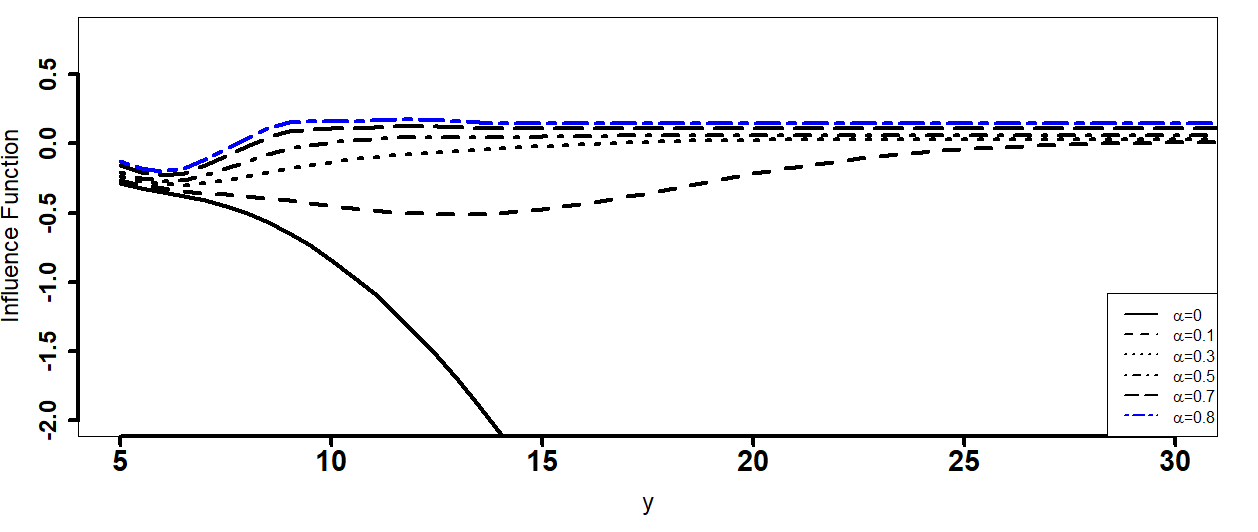}
 		\label{FIG:PIF_test_stat_latest1}\\
 		
 	\caption{Power influence function (PIF) of the robust Wald-type test statistic  with $\boldsymbol{\theta}=(3,2,2)$ for different $\alpha$.}
 	\label{FIG:PIF_latest1}
 \end{figure}

 The robustness of the Wald-type test based on the statistic (\ref{EQ:TestStat_beta}) for testing (\ref{EQ:Hyp_beta})
 can be studied through the second order influence function of the test statistic and the PIF. From the general formulas presented in Section \ref{SEC:Test_Gen}, we can easily calculate these measures 
 in the present case of testing (\ref{EQ:Hyp_beta}) as given by 
 \begin{eqnarray}
 	IF_{2,i_{0}}\left(y_{i_{0}},W_{n},{\boldsymbol{\theta}_{0}}\right)
 	&=&2 IF_{i_{0}}\left(y_{i_{0}},T_{\alpha}^{(\beta)},{\boldsymbol{\theta}_{0}}\right)^{2}/\Sigma_{\alpha}^{(11)}(\boldsymbol{\theta}_0).
 	\label{EQ:IF2i0_test_beta}
 	\\
 	IF_{2}\left(\boldsymbol{y},W_{n},{\boldsymbol{\theta}_{0}}\right)
 	&=&2 IF\left(\boldsymbol{y},T_{\alpha}^{(\beta)},{\boldsymbol{\theta}_{0}}\right)^{2}/\Sigma_{\alpha}^{(11)}(\boldsymbol{\theta}_0).
 	\label{EQ:IF2all_test_beta}
 	\\
 	PIF\left(\boldsymbol{y}, W_{n}, {\boldsymbol{\theta}_{0}}\right) 
 	&=&C_{1}^{*}\left({d}^{2}/\Sigma_{\alpha}^{(11)}(\boldsymbol{\theta}_0)\right)
 	IF\left(\boldsymbol{y}, T_{\alpha}^{(\beta)}, {\boldsymbol{\theta}_{0}}\right)d/\Sigma_{\alpha}^{(11)}(\boldsymbol{\theta}_0),
 	\label{EQ:PIF_beta}
 \end{eqnarray} 
 where $T_{\alpha}^{(\beta)}$ is the MDPDE functional corresponding to $\beta$ and 
 $\boldsymbol{y}=y \boldsymbol{1}$. For illustration, we have considered the model $y=x\beta+\epsilon$ with $ \epsilon \sim SN(0,\sigma,\gamma)$ and presented the plots of these $IF_2$ and $PIF$ for $6$ different values of $\alpha$ at $\boldsymbol{\theta}_{0}=(3,1,2)^T$ and $d=0.01$ in Figures \ref{FIG:IF2_test_stat_latest1} and \ref{FIG:PIF_latest1}. It is clear from these figures that the MDPDE based Wald-type test statistic (\ref{EQ:TestStat_beta}) has bounded second order IF and PIF for all $\alpha>0$.

 \section{Simulation studies}
 \label{SEC:Simulation}
 In this section, we present some interesting findings from extraneous simulation studies in order to examine the claimed robustness of the skewed regression model as well as the robust Wald-type test. First, we calculate the empirical bias and MSE (mean squared error) of the estimates of the model parameters based on simulated datasets. Next, we conduct robust Wald-type tests related to regression coefficients and calculate the power and level of the test. We use the R function \textsf{optim} to find the MDPDE of the model parameters.

 \textcolor{black}{We would like to mention that we have found only two robust estimation methods available for the skew-normal error model but both of these model set-ups are completely different from ours. 
 	The first one by Morales et al. (2017)\cite{Morales et al.:2017} considered a generalized class of skew-normal distributions under a quantile regression model 
 	-- rather than the linear mean regression model considered here. Although the second paper by Cao et al. (2020)\cite{Cao:2020} considered the linear mean regression model with skewed errors, 
 	they have used a different parametrization of the skew-normal distribution as their error distribution. 
 	We have indeed compared these two methods with the proposed MDPDE in our simulation studies and have observed that they have significantly poorer performance compared to the MDPDEs. 
 	However, since the model set-ups are not exactly the same, these comparisons are not ``apple-to-apple" comparisons, and so we have not added them in our paper.}
 \subsection{Performance of the MDPDE}
 \label{SEC:Bias_MSE}
 
 We consider the linear regression model 
 \begin{equation}
 	Y_{i}=\beta_{0}+\beta_{1}x_{1i}+\beta_{2}x_{2i}+\epsilon_{i}
 	\label{EQ:4.1}
 \end{equation}
where $\epsilon_{i} \sim SN(0,1,2)$. We have taken $(\beta_{0},\beta_{1},\beta_{2})=(1,2,3)$. Besides, we simulate $x_1$ from $N(1,1)$ and $x_2$ from $N(-1,1)$. Under these specifications, we simulate a random sample of size $n=100$ observations. Based on $500$ replications, we have tabulated the empirical biases and MSEs for $6$ different $\alpha$ in Table \ref{TAB:Sim}. We have used the R package \textsf{sn} to simulate random samples from the skew normal distribution.

Next, we consider $5\%$ contamination in the data. In this case, we simulate $5\%$ of the errors from $SN(-10,1,2)$ and rest from $SN(0,1,2)$. Again, based on $500$ replications, we calculate the bias and MSE which are tabulated in Table \ref{TAB:Sim}. The process is then repeated for sample size $n=200$ and the bias and MSEs are summarized in the same table.

\begin{table}[!h]
	\caption{Empirical bias and MSE of the estimates of regression parameters under pure data as well as $5\%$ and $10\%$ contaminated data of size $n$ [$\epsilon$ denotes the contamination proportion]}
	\centering
	\resizebox{1\textwidth}{!}{
		\begin{tabular}{c c | c c c c c c | c c c c c c }
			\hline
			  & & \multicolumn{6}{c|}{Parametric MDPDE at different $\alpha$} & \multicolumn{6}{|c}{Parametric MDPDE at different $\alpha$}\\
			 $\epsilon$ & & 0(MLE) & 0.1 & 0.3 & 0.5 & 0.7 & 1 & 0(MLE)  & 0.1 & 0.3 & 0.5 & 0.7 & 1\\
			\hline
			 & & \multicolumn{6}{c|}{Bias} & \multicolumn{6}{c}{MSE}\\
			\hline
			\multicolumn{14}{c}{$\boldsymbol{n=100}$}\\
			$0\%$	&	$\beta_0$	&	0.0554	&	0.0604	&	0.0687	&	0.0872	&	0.0904	&	0.0954	&	0.0668	&	0.0807	&	0.0843	&	0.0879	&	0.0893	&	0.0902	\\
				&	$\beta_1$	&	0.0001	&	-0.0003	&	-0.0004	&	-0.0004	&	-0.0019	&	-0.0025	&	0.0047	&	0.0052	&	0.0055	&	0.0056	&	0.0072	&	0.0085	\\
				&	$\beta_2$	&	0.0037	&	0.0038	&	-0.0054	&	-0.0055	&	-0.0055	&	-0.0061	&	0.0043	&	0.0047	&	0.0055	&	0.0061	&	0.0065	&	0.0074	\\
			&	$\sigma$	&	-0.0222	&	-0.0225	&	-0.0243	&	-0.0467	&	-0.0551	&	-0.0657	&	0.0157	&	0.0162	&	0.0173	&	0.0178	&	0.0187	&	0.0225	\\
			&	$\gamma$	&	0.0098	&	0.0545	&	0.1370	&	0.2802	&	0.3501	&	0.5870	&	0.1667	&	0.3844	&	0.6104	&	0.8000	&	1.0094	&	1.2017	\\\hline
			$5\%$	&	$\beta_0$	&	-0.1389	&	-0.1189	&	0.1090	&	0.1003	&	0.0973	&	0.0981	&	0.1564	&	0.1429	&	0.1368	&	0.1249	&	0.1001	&	0.1024	\\
				&	$\beta_1$	&	0.0075	&	0.0064	&	0.0051	&	-0.0039	&	-0.0030	&	-0.0034	&	0.0117	&	0.0105	&	0.0099	&	0.0093	&	0.0086	&	0.0090	\\
				&	$\beta_2$	&	-0.0089	&	-0.0075	&	-0.0064	&	-0.0059	&	-0.0057	&	-0.0063	&	0.0108	&	0.0100	&	0.0096	&	0.0088	&	0.0076	&	0.0080	\\
				&	$\sigma$	&	-0.0782	&	-0.0725	&	-0.0658	&	-0.0598	&	-0.0575	&	-0.0659	&	0.0400	&	0.0348	&	0.0313	&	0.0287	&	0.0232	&	0.0243	\\
				&	$\gamma$	&	0.9028	&	0.8635	&	0.7318	&	0.6042	&	0.4828	&	0.5385	&	1.7183	&	1.6048	&	1.4652	&	1.3378	&	1.1494	&	1.2246	\\\hline
			$10\%$	&	$\beta_0$	&	-0.2013	&	-0.1648	&	-0.1313	&	0.1261	&	0.0980	&	0.0996	&	0.1824	&	0.1638	&	0.1448	&	0.1370	&	0.1124	&	0.1206	\\
				&	$\beta_1$	&	0.0086	&	0.0077	&	0.0066	&	-0.0048	&	-0.0037	&	-0.0042	&	0.0219	&	0.0155	&	0.0129	&	0.0108	&	0.0094	&	0.0101	\\
				&	$\beta_2$	&	-0.0103	&	-0.0095	&	-0.0086	&	-0.0071	&	-0.0063	&	-0.0072	&	0.0166	&	0.0148	&	0.0115	&	0.0100	&	0.0088	&	0.0095	\\
				&	$\sigma$	&	-0.0938	&	-0.0873	&	-0.0795	&	-0.0721	&	-0.0653	&	-0.0718	&	0.0528	&	0.0447	&	0.0385	&	0.0305	&	0.0281	&	0.0296	\\
				&	$\gamma$	&	0.9383	&	0.9031	&	0.7982	&	0.6418	&	0.5435	&	0.5828	&	1.8214	&	1.6925	&	1.5597	&	1.4043	&	1.2295	&	1.2792	\\
		\hline\hline
		\multicolumn{14}{c}{$\boldsymbol{n=200}$}\\
			$0\%$	&	$\beta_0$	&	0.0414	&	0.0479	&	0.0544	&	0.0658	&	0.0770	&	0.0824	&	0.0548	&	0.0677	&	0.0753	&	0.0810	&	0.0855	&	0.0882	\\
	&	$\beta_1$	&	0.0001	&	-0.0002	&	-0.0003	&	-0.0004	&	-0.0016	&	-0.0022	&	0.0036	&	0.0044	&	0.0050	&	0.0054	&	0.0065	&	0.0077	\\
		&	$\beta_2$	&	0.0029	&	0.0033	&	-0.0040	&	-0.0046	&	-0.0050	&	-0.0057	&	0.0038	&	0.0042	&	0.0049	&	0.0056	&	0.0062	&	0.0069	\\
				&	$\sigma$	&	-0.0199	&	-0.0205	&	-0.0216	&	-0.0377	&	-0.0451	&	-0.0547	&	0.0101	&	0.0108	&	0.0114	&	0.0120	&	0.0124	&	0.0130	\\
			&	$\gamma$	&	0.0090	&	0.0425	&	0.1060	&	0.2285	&	0.3011	&	0.4861	&	0.1067	&	0.2484	&	0.4404	&	0.6199	&	0.7987	&	0.9171	\\\hline
			$5\%$	&	$\beta_0$	&	-0.1162	&	-0.1028	&	-0.0931	&	0.0813	&	0.0732	&	0.0843	&	0.1503	&	0.1392	&	0.1234	&	0.1093	&	0.0902	&	0.0993	\\
				&	$\beta_1$	&	0.0058	&	0.0052	&	0.0043	&	-0.0031	&	-0.0018	&	-0.0024	&	0.0101	&	0.0096	&	0.0084	&	0.0073	&	0.0066	&	0.0071	\\
				&	$\beta_2$	&	-0.0074	&	-0.0065	&	-0.0058	&	-0.0047	&	-0.0039	&	-0.0046	&	0.0095	&	0.0087	&	0.0080	&	0.0071	&	0.0053	&	0.0067	\\
				&	$\sigma$	&	-0.0681	&	-0.0591	&	-0.0508	&	-0.0417	&	-0.0341	&	-0.0425	&	0.0261	&	0.0193	&	0.0154	&	0.0113	&	0.0098	&	0.0125	\\
				&	$\gamma$	&	0.7821	&	0.7032	&	0.6090	&	0.5199	&	0.4181	&	0.4612	&	1.2762	&	1.1549	&	0.9962	&	0.8574	&	0.7877	&	0.8482	\\\hline
			$10\%$	&	$\beta_0$	&	-0.1423	&	-0.1283	&	-0.1002	&	0.0922	&	0.0842	&	0.0913	&	0.1663	&	0.1519	&	0.1392	&	0.1204	&	0.1022	&	0.1195	\\
				&	$\beta_1$	&	0.0069	&	-0.0063	&	-0.0052	&	-0.0039	&	-0.0025	&	-0.0031	&	0.0184	&	0.0143	&	0.0115	&	0.0096	&	0.0082	&	0.0094	\\
				&	$\beta_2$	&	-0.0094	&	-0.0085	&	-0.0073	&	-0.0059	&	-0.0052	&	-0.0064	&	0.0119	&	0.0103	&	0.0094	&	0.0083	&	0.0072	&	0.0088	\\
				&	$\sigma$	&	-0.0861	&	-0.0754	&	-0.0683	&	-0.0554	&	-0.0431	&	-0.0529	&	0.0374	&	0.0295	&	0.0214	&	0.0171	&	0.0159	&	0.0187	\\
			&	$\gamma$	&	0.8412	&	0.8062	&	0.6950	&	0.5736	&	0.4610	&	0.5013	&	1.3166	&	1.2239	&	1.0952	&	0.9484	&	0.8294	&	0.8618	\\
			\hline
		\end{tabular}
	}
	\label{TAB:Sim}
\end{table}

From Table \ref{TAB:Sim} we observe that, for pure data, the empirical biases and MSEs have increasing trends with the increment of the tuning parameter $\alpha$. The biases and MSEs also decrease with the sample size $n$. Under contamination these quantities are high for small values of $\alpha$; however, they steadily decline with increasing $\alpha$ (in absolute value for the bias) and eventually stabilize at values close to those of the pure data case (although there appears to be a slight upturn somewhere between $\alpha = 0.7$ and $\alpha = 1$). 

\subsection{Performance of the robust Wald-type test}
\label{SEC:Level_Power}
Here we take the same set up of the regression model as in Section \ref{SEC:Bias_MSE}. Here we want to test  $H_{0}$: $\beta_{1}=2$ against $H_{1,n}$: $\beta_{1} =\beta_{1,n}$, where $\beta_{1,n}=2 + n^{-1/2}d$,
for which the Wald-type test statistic $W_{n}$ has a form similar to that in (\ref{EQ:TestStat_beta}). We first simulate random samples of sizes $n=100,~200$  and 
perform the MDPDE based Wald-type test for different $\alpha$, including the classical Wald test at $\alpha=0$.
Based on $500$ replications, we then compute the empirical levels of the tests, measured as 
the proportion of test statistics exceeding the chi-square critical value among the $500$ replications.
Subsequently, to compute the empirical power of the tests, we repeat the above exercise 
but now we calculate the proportion of times \textcolor{black}{the contiguous alternative} $H_{1,n}$: $\beta_{1}=\beta_{1,n}$ is rejected, where $\beta_{1,n}=2 + n^{-\frac{1}{2}}1.5$. For calculating the empirical power, we have considered the same linear regression model as used in case of level calculation except that we have taken the value of $\beta_1$ as $\beta_{1,n}=2 + n^{-\frac{1}{2}}1.5$. Finally, to illustrate the claimed robustness, we recalculate the level and power of the Wald-type tests
after contamination $100\epsilon\%$ of each sample in the previous simulation exercises with $\epsilon=0.05, 0.1$.
The contaminated observations generated using the error distribution as $SN(-5,1,2)$ and $SN(-3,1,2)$ respectively, for the level and power calculations. 
In Table \ref{TAB:Level_Power}, we report all the resulting empirical levels and powers 
obtained from different simulation scenarios.

\begin{table}[!h]
	\caption{Empirical levels and powers of robust Wald-type test statistic under pure data as well as $5\%$ and $10\%$ contaminated data [$\epsilon$ denotes the contamination proportion]}
	\centering
		\begin{tabular}{c c  | c c c c c c }
			\hline
			Sample&  &  \multicolumn{6}{c}{Parametric MDPDE at different $\alpha$}\\
			size& $\epsilon$ &  0(MLE) & 0.1 & 0.3 & 0.5 & 0.7 & 1 \\
			\hline
			& &  \multicolumn{6}{c}{Level} \\\hline
			$n=100$	&	$0\%$	&	0.072	&	0.076	&	0.084	&	0.095	&	0.106	&	0.119	\\
			&	$5\%$	&	0.429	&	0.325	&	0.219	&	0.163	&	0.112	&	0.091	\\
			&	$10\%$	&	0.467	&	0.359	&	0.261	&	0.185	&	0.134	&	0.098	\\\hline
			$n=200$	&	$0\%$	&	0.054	&	0.057	&	0.062	&	0.071	&	0.08	&	0.094	\\
			&	$5\%$	&	0.454	&	0.331	&	0.193	&	0.129	&	0.094	&	0.076	\\
			&	$10\%$	&	0.482	&	0.368	&	0.247	&	0.161	&	0.11	&	0.089	\\
			\hline
			\hline
			& &  \multicolumn{6}{c}{Power} \\\hline
			$n=100$	&	0\%	&	1	&	1	&	1	&	1	&	0.996	&	0.963	\\
			&	$5\%$	&	0.403	&	0.498	&	0.615	&	0.707	&	0.829	&	0.952	\\
			&	$10\%$	&	0.395	&	0.476	&	0.575	&	0.696	&	0.756	&	0.934	\\\hline
			$n=200$	&	$0\%$	&	1	&	1	&	1	&	1	&	1	&	1	\\
			&	$5\%$	&	0.416	&	0.597	&	0.713	&	0.886	&	1	&	1	\\
			&	$10\%$	&	0.409	&	0.567	&	0.694	&	0.842	&	0.92	&	1	\\
			\hline
		\end{tabular}
	\label{TAB:Level_Power}
\end{table}

 From Table \ref{TAB:Level_Power}, we observe that, for pure data, the empirical levels increase with the increment in $\alpha$. It can also be observed that, by the time the sample size rises to 200, the observed levels are close to the nominal level of 0.05 for small values of $\alpha$; however it gets inflated for larger $\alpha$. For contaminated data the levels are high for large $\alpha$, but steadily drop with increasing $\alpha$. In case of power the contamination leads to a significant drop in power for smaller $\alpha$, but at larger values of $\alpha$ the power is much better protected.

 \section{Real data examples}
 \label{SEC:Real_data}
 In this section, we apply our MDPDE based skew normal regression methods to analyze two real life datasets. For each dataset, we calculate the estimates along with the standard errors (SEs) of the regression coefficients ($\boldsymbol{\beta}$) as well as the scale parameter ($\sigma$) and shape parameter ($\gamma$). Apart from these, we also conduct robust Wald-type tests for testing the significance of the covariates (involving the $\boldsymbol{\beta}$ parameters) and the symmetry of the model (involving the $\gamma$ parameter). 
 
 \subsection{Australian Institute of Sports (AIS) data}
 \label{SEC:AIS}
 These data consist of the health measurements of $706$ Australian athletes from $12$ different sports which were collected at the Australian Institute of Sports (AIS) in 1990. The data were first analyzed by Telford and Cunningham (1991) \cite{Telford/Cunningham:1991}. A subset of these data (consisting of $207$ observations) are now available in the R package \textsf{sn} which we use in the following analysis.

There are several biomedical variables related to the health conditions of the athletes included in this dataset. For our analysis, we model the plasma ferritin concentration in blood (denoted as ``Fe") with the body mass index (BMI) and the lean body mass (LBM) as the covariates using the  regression equation
$$
\text{Fe}_i=\beta_{0}+\beta_{1}\text{BMI}_i+\beta_{2}\text{LBM}_i+\epsilon_{i}, ~~~i=1, \ldots, 207,
$$ 
where $\epsilon_{i}\sim SN(0,\sigma,\gamma)$, independently, for each $i$. 
We summarize the results corresponding \textcolor{black}{to the minimum DPD fits} in Table \ref{TAB:AIS2}, 
where we also provide the p-values of the robust Wald-type tests for testing \textcolor{black}{the null hypothesis that the individual parameters are} zero.

 \begin{table}[!h]
 	\caption{Parameter estimates along with standard error (SE) and p-values for AIS data at different $\alpha$}
 	\centering
 	\resizebox{1\textwidth}{!}{
 	\begin{tabular}{c | c c c c c c c c}
 		\hline
 		Parameter & \multicolumn{8}{c}{ MDPDE at different $\alpha$}\\
 		& 0(MLE)& 0.05 & 0.1 & 0.3 & 0.5 & 0.7 & 0.8 &1\\
 		\hline
 		Intercept	&	$-46.3478$	&	$-46.3477$ & $-46.3477$	
 		&	$-46.0384$	&	$-45.8145$	&	$-45.8965$	&	$-46.2416$
 		& $-46.3238$	\\
 		SE	&	(13.2494) & (13.1964)	&	(13.1176)	&	(13.6866)	&	(14.2514)	&	(14.5397)	& (14.1700)	& (14.6015)	\\
 		p-value	&	(0.0005)	&	(0.0005) &(0.0004)	 &	(0.0008)	&	(0.0013)	&	(0.0016)	& (0.0011) &	(0.0015)	\\\hline
 		
 		BMI	&	2.3485	& 2.3484 &	2.3482	 &	2.2346	&	2.1372	&	1.9937	& 1.8058  &	1.6145	\\
 		SE	&	(0.8049)	& (0.8002) &	(0.7969)	 &	(0.8317)	&	(0.8667)	& 	(0.8854) & (0.8637)	&	(0.8910)	\\
 		p-value	&	(0.0035)	& (0.0034) &	(0.0032) 	&	(0.0072)	&	(0.0137)	&	(0.0243) & (0.0365)	&	(0.0603)	\\ \hline
 		
 		LBM	&	0.2280	& 0.2286 &	0.2290	& 0.2745	&	0.3139	&	0.3735 & 0.4536	&	0.5331	\\
 		SE	&	(0.1764)	& (0.1753) &	(0.1746) 	&	(0.1822)	&	(0.1899)	&	(0.1940)	& (0.1892) &	(0.1952)	\\
 		p-value	&	(0.1960) & (0.1918)	&	(0.1897)	&	(0.1321)	&	(0.0983)	&	(0.0542)	&  (0.0165) &	(0.0093)	\\ \hline
 		
 		$\sigma$	&	70.8381	& 70.1732&	69.6698	&	66.4086	&	63.7539	&	62.4061	& 62.0975 & 61.2081	\\
 		SE	&	(3.8689) &(3.8601)	&	(3.8562) &(4.0110)	&	(4.2578)	&	(4.5166)	& (4.5966) &	(4.8078)	\\ \hline
 		
 		$\gamma$	&	10.9054	& 10.9052&	10.9048	&	9.7672	&	9.2024	&	9.5091	& 10.5502 &	10.7931	\\
 		SE	&	(3.5252) & (3.5419)	&	(3.5635)	 &	(3.2956)	&	(3.3856)	&	(3.9831)	&  (4.8686)&	(5.6060)	\\
 		p-value	&	(0.0020) &(0.0022)	&	(0.0022)	 &	(0.0030)	&	(0.0066)	&	(0.0199)	& (0.0382)	& (0.0642)	\\ 
 		\hline
 	\end{tabular}}
 	\label{TAB:AIS2}
 \end{table}
 
  \begin{table}[!h]
 	\caption{Parameter estimates along with standard error (SE) and p-values for AIS data for different $\alpha$ after excluding $28$ outlying observations}
 	\centering
 	\resizebox{1\textwidth}{!}{
 		\begin{tabular}{c | c  c c c c c c c}
 			\hline
 			Parameter & \multicolumn{8}{c}{MDPDE at different $\alpha$}\\
 			& 0(MLE) &  0.05 & 0.1 & 0.3 & 0.5 & 0.7 & 0.8 & 1\\
 			\hline
 			Intercept	&	-46.0909	&	-46.0908	&	-46.0907	&	-45.9762	&	-45.7821	&	-45.8677	&	-46.1818	&	-46.2655	\\
 			SE	&	(12.8667)	&	(12.8114)	&	(12.7693)	&	(12.9808)	&	(13.8719)	&	(14.0155)	&	(14.0680)	&	(14.5687)	\\
 			p-value	&	(0.0003)	&	(0.0003)	&	(0.0003)	&	(0.0004)	&	(0.0010)	&	(0.0011)	&	(0.0011)	&	(0.0016)	\\ \hline
 			BMI	&	1.4469	&	1.4469	&	1.4469	&	1.4521	&	1.3445	&	1.2700	&	1.4001	&	1.4158	\\
 			SE	&	(0.7719)	&	(0.7749)	&	(0.7777)	&	(0.7851)	&	(0.8095)	&	(0.8393)	&	(0.8533)	&	(0.8848)	\\
 			p-value	&	(0.0609)	&	(0.0619)	&	(0.0628)	&	(0.0644)	&	(0.0717)	&	(0.0783)	&	(0.0808)	&	(0.0896)	\\ \hline
 			LBM	&	0.3565	&	0.3567	&	0.3570	&	0.3575	&	0.3970	&	0.4187	&	0.4862	&	0.5187	\\
 			SE	&	(0.1723)	&	(0.1720)	&	(0.1710)	&	(0.1739)	&	(0.1760)	&	(0.1814)	&	(0.1890)	&	(0.1960)	\\
 			p-value	&	(0.0039)	&	(0.0038)	&	(0.0037)	&	(0.0040)	&	(0.0043)	&	(0.0058)	&	(0.0061)	&	(0.0081)	\\ \hline
 			$\sigma$	&	64.8210	&	64.6987	&	63.9059	&	61.6240	&	60.0905	&	59.1705	&	60.0564	&	60.7216	\\
 			SE	&	(3.7631)	&	(3.7629)	&	(3.7597)	&	(3.9168)	&	(4.1880)	&	(4.5009)	&	(4.5734)	&	(4.6214)	\\ \hline
 			$\gamma$	&	10.1477	&	10.1476	&	10.1473	&	9.9492	&	9.2560	&	9.5777	&	10.1079	&	10.6131	\\
 			SE	&	(3.5649)	&	(3.6755)	&	(3.7015)	&	(3.8120)	&	(3.9318)	&	(4.0789)	&	(4.2386)	&	(4.8999)	\\
 			p-value	&	(0.0326)	&	(0.0428)	&	(0.0509)	&	(0.0540)	&	(0.0572)	&	(0.0615)	&	(0.0638)	&	(0.0690)	\\ \hline
 			
 	\end{tabular}}
 	\label{TAB:AIS2_out_del1}
 \end{table}
 
\begin{figure}[!h]
 	\color{black}
	\centering
	
	\subfloat[QQ plot for OLS]{
		\includegraphics[width=0.3\textwidth]{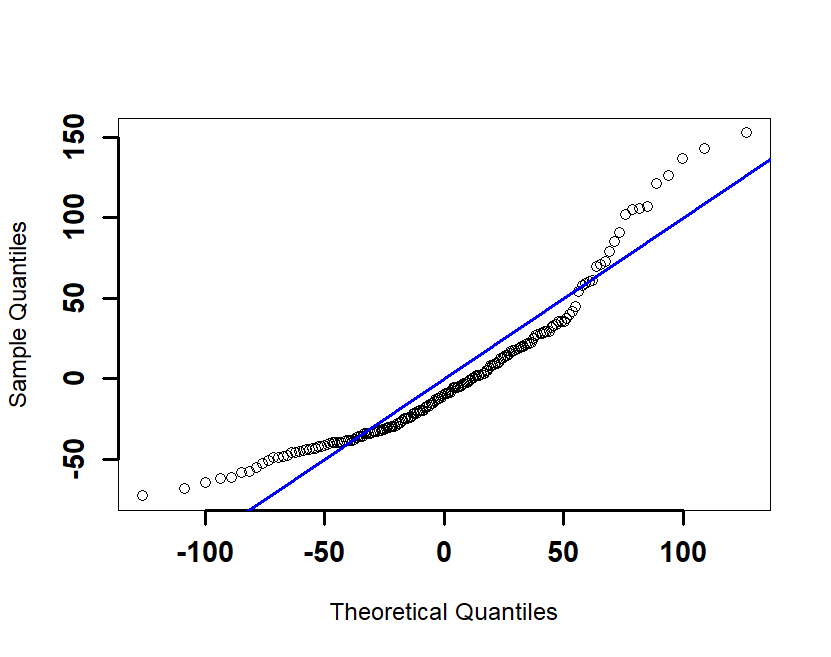}
		\label{FIG:AIS_qq_ols}}
	\subfloat[QQ plot for $\alpha=0$]{
		\includegraphics[width=0.3\textwidth]{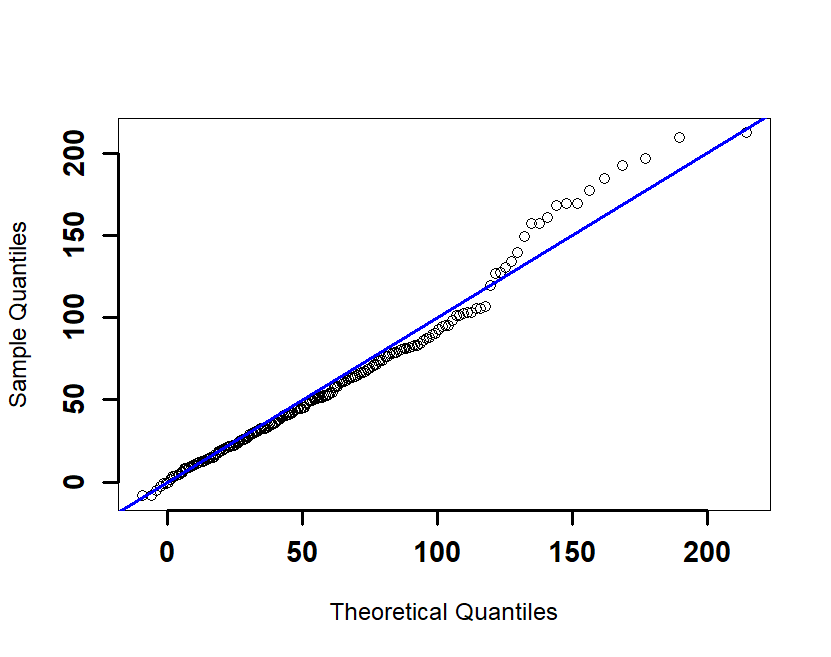}
		\label{FIG:AIS_qq_alp_0}}
	\subfloat[QQ plot for $\alpha=0.05$]{
		\includegraphics[width=0.3\textwidth]{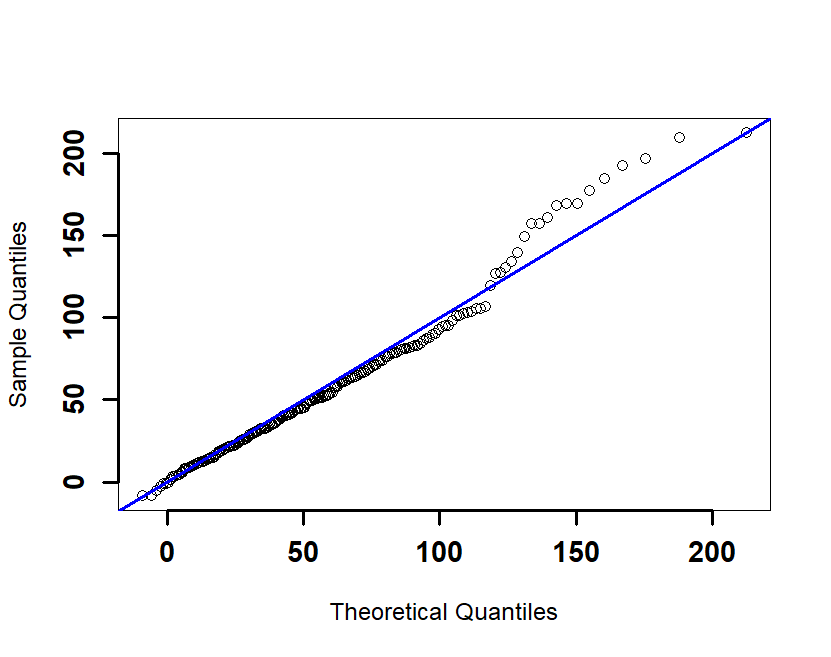}
		\label{FIG:AIS_qq_alp_005}}
	\\
	\subfloat[QQ plot for $\alpha=0.1$]{
		\includegraphics[width=0.3\textwidth]{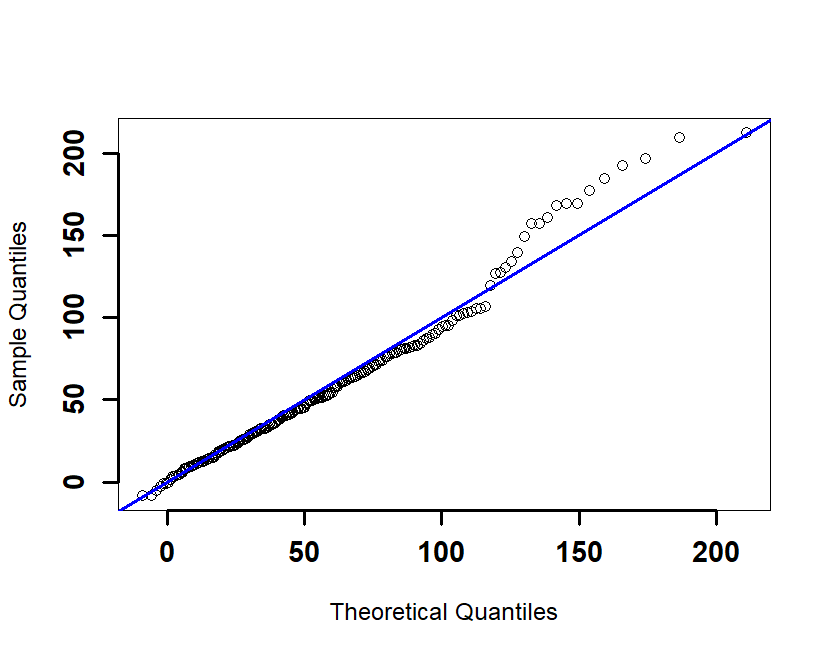}
		\label{FIG:AIS_qq_alp_01}}
	\subfloat[QQ plot for $\alpha=0.3$]{
		\includegraphics[width=0.3\textwidth]{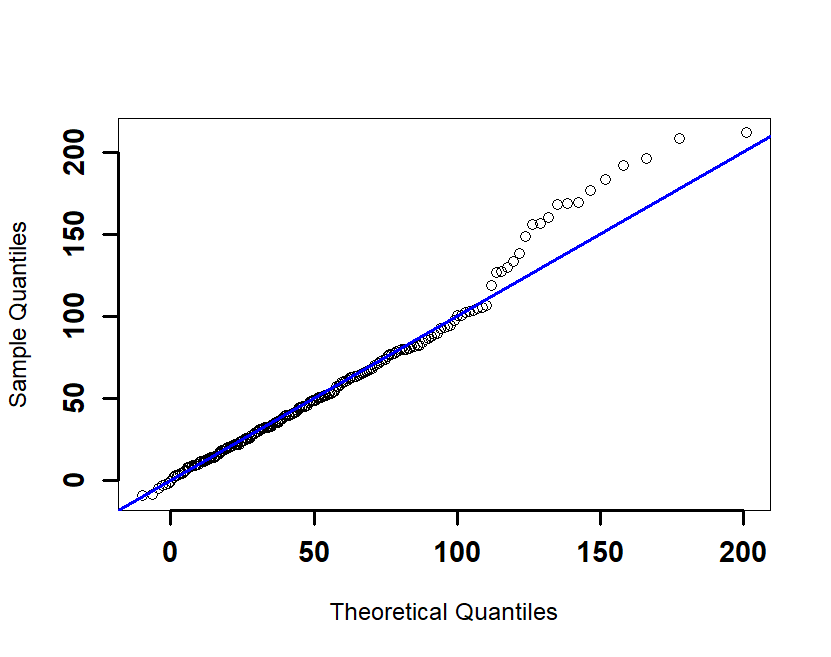}
		\label{FIG:AIS_qq_alp_03}}
	\subfloat[QQ plot for $\alpha=0.5$]{
		\includegraphics[width=0.3\textwidth]{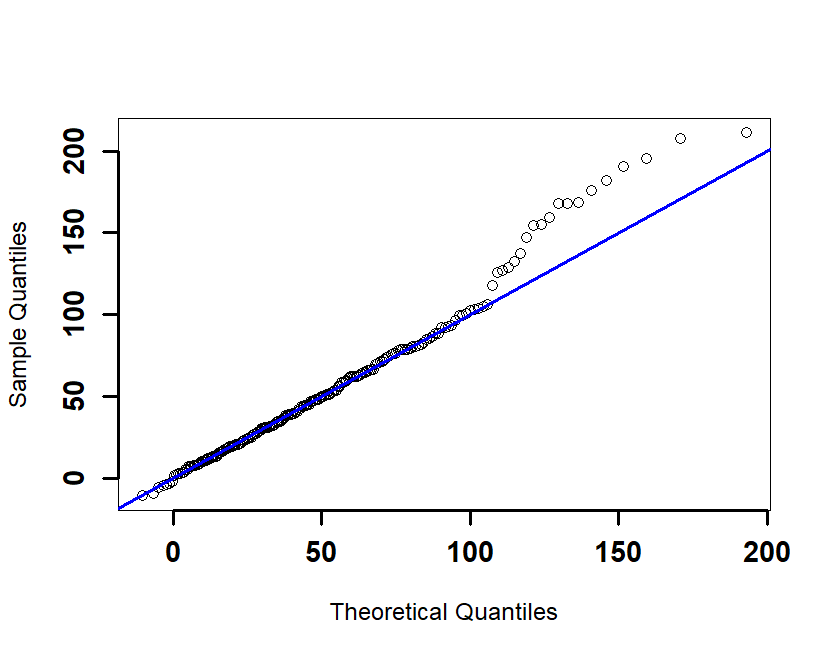}
		\label{FIG:AIS_qq_alp_05}}
	\\
	\subfloat[QQ plot for $\alpha=0.7$]{
		\includegraphics[width=0.3\textwidth]{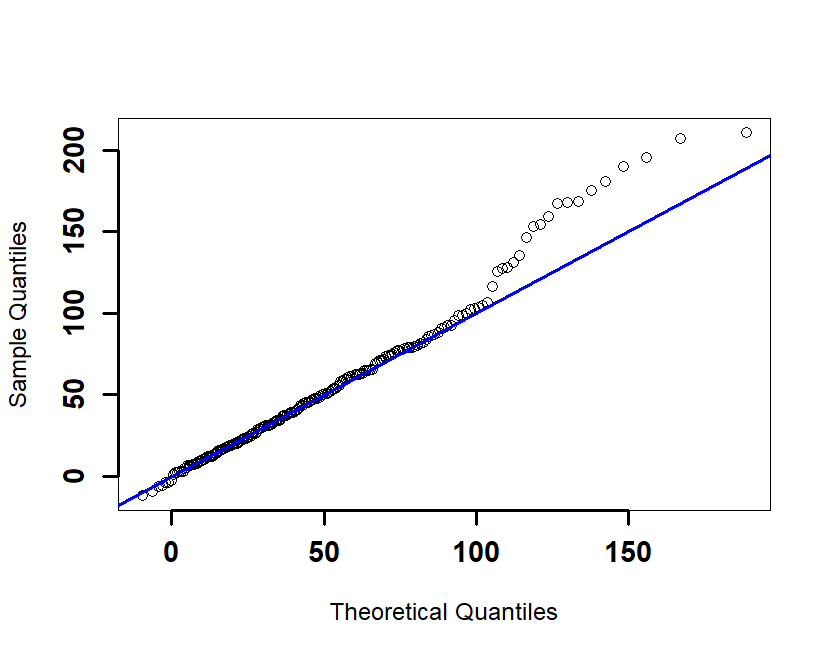}
		\label{FIG:AIS_qq_alp_017}}
	\subfloat[QQ plot for $\alpha=0.8$]{
		\includegraphics[width=0.3\textwidth]{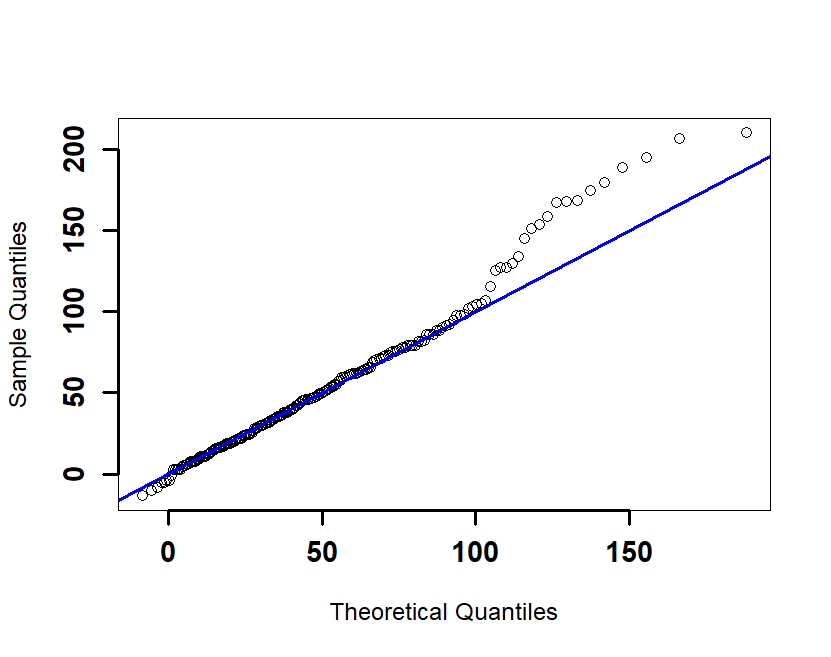}
		\label{FIG:AIS_qq_alp_08}}
	\subfloat[QQ plot for $\alpha=1$]{
		\includegraphics[width=0.3\textwidth]{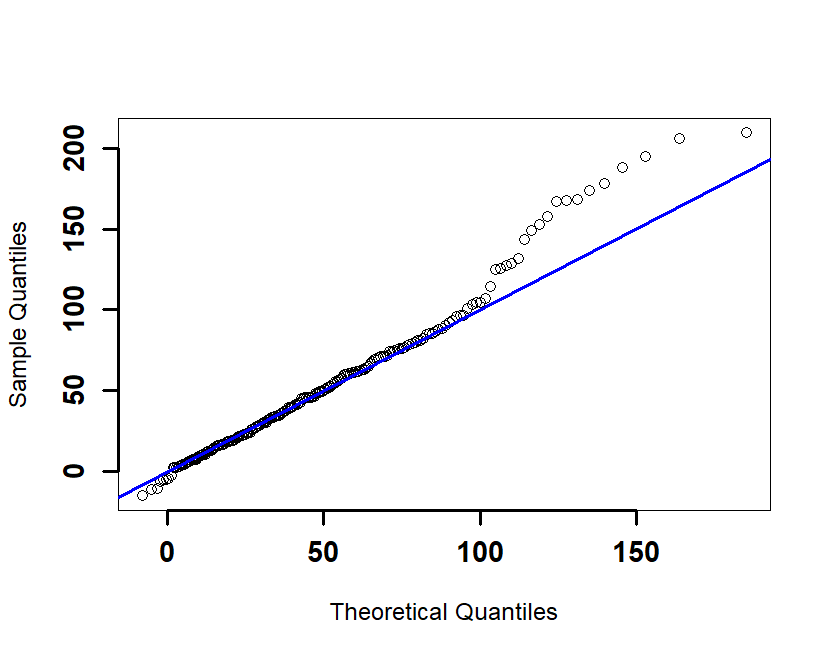}
		\label{FIG:AIS_qq_alp_1}}
	\\
	
	\caption{ 	\color{black} QQ plot of residuals for AIS data at different values of $\alpha$ under the regression model with SN error. The plot in (a) represents the least squares fit for normal errors.}
	\label{FIG:AIS_QQ}
\end{figure}

\begin{figure}[!h]
	 	\color{black}
	\centering
	
	\subfloat[QQ plot for OLS]{
		\includegraphics[width=0.3\textwidth]{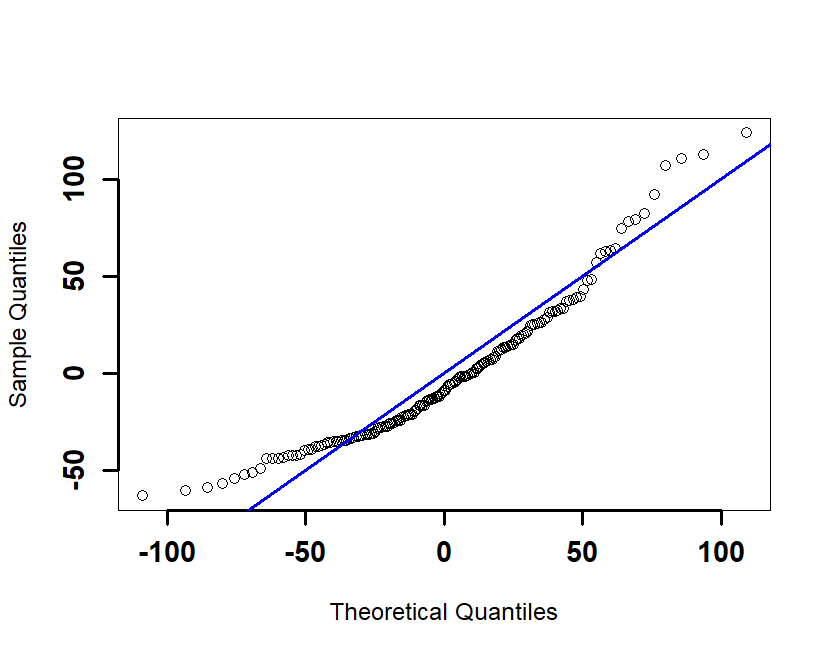}
		\label{FIG:AISOD_qq_ols}}
	\subfloat[QQ plot for $\alpha=0$]{
		\includegraphics[width=0.3\textwidth]{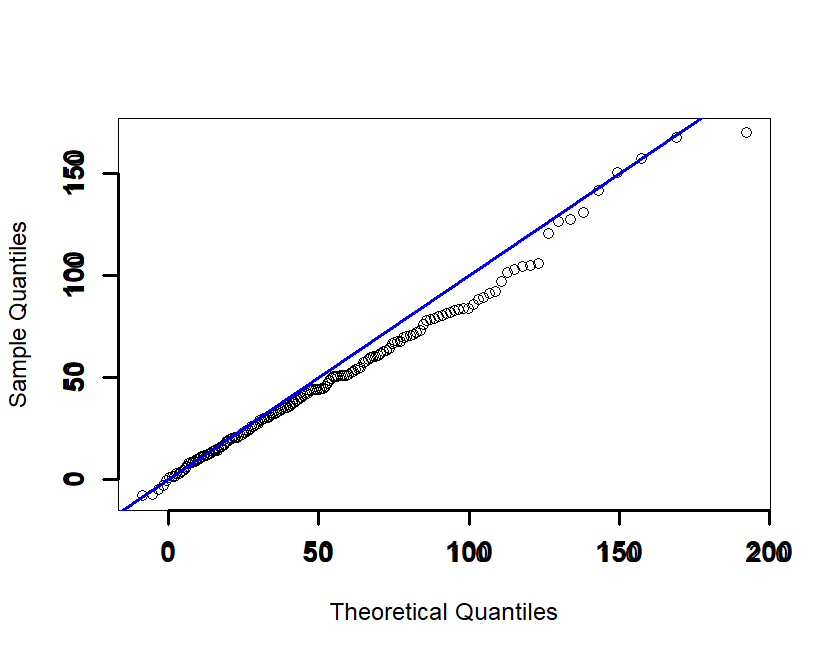}
		\label{FIG:AISOD_qq_alp_0}}
	\subfloat[QQ plot for $\alpha=0.05$]{
		\includegraphics[width=0.3\textwidth]{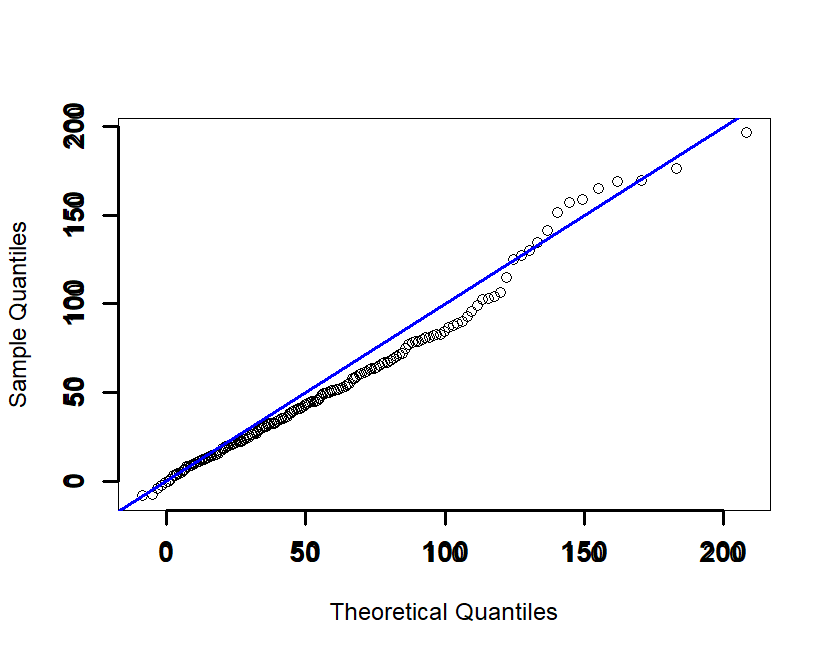}
		\label{FIG:AISOD_qq_alp_005}}
	\\
	\subfloat[QQ plot for $\alpha=0.1$]{
		\includegraphics[width=0.3\textwidth]{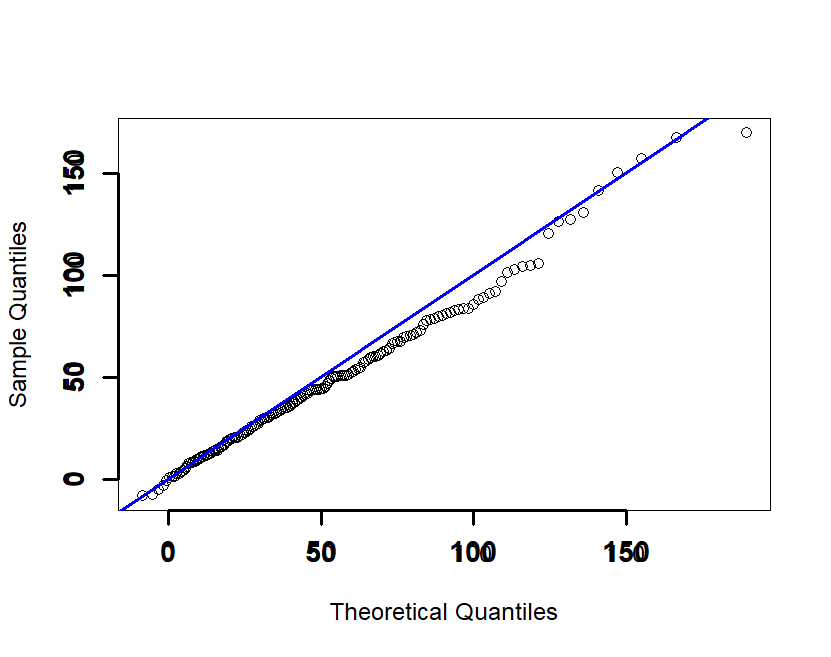}
		\label{FIG:AISOD_qq_alp_01}}
	\subfloat[QQ plot for $\alpha=0.3$]{
		\includegraphics[width=0.3\textwidth]{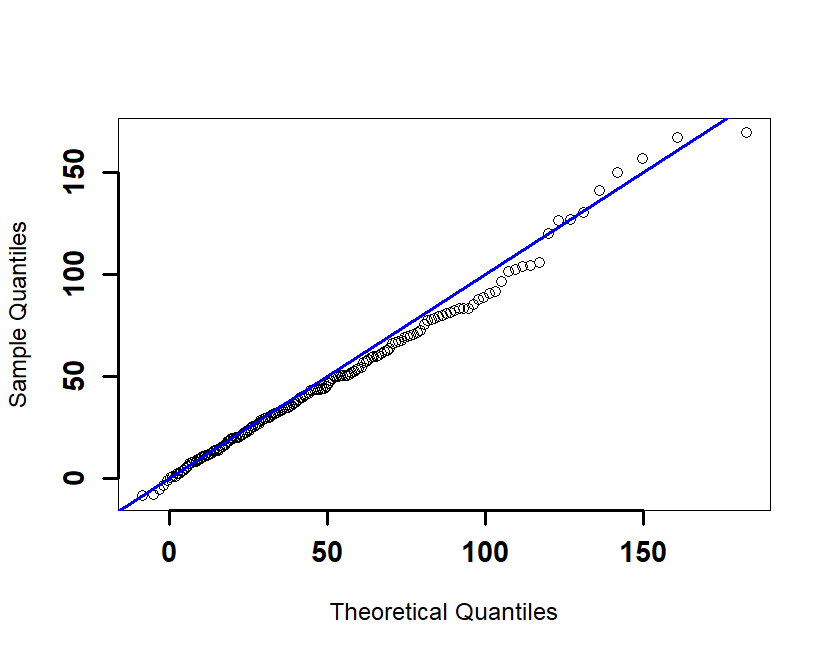}
		\label{FIG:AISOD_qq_alp_03}}
	\subfloat[QQ plot for $\alpha=0.5$]{
		\includegraphics[width=0.3\textwidth]{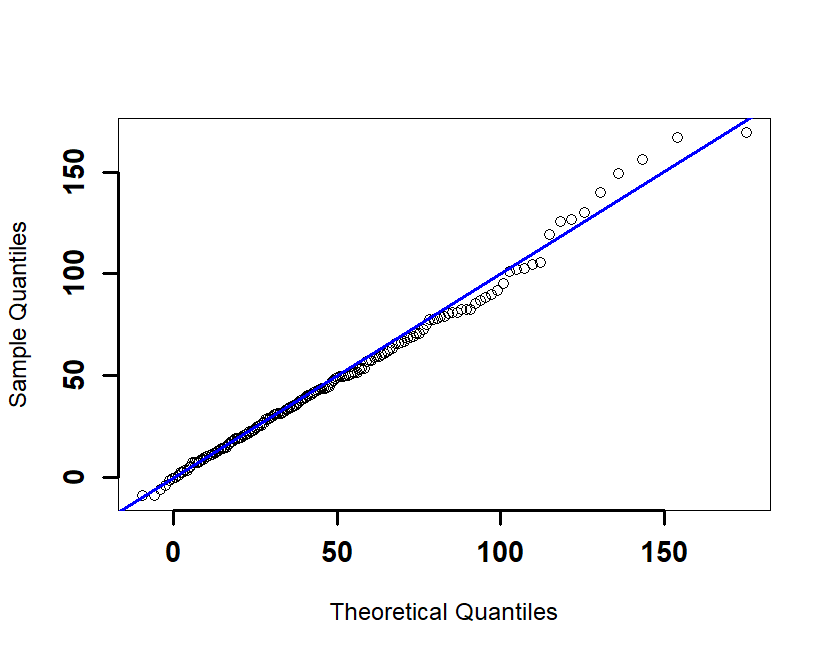}
		\label{FIG:AISOD_qq_alp_05}}
	\\
	\subfloat[QQ plot for $\alpha=0.7$]{
		\includegraphics[width=0.3\textwidth]{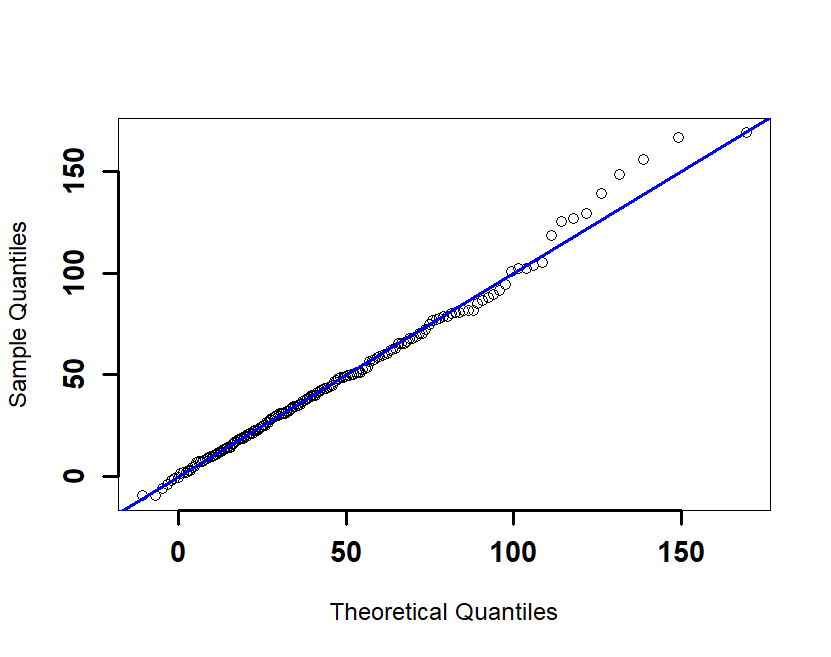}
		\label{FIG:AISOD_qq_alp_017}}
	\subfloat[QQ plot for $\alpha=0.8$]{
		\includegraphics[width=0.3\textwidth]{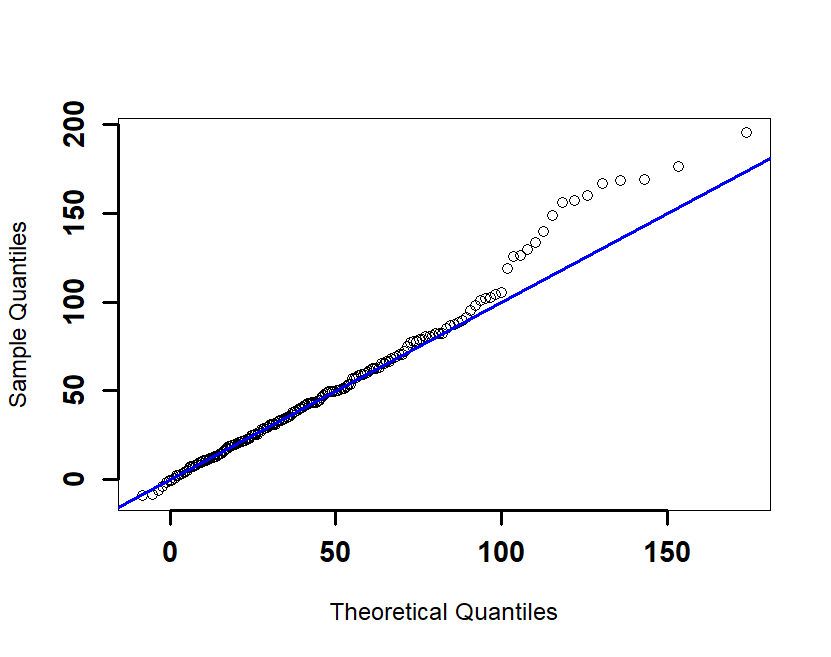}
		\label{FIG:AISOD_qq_alp_08}}
	\subfloat[QQ plot for $\alpha=1$]{
		\includegraphics[width=0.3\textwidth]{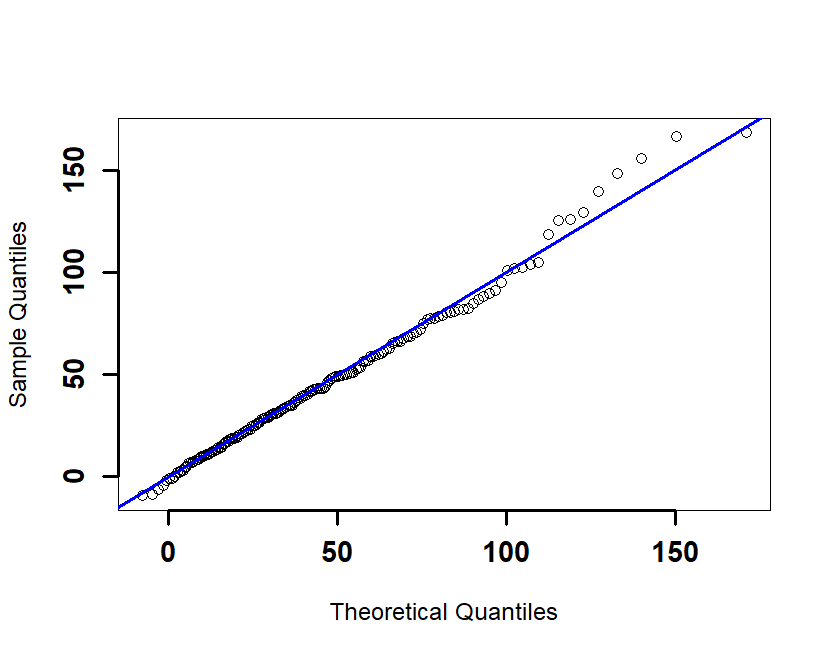}
		\label{FIG:AISOD_qq_alp_1}}
	\\
	
	\caption{ 	\color{black} QQ plot of residuals for AIS data after excluding $28$ observations at different values of $\alpha$ under the regression model with SN error. Figure 7(a) represents the QQ plot of the OLS errors for the normal model, which, even for the outlier deleted case provides a clearly unsatisfactory fit.}
	\label{FIG:AIS_QQ_OUT_DEL}
\end{figure}

 To explain how the analysis based on the procedures corresponding to larger values of the tuning parameter $\alpha$ provide more stable inference, we offer the following description. 
 Together with the results of the full data (containing $207$ observations) analysis, we also provide the analysis based on the outlier deleted data (containing \textcolor{black}{$179$} observations) 
 \textcolor{black}{after removing $28$ outlying observations, identified on the basis of the QQ plot of the residuals under the model fitted with the full data (presented in Figure \ref{FIG:AIS_QQ}).
 The results of the outlier deleted data analysis are presented in Table \ref{TAB:AIS2_out_del1}, and the QQ-plots of the associated residuals are presented in Figure \ref{FIG:AIS_QQ_OUT_DEL}.}

 Contrasting the results of Tables \ref{TAB:AIS2} and \ref{TAB:AIS2_out_del1} provides instructive insights. For the full data analysis of the BMI variable, the p-values indicate that for smaller $\alpha$ the covariate BMI is highly significant, although the significance disappears (or at best becomes an extremely tentative one) for larger $\alpha$ (Table \ref{TAB:AIS2}). Thus different values of $\alpha$ may lead to very different inference on the BMI variable in this case. However, for the outlier deleted data case, the p-values for the same test are \textcolor{black}{increasing} over $\alpha$, all indicating \textcolor{black}{lack of significance (Table \ref{TAB:AIS2_out_del1})}. For the case $\alpha = 0$, the p-values of this test are widely disparate between the full data and outlier deleted cases ($0.0035$ versus \textcolor{black}{$0.0609$}). On the other hand, for $\alpha = 1$, the p-value is only minimally affected by the presence or absence of outliers ($0.0603$ versus \textcolor{black}{$0.0896$}). Clearly the significance of the BMI variable in the $\alpha = 0$ case is driven by the $28$ excluded observations, which is also the case for other low values of $\alpha$; but the effect of the outliers is minimal for large $\alpha$, demonstrating the stability of the procedures for such $\alpha$. On the whole, we repose our faith on the analysis based on large values of $\alpha$ in this case, as they are stable (minimally affected by the presence or absence of outliers) and closely match the consistent results of the analysis for the outlier deleted case, where the conclusions are consistent over different values of $\alpha$ (small and large).

\begin{figure}[!h]
 	\color{black}
 	\centering
 		
 		\subfloat[RD for Intercept]{
 			\includegraphics[width=0.33\textwidth]{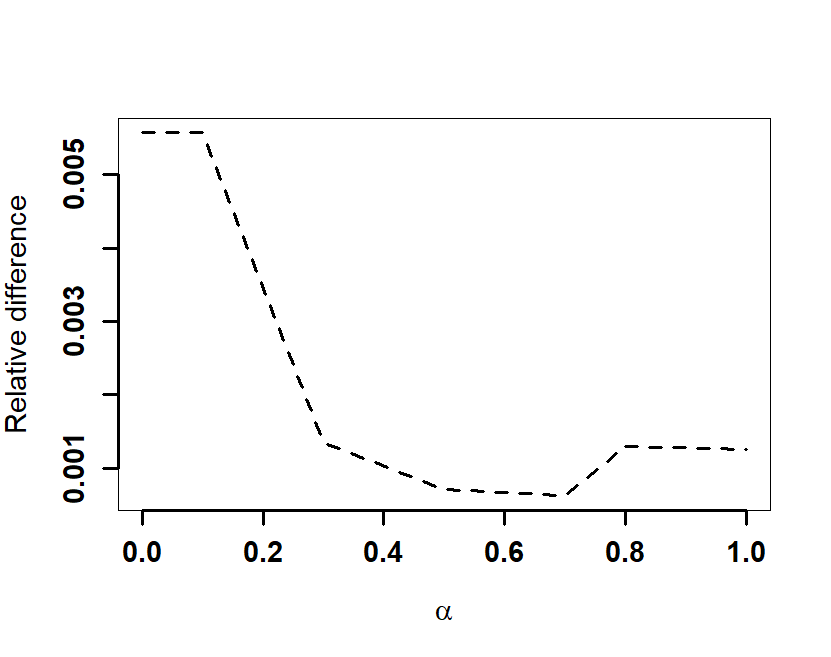}
 			\label{FIG:RD_ais_intercept}}
 		\subfloat[RD for BMI]{
 			\includegraphics[width=0.33\textwidth]{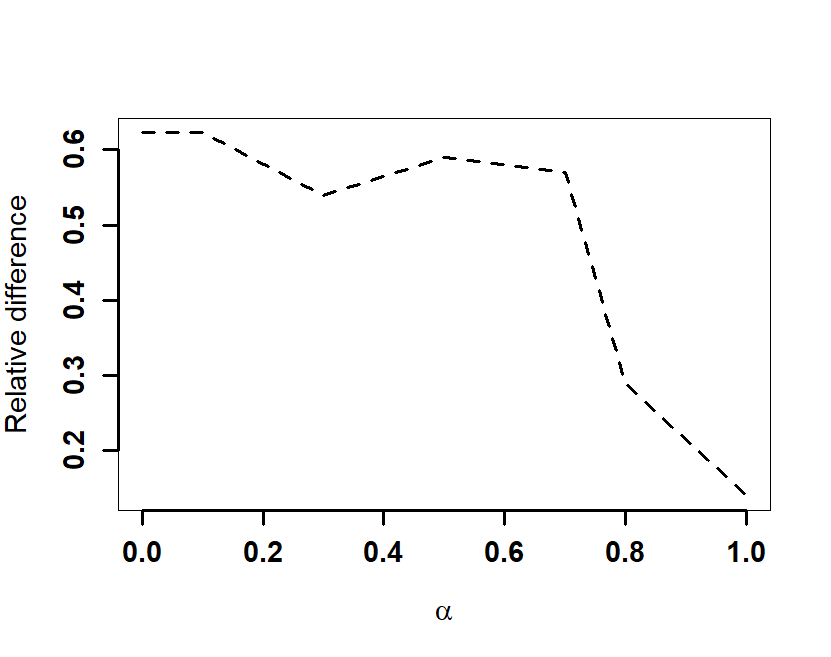}
 			\label{FIG:RD_ais_bmi}}
 		\subfloat[RD for LBM]{
 			\includegraphics[width=0.33\textwidth]{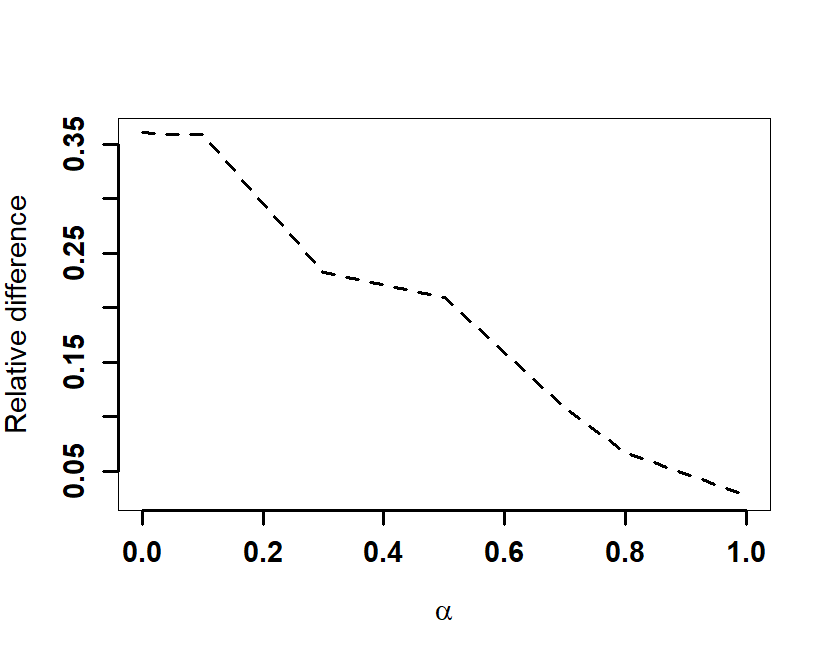}
 			\label{FIG:RD_ais_lbm}}\\
 		\subfloat[RD for $\sigma$]{
 			\includegraphics[width=0.33\textwidth]{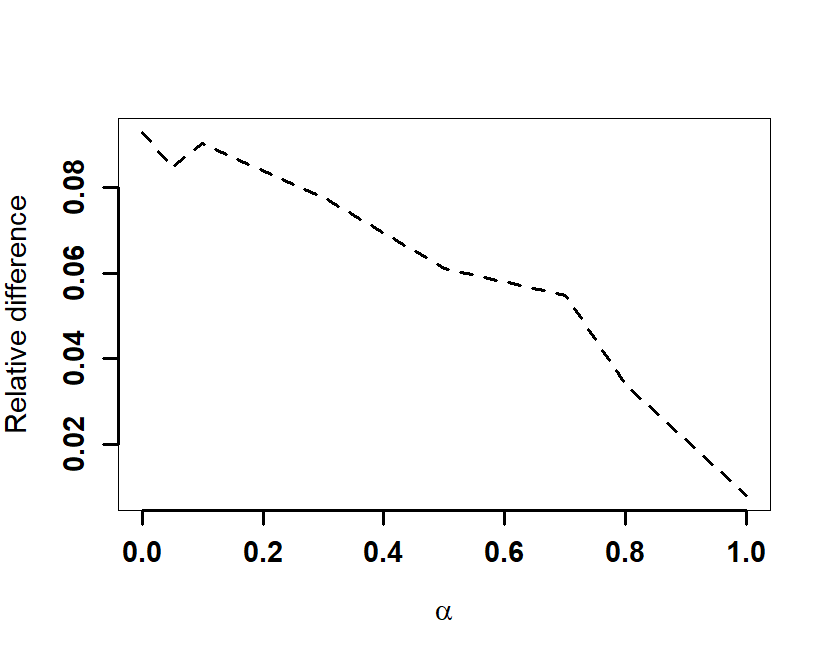}
 			\label{FIG:RD_ais_sigma}}
 		\subfloat[RD for $\gamma$]{
 			\includegraphics[width=0.33\textwidth]{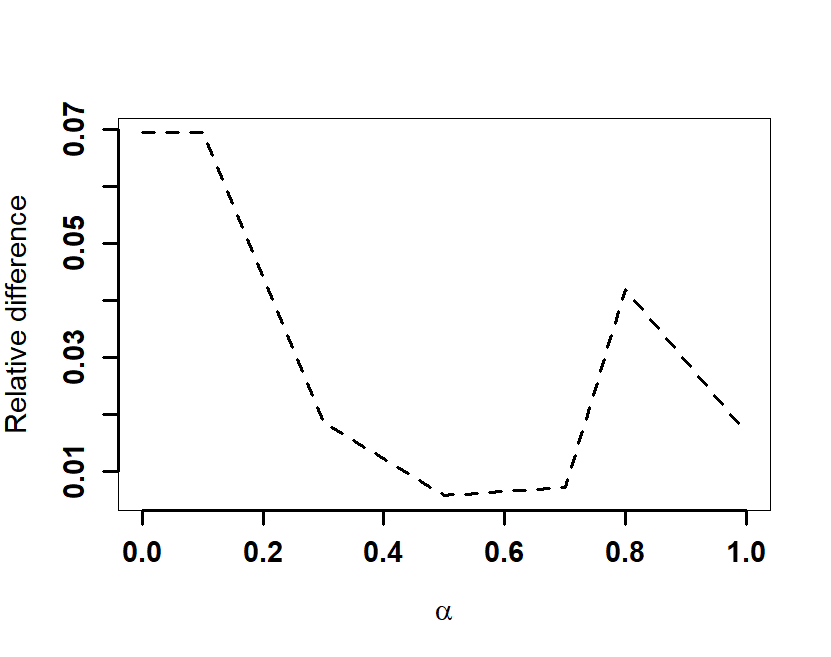}
 			\label{FIG:RD_ais_gamma}}\\

 	\caption{\color{black} Relative differences (RD) between the MDPDEs obtained with and without outliers for the AIS data across different values of $\alpha$.}
 	\label{FIG:AIS_RD_NEW}
 \end{figure}

 \textcolor{black}{Further, we have also computed the relative changes in the estimators obtained from the data with and without outliers, which are presented in Figure \ref{FIG:AIS_RD_NEW}. Here, these relative differences are computed as $|\frac{\boldsymbol{\widehat{\theta}}_{\text{full}}-\boldsymbol{\widehat{\theta}}_{\text{clean}}}{\boldsymbol{\widehat{\theta}}_{\text{full}}}|$, where $\boldsymbol{\widehat{\theta}}_{\text{full}}$ and $\boldsymbol{\widehat{\theta}}_{\text{clean}}$ denote the estimates obtained from the full and the outlier-deleted (clean) data. 
 	From Figure \ref{FIG:AIS_RD_NEW}, we can see that these relative differences decrease as $\alpha$ increases except for ``Intercept" and $\gamma$, where the relative differences slightly increase for $\alpha$ in $(0.7,0.9)$. On the whole, it indicates the robustness of our method for larger $\alpha$, where the estimates of the parameters become closer for both cases (with and without outliers).}

 \textcolor{black}{From the QQ plot of full data in Figure \ref{FIG:AIS_QQ}, we can see that the ordinary least squares (OLS) estimator based on the ordinary normal model does not fit the data well 
 	at all as the majority of the points including those in the tails stray off the straight line (Figure 5a), indicating that the ordinary normal distribution does not properly model the errors for these data. 
 	However, when we use the SN regression model, the associated QQ plots clearly demonstrate that this error structure is a much more realistic model for the AIS data giving a perfect fit for most of the data points except a bunch of outliers in the right tail. When these outliers, thus identified, are deleted from the dataset, the QQ-plots of the fitted models (in Figure \ref{FIG:AIS_QQ_OUT_DEL}) show excellent conformity with the target straight line indicating that these outlier-deleted data are very well-modeled by the SN regression structure.}

 \begin{figure}[!b]
 	\centering
 		\subfloat[p-value for BMI]{
 			\includegraphics[width=0.4\textwidth]{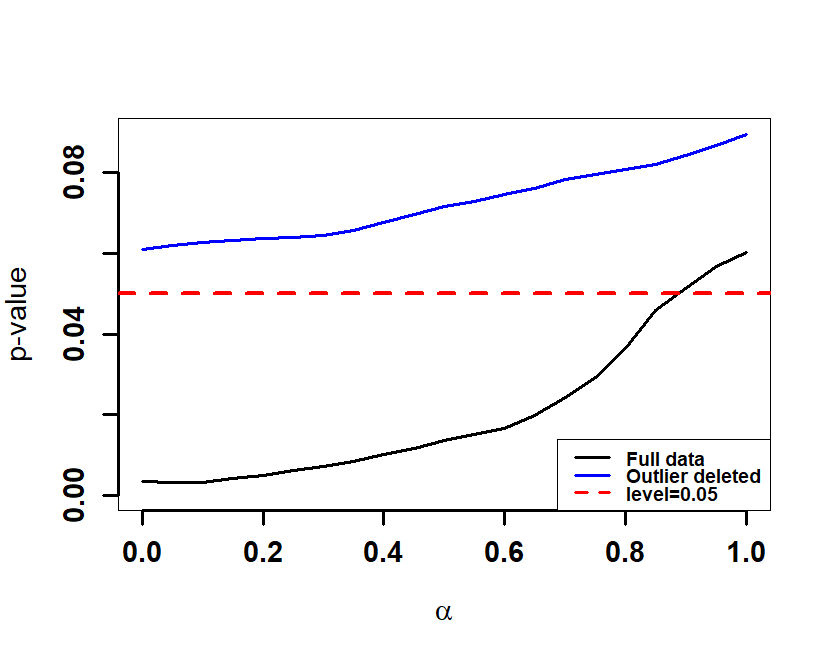}
 			\label{FIG:pvalue_BMI}}
 		\subfloat[p-value for LBM]{
 			\includegraphics[width=0.4\textwidth]{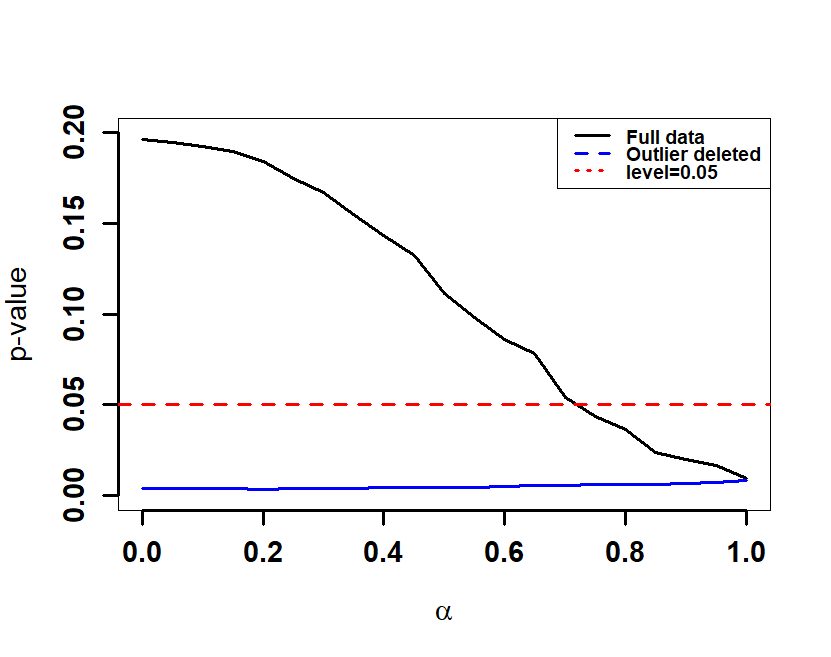}
 			\label{FIG:pvalue_LBM}}\\
 		
 		\subfloat[p-value for $\gamma$]{
 			\includegraphics[width=0.4\textwidth]{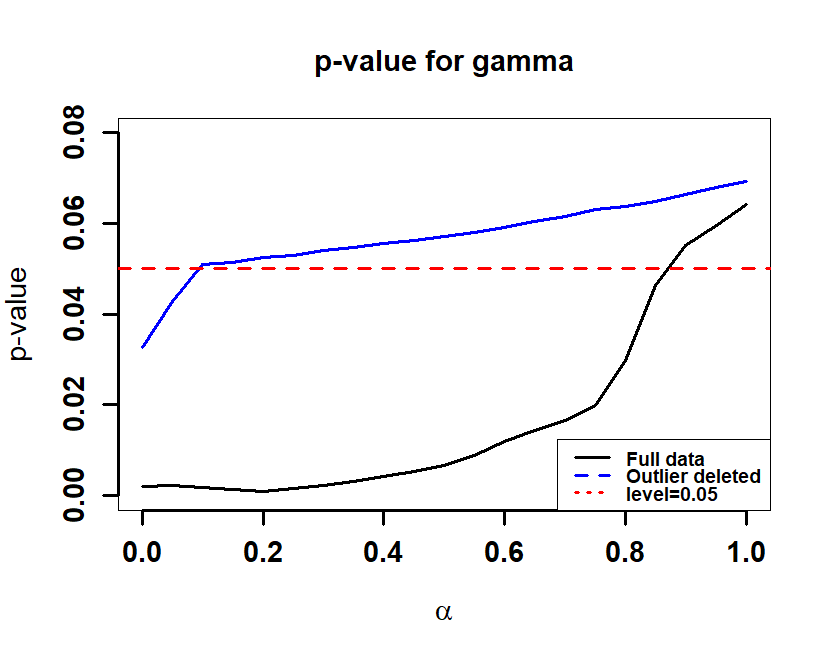}
 			\label{FIG:pvalue_AIS_gamma}}
 	\caption{p-values for the tests of significance of covariates and test of symmetry in case of AIS data. 
 		[Solid black line: Full data, Solid black line: Outlier deleted data, Dashed red line: level of significance.]}
 	\label{FIG:AIS2_pvalue}
 \end{figure}

The p-values of the three tests (for the significance of the BMI, the LBM and the test of symmetry) are graphically presented in Figure \ref{FIG:AIS2_pvalue} to give a better visual representation. For the LBM variable, on the other hand, insignificant p-values are pushed to significance through the removal of outliers for lower values of $\alpha$, whereas for large values of $\alpha$ the p-values remain stable, exhibiting significance with or without the outliers. All these results, taken together, indicate that the analysis based on larger values of $\alpha$ provide protection against both kinds of errors -- false declaration of significance when there is none, and failure to detect true significance -- and should therefore be the trusted version of our analysis. The results of the significance test of the $\gamma$ parameter show exactly similar trends. A strong significance at low $\alpha$ is pushed to weak significance through the removal of the outliers. At larger $\alpha$ (\textcolor{black}{for $\alpha>0.1$}) the p-value remains stable, demonstrating a lack of significance/weak significance with or without the outliers.

\subsection{Crime data}
\label{SEC:CRIME}
In this  example we will once again demonstrate how the presence or absence of a handful of outlying observations can drive the decision for small $\alpha$, but for large $\alpha$ the outliers have little or no effect. We consider the Crime data (consisting of $51$ observations from each of $51$ states of the United States of America, measured in 1993) presented in Agresti and Finlay (1997) \cite{Agresti/Finlay:1997}. The dataset is loaded in R studio from the link: \url{https://stats.idre.ucla.edu/stat/data/crime.dta}. 

Among the available variables within this dataset, we use the number of murders per $1,000,000$ people in each state (denoted as ``murder") as our response variable and the percentage of people living under the poverty line (denoted as ``poverty") and percentage of populations who are single parents (denoted as ``single") as our explanatory variables. Thus, we fit the model 
$$\text{murder}_{i}=\beta_{0}+\beta_{1}\text{poverty}_{i}+\beta_{2}\text{single}_{i}+\epsilon_{i},~~~i=1, \ldots, 51,
$$ 
where the subscript $i$ represents the observation from $i$th state and $\epsilon_{i}$ is the error component having SN distribution as in (\ref{EQ:lm}).

\begin{figure}[h]
	\color{black}
	\centering
	
	\subfloat[RD for Intercept]{
		\includegraphics[width=0.33\textwidth]{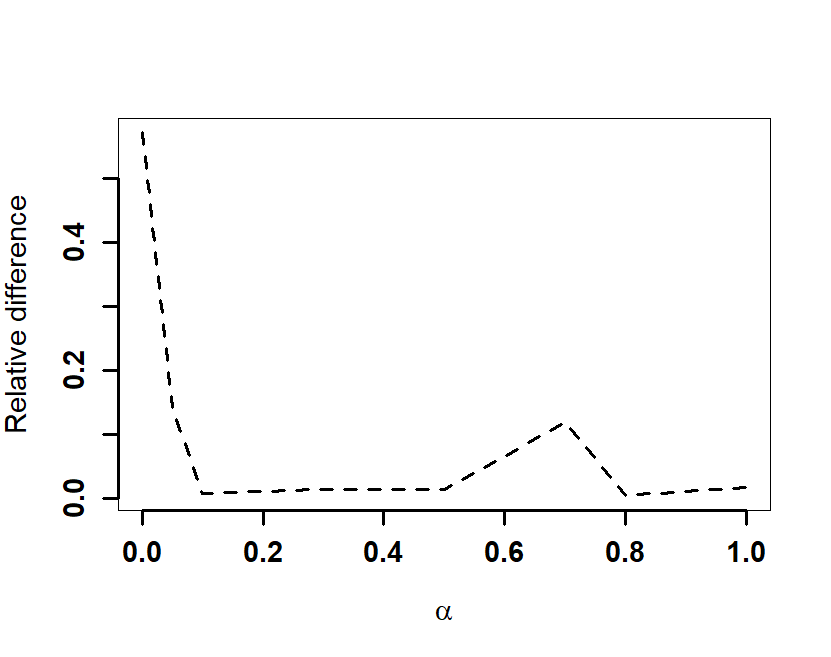}
		\label{FIG:RD_crime_intercept}}
	\subfloat[RD for Poverty]{
		\includegraphics[width=0.33\textwidth]{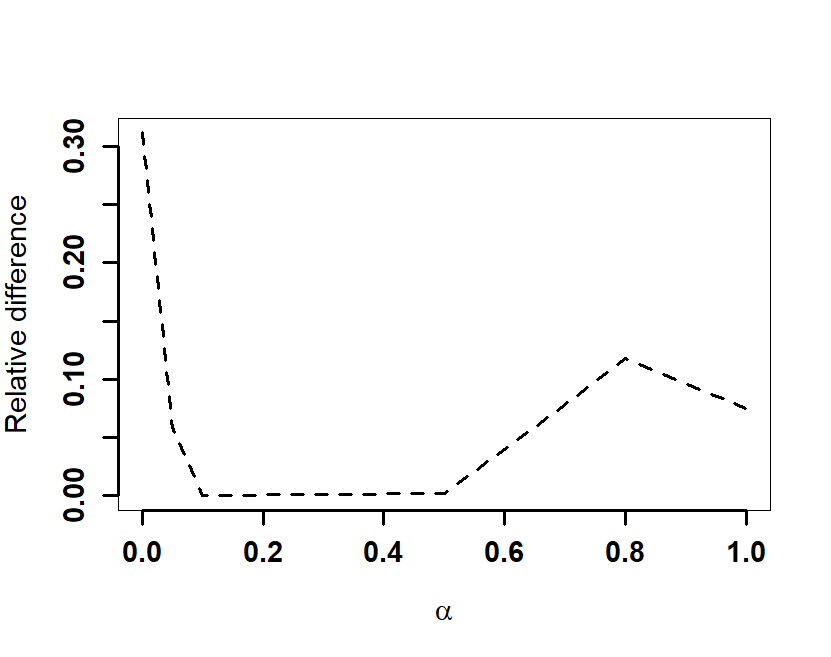}
		\label{FIG:RD_crime_poverty}}
	\subfloat[RD for Single]{
		\includegraphics[width=0.33\textwidth]{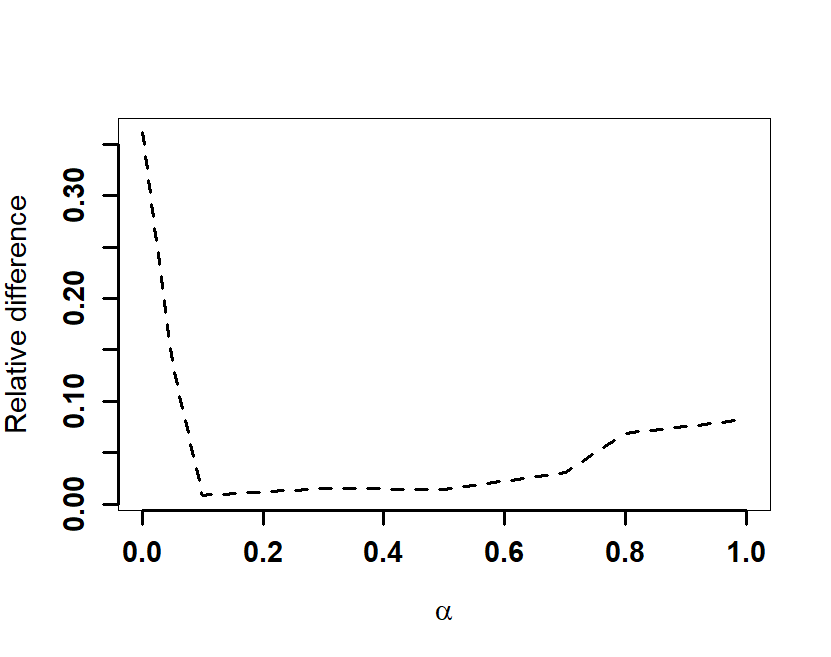}
		\label{FIG:RD_crime_single}}\\
	\subfloat[RD for $\sigma$]{
		\includegraphics[width=0.33\textwidth]{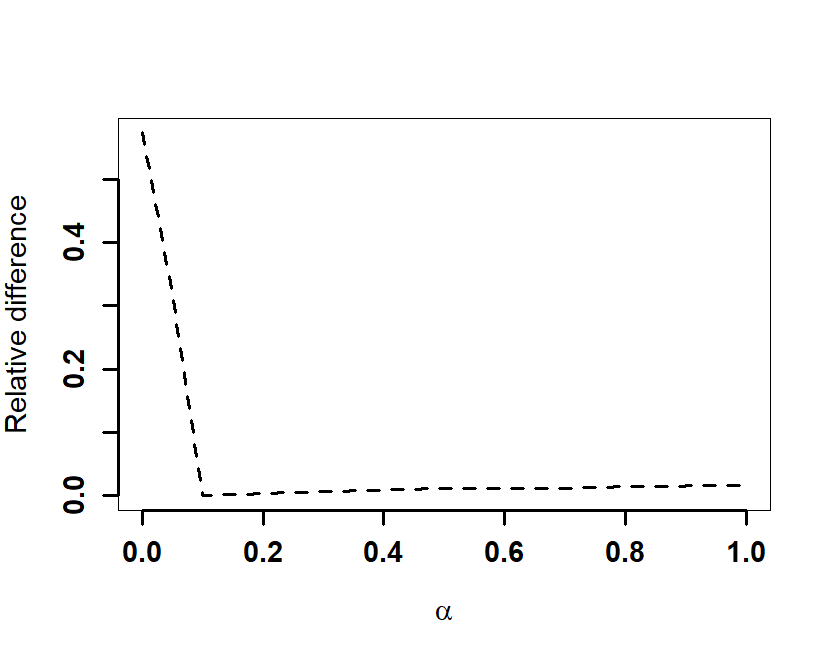}
		\label{FIG:RD_crime_sigma}}
	\subfloat[RD for $\gamma$]{
		\includegraphics[width=0.33\textwidth]{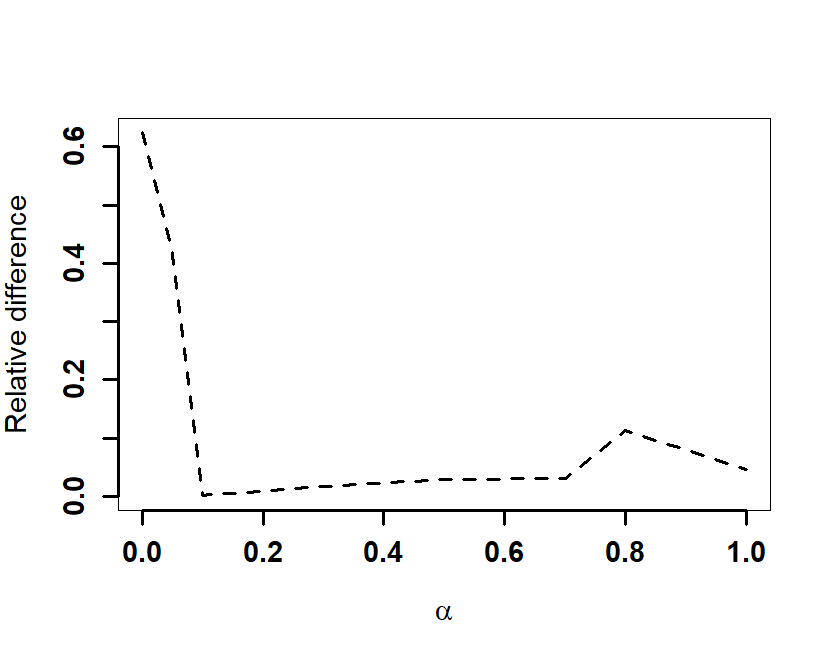}
		\label{FIG:RD_crime_gamma}}\\
	
	\caption{\color{black} Relative differences (RD) between the MDPDEs obtained with and without outliers for the Crime data across different values of $\alpha$.}
	\label{FIG:Crime_RD}
\end{figure}

The results obtained from the full data are presented in Table \ref{TAB:Crime}. The last ($51$st) observation is a clear outlier by any standards as observed from the residuals of the regression for any $\alpha$. The results of the outlier deleted regression (based on the remaining $50$ observations) are presented in Table \ref{TAB:Crime_out_del} and \textcolor{black}{the relative differences between these estimates are plotted in Figure \ref{FIG:Crime_RD}}.

\begin{figure}[!h]
	\centering
		\subfloat[p-value for Poverty]{
			\includegraphics[width=0.5\textwidth]{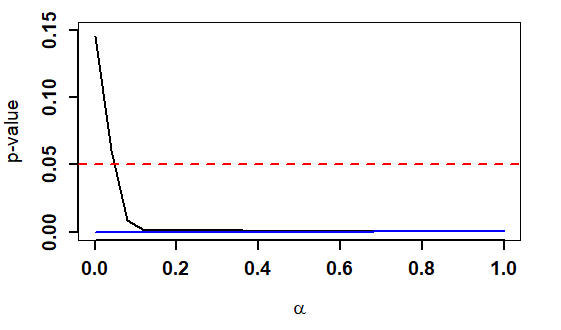}
			\label{FIG:pvalue_poverty}}
		\subfloat[p-value for $\gamma$]{
			\includegraphics[width=0.5\textwidth]{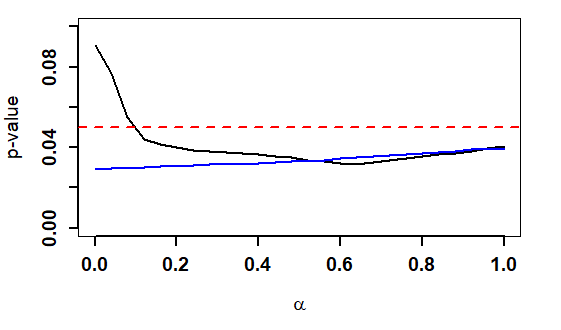}
			\label{FIG:pvalue_crime_gamma}}
	\caption{p-values for the tests of significance of ``Poverty" and test of symmetry in case of Crime data. [Solid black line: Full data, Solid black line: Outlier deleted data, Horizontal dashed red line: level of significance.]}
	\label{FIG:Crime_pvalue}
\end{figure}

There are many similarities between the comparative analysis of Tables \ref{TAB:Crime} and \ref{TAB:Crime_out_del} and the comparative analysis of Tables \ref{TAB:AIS2} and \ref{TAB:AIS2_out_del1}. An additional observation is that stable results are obtained in Table \ref{TAB:Crime} even for very small (positive) values of $\alpha$. In case of the ``poverty" variable, for example, the insignificance at $\alpha = 0$ is turned into strong significance by the removal of the single outlier; however, for full data even a small positive value of $\alpha$ such as $0.05$ exhibits a large drop in p-value compared to $\alpha = 0$, and the removal of the outlier has no effect on the decision in this case. \textcolor{black}{From Figure \ref{FIG:Crime_RD}, we can see that the differences decrease as $\alpha$ increase. 
	Though, the differences increase for $\alpha \ge 0.7$, the changes in the magnitude does not appear significant. 
	Also, in case of $\sigma$, these values stabilize even for smaller $\alpha$.} Please see also the plots of p-values in Figure \ref{FIG:Crime_pvalue}.
This phenomenon is consistent with the observations of other authors (e.g., Basak et al. (2021) \cite{Basak/etc:2020}) where it has been observed that if the outliers are widely discrepant with the main data cloud, even small (positive) values of $\alpha$ can discount them.

\begin{table}[!h]
	\caption{Parameter estimates along with standard error (SE) and p-values for Crime data for different $\alpha$}
	\centering
	\resizebox{1\textwidth}{!}{
		\begin{tabular}{c | c  c c c c c c c}
			\hline
			Parameter & \multicolumn{8}{c}{MDPDE at different $\alpha$}\\
			& 0(MLE) & 0.05 & 0.1 & 0.3 & 0.5 & 0.7 & 0.8 & 1\\
			\hline
			Intercept	&	$-23.5159$	&  $-17.0934$ &	$-14.8622$	&	$-14.7993$	&	$-14.8492$	&	$-14.5321$	&	$-17.4511$ & $-19.0914$\\
			SE	&	(2.4342)	&  (1.9853)&	(1.6189)	&	(1.6701)	&	(1.7170)	&	(1.6926)	& (1.4126) &	(0.9116)	\\
			p-value	&	(4.43E-22)	&  (8.45E-21)&	(4.28E-20)	&	(7.91E-19)	&	(5.23E-18)	&	(9.02E-18)	& (1.24E-21) &	(1.10E-23)	\\ \hline
			
			Poverty	&	0.2805	& 0.3495&	0.3708	&	0.3730	&	0.3725	&	0.3681	&  0.4604 &	0.4743	\\
			SE	&	(0.1148) & (0.1083)	&	(0.1052)	&	(0.0779)	&	(0.0804)	&	(0.0798)	& (0.067) &	(0.0436)	\\
			p-value	&	(0.1459)	& (0.0007)	& (8.20E-04)	&	(6.20E-04)	&	(1.68E-04)	&	(3.64E-05)	& (3.73E-05) &	(3.93E-05)	\\ \hline
			
			Single	&	2.0139	&  1.4853 &	1.2701	&	1.2565	&	1.2607	&	1.2427	&	1.4403 & 1.5882	\\
			SE	&	(0.2482)	&  (0.1973)&	(0.1625)	&	(0.1683)	&	(0.1738)	&	(0.1724)	& (0.1447) &	(0.0943)	\\
			p-value	&	(4.85E-04)	& (3.95E-04) &	(3.73E-04)	&	(5.48E-04)	&	(8.35E-04)	&	(4.08E-04)	&  (4.72E-04)&	(5.60E-04)	\\ \hline
			$\sigma$	&	8.6706	& 5.4882 &	3.7718	&	3.9108	&	3.9956	&	4.0052	& 3.7316  &	3.6592	\\
			SE	&	(0.9717)	&  (0.6713)&	(0.5321)	&	(0.5739)	&	(0.6249)	&	(0.6558)	& (0.5924)	&(0.5411)	\\ \hline
			$\gamma$	&	8.6820	& 5.7169 &	3.3957	&	3.7598	&	4.1798	&	4.9330	& 6.7388	&8.2260	\\
			SE	&	(5.1359)	& (3.0184) &	(1.6423)	&	(1.9707)	&	(2.4778)	&	(3.3800)	&	(4.2761) &(5.5665)	\\
			p-value	&	(0.0909)	& (0.0413) &	(0.0387)	&	(0.0364)	&	(0.0316)	&	(0.0344)	& (0.0382) & (0.0406)	\\
			\hline
	\end{tabular}}
	\label{TAB:Crime}
\end{table}

\begin{table}[!h]
	\caption{Parameter estimates along with standard error (SE) and p-values for Crime data for different $\alpha$ after excluding the outlier}
	\centering
	\resizebox{1\textwidth}{!}{
		\begin{tabular}{c | c  c c c c c c c}
			\hline
			Parameter & \multicolumn{8}{c}{MDPDE at different $\alpha$}\\
			& 0(MLE) & 0.05 & 0.1 & 0.3 & 0.5 & 0.7 & 0.8 & 1\\
			\hline
			Intercept	&	$-14.9633$ & $-14.9713$	&  $-14.9776$ &	$-15.0102$	&	$-15.0459$	&	$-16.4709$	& $-17.5197$ &	$-18.7928$\\
			SE	&	(1.2255) & (1.2392)	&  (1.2483)&	(1.3284)	&	(1.4002)	&	(1.5119)	& (1.6736) &	(1.6919)\\
			p-value	&	(4.15E-25)	&  (7.23E-25)&	(9.94E-25)	&	(2.31E-24)	&	(5.42E-24)	&	(7.11E-24)	& (8.54E-24) &	(1.03E-23)\\ \hline
			Poverty	&	0.3680 & 0.3701	& 0.3707 &	0.3734	&	0.3733	&	0.3971	& 0.4061 &	0.4389\\
			SE	&	(0.0343) & (0.0351) & (0.0355)	&	(0.0370)	&	(0.0395)	&	(0.0409)	& (0.0427) &	(0.0458)\\
			p-value &(1.19E-05) & (1.39E-05) &(1.47E-05) & (2.19E-05) & (2.39E-05) & (2.86E-05) & (3.23E-05) & (3.71E-05)  \\
			\hline
			Single	&	1.2859 & 1.2815	&  1.2805 &	1.2760	&	1.2790	&	1.2808	& 1.3416 &	1.4582\\
			SE	&	(0.1061) & (0.1079)	&  (0.1085)&	(0.1166)	&	(0.1341)	&	(0.1464)	& (0.1679) &	(0.1940)\\
			p-value &(1.81E-05)&(2.35E-05) &(2.68E-05)&(4.26E-05)&(4.63E-05)&(5.72E-05)&(6.49E-05)&(7.84E-05)\\ \hline
			$\sigma$	&	3.6931 & 3.7411	& 3.7710 &	3.8875	&	3.9528	&	3.9590	& 3.7868 &	3.6008\\
			SE	&		(0.5163)& (0.5208) & (0.5272)	&  (0.5369)&	(0.5792)	&	(0.6298)	& (0.5768) &	(0.6497)\\ \hline
			$\gamma$	&	3.2595 & 3.3314	& 3.4009 &	3.6961	&	4.0558	&	5.0839	& 5.9720 &	7.8440	\\
			SE	&	(1.5748) & (1.6074)	& (1.6611) &	(1.9551)	&	(2.4217)	&	(3.5364)	& (3.7488) &	(4.3472)\\
			p-value &  (0.0292)& (0.0307) & (0.0312) & (0.0325) & (0.0338) & (0.0353) & (0.0376) &  (0.0397)\\
			\hline
	\end{tabular}}
	\label{TAB:Crime_out_del}
\end{table}

\begin{figure}[!h]
	\color{black}
	\centering
		
		\subfloat[QQ plot for OLS]{
			\includegraphics[width=0.3\textwidth]{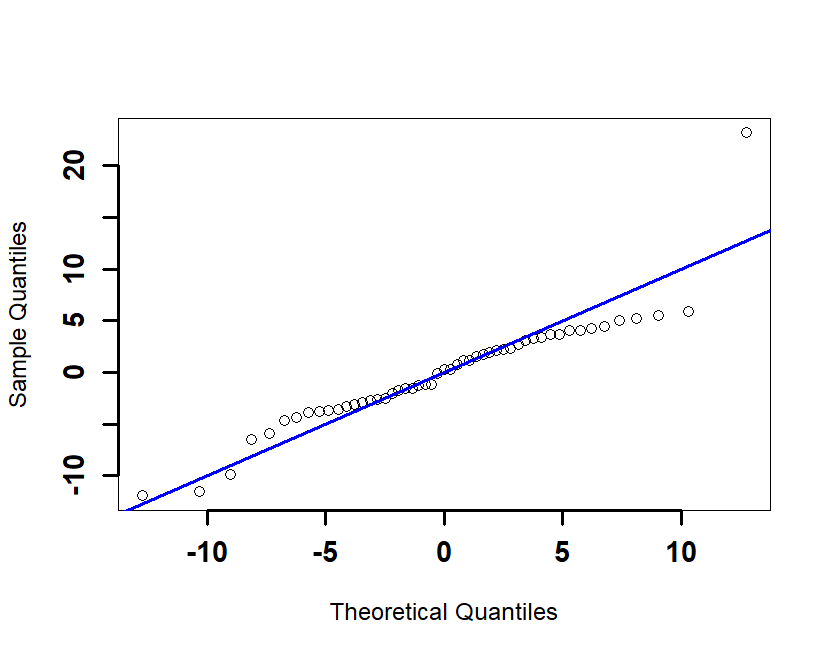}
			\label{FIG:Crime_qq_ols}}
		\subfloat[QQ plot for $\alpha=0$]{
			\includegraphics[width=0.3\textwidth]{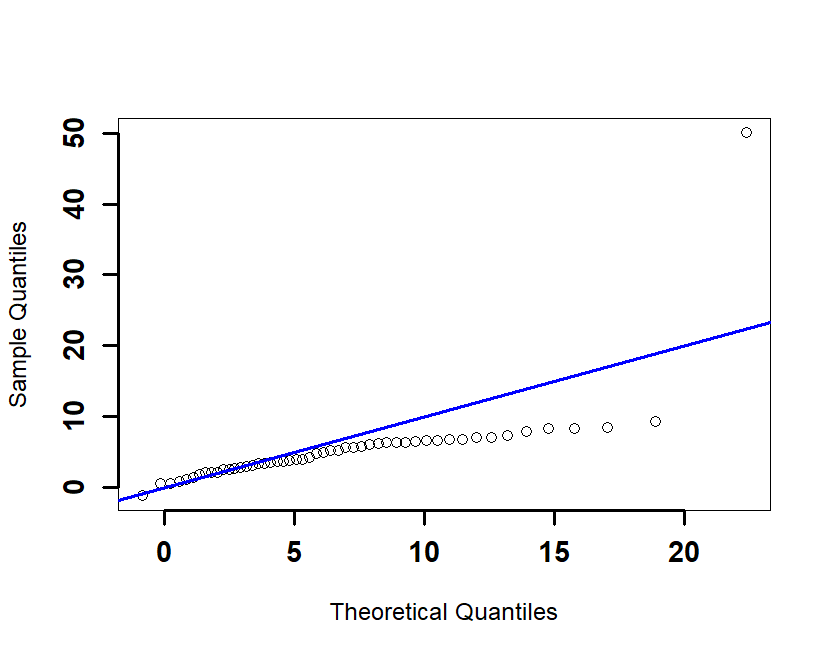}
			\label{FIG:Crime_qq_alp_0}}
		\subfloat[QQ plot for $\alpha=0.05$]{
			\includegraphics[width=0.3\textwidth]{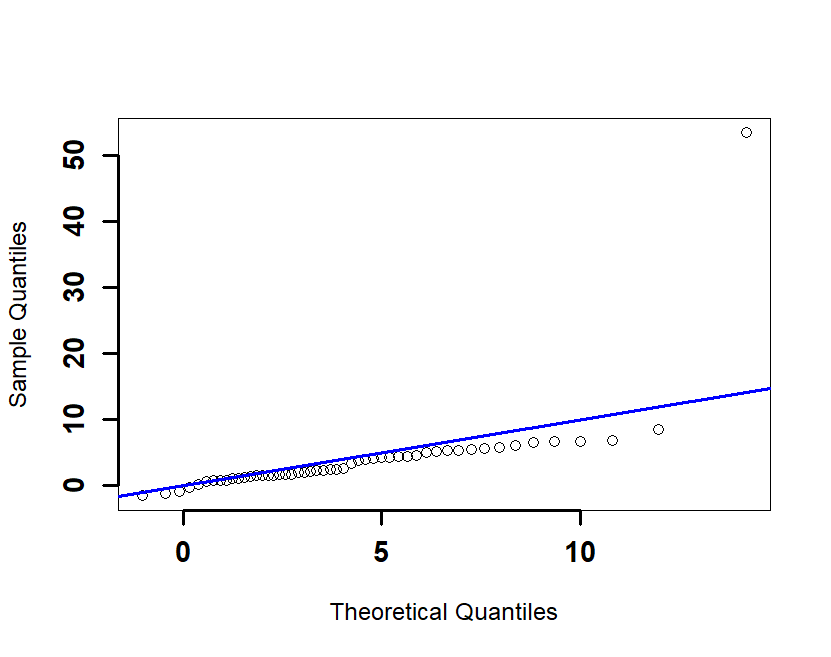}
			\label{FIG:Crime_qq_alp_005}}
		\\
		\subfloat[QQ plot for $\alpha=0.1$]{
			\includegraphics[width=0.3\textwidth]{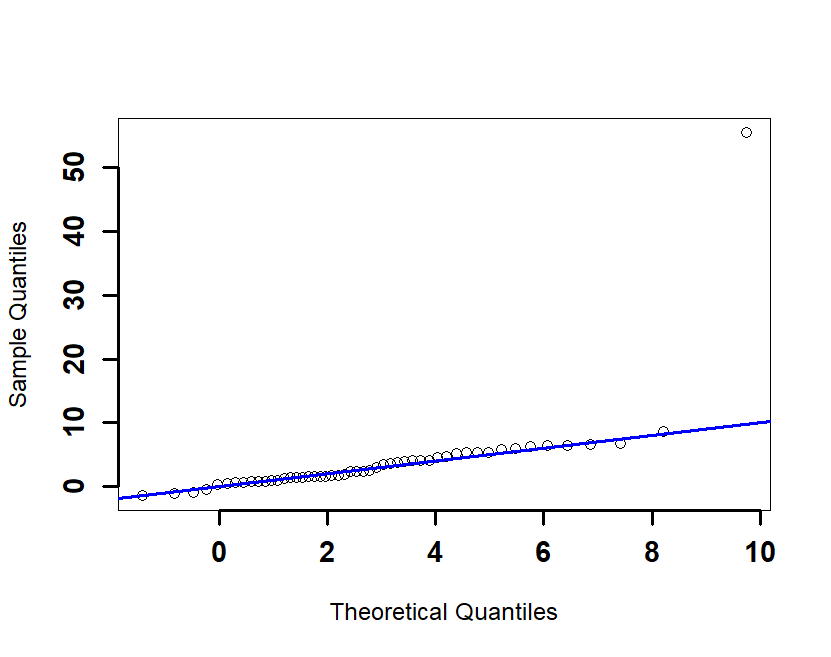}
			\label{FIG:Crime_qq_alp_01}}
		\subfloat[QQ plot for $\alpha=0.3$]{
			\includegraphics[width=0.3\textwidth]{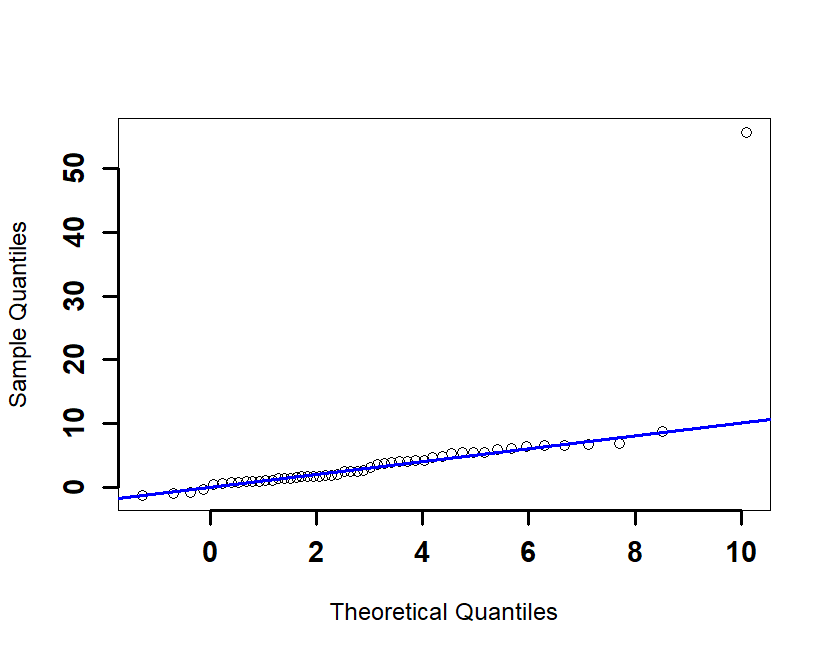}
			\label{FIG:Crime_qq_alp_03}}
		\subfloat[QQ plot for $\alpha=0.5$]{
			\includegraphics[width=0.3\textwidth]{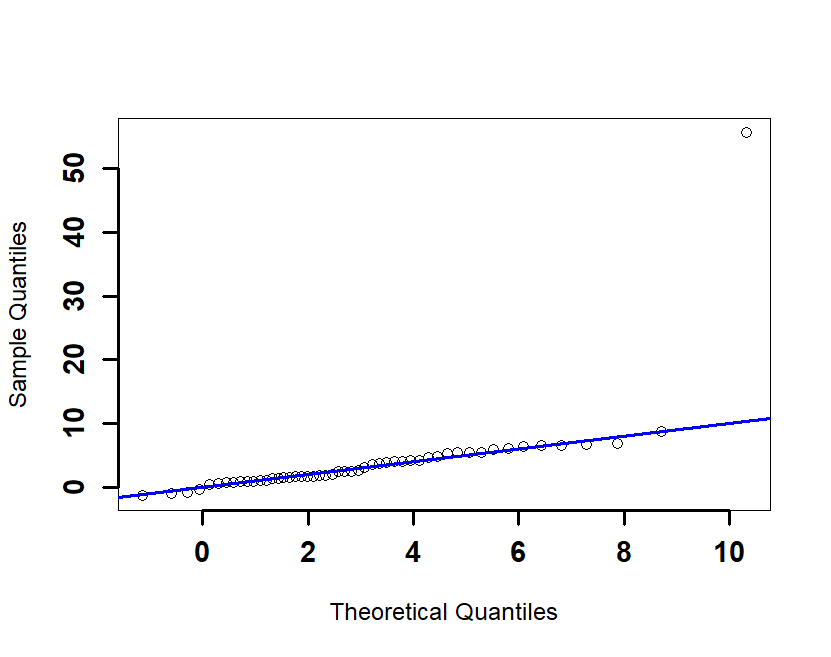}
			\label{FIG:Crime_qq_alp_05}}
		\\
		\subfloat[QQ plot for $\alpha=0.7$]{
			\includegraphics[width=0.3\textwidth]{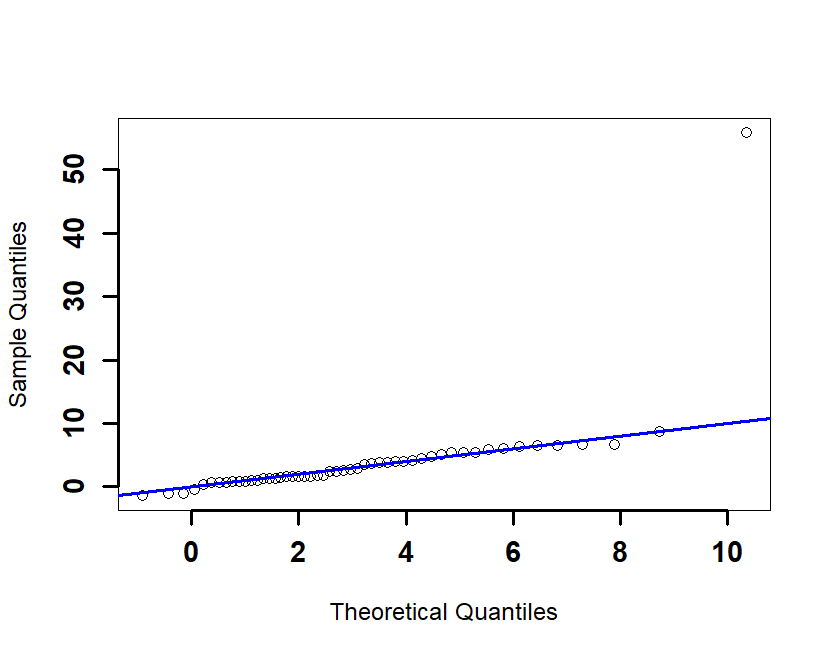}
			\label{FIG:Crime_qq_alp_07}}
		\subfloat[QQ plot for $\alpha=0.8$]{
			\includegraphics[width=0.3\textwidth]{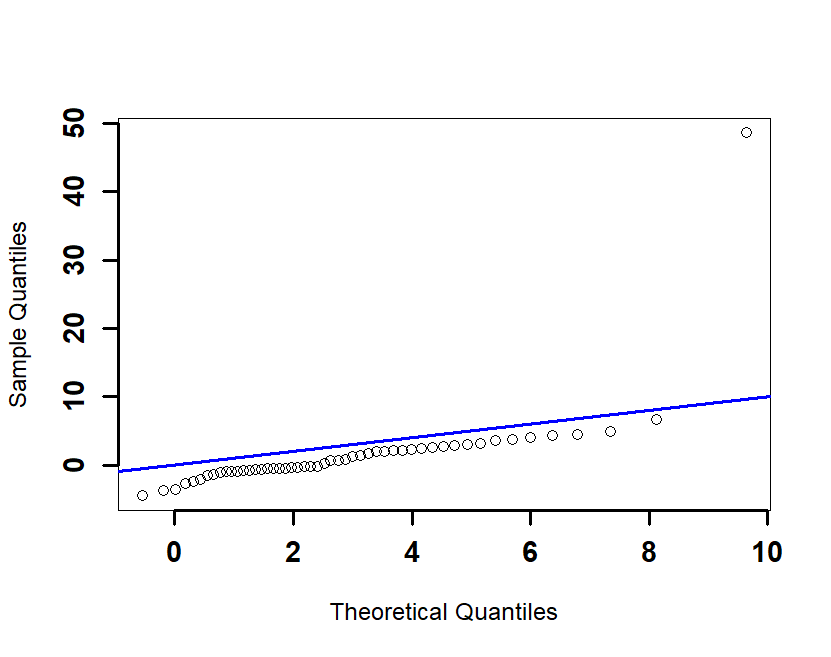}
			\label{FIG:Crime_qq_alp_08}}
		\subfloat[QQ plot for $\alpha=1$]{
			\includegraphics[width=0.3\textwidth]{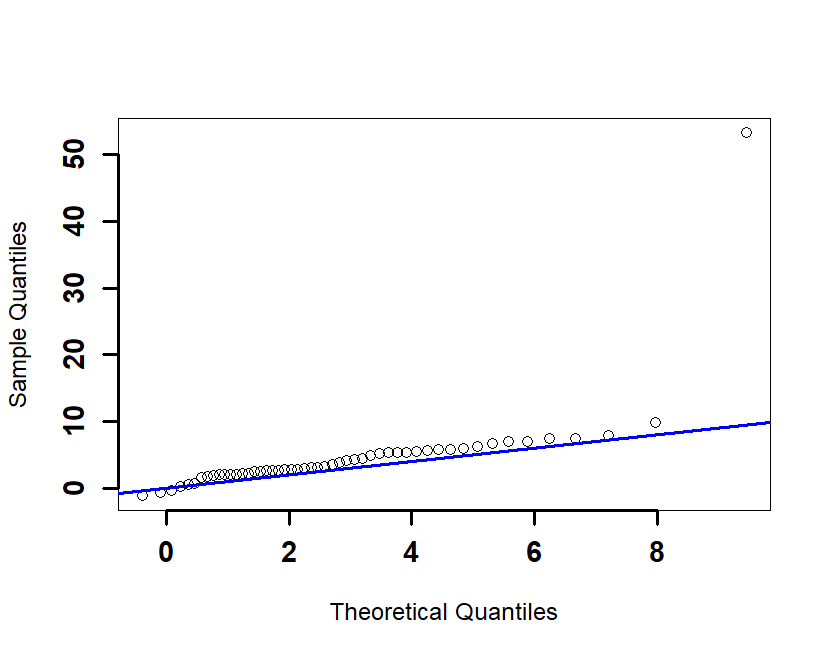}
			\label{FIG:Crime_qq_alp_1}}
		\\
	\caption{ \color{black} QQ plot of residuals for Crime data at different values of $\alpha$ under the regression model with SN error. The plot in (a) represents the least squares fit for normal errors.}
	\label{FIG:Crime_QQ}
\end{figure}

\newpage
\textcolor{black}{Finally, for this example also, we have drawn the QQ plot of the skew-normal model of the residual values in order to determine the suitability of the SN model; 
	these graphs are presented in Figure \ref{FIG:Crime_QQ}. 
	We can clearly observe that all the data points, except only for the one outlier, 
	fall near the straight line ($y=x$) for $\alpha \in [0.1, 0.5]$, justifying the appropriateness of our model assumptions for this particular dataset.}

\section{Choice of the robustness tuning parameter}
\label{Sec 6}
The tuning parameter $\alpha$ used in the proposed MDPDE based skewed regression model controls the trade-off between the asymptotic efficiency and the robustness of the procedure. At $\alpha= 0$, the MDPDE coincides with the MLE, which is the most efficient estimator for pure data but may have degraded performance under model misspecification and data contamination. As $\alpha$ increases, so does the stability of the estimator, but at the cost of an efficiency loss under pure data. It is therefore very important to choose the appropriate tuning parameter depending on the situation to get optimal performance. However, the experimenter would not know apriori how much anomaly is present in the data in relation to the assumed model. So a fixed choice of $\alpha$ cannot provide optimal performance in all situations. Thus it will be very useful to have an automatic data-specific method of selecting the tuning parameter that will provide the most appropriate treatment for the data at hand.

Several authors have considered data-specific optimal choices of the tuning parameter $\alpha$. Warwick and Jones (2005) \cite{Warwick/Jones:2005} explored this problem  under the IID set up, where they suggested the minimization of an empirical estimate of the asymptotic summed mean square error (AMSE) over $\alpha$ to choose the optimal tuning parameter. This approach was extended to the non-homogeneous setup by Ghosh and Basu (2015) \cite{Ghosh/Basu:2015}. While this approach generally provides a reasonable solution, it is dependent on the choice of an initial pilot estimator. Hong and Kim (2001) \cite{Hong/Kim:2001} had earlier proposed a method for the optimal tuning parameter selection, where they used the variance component only (rather than the entire mean squared error). As this method does not consider the bias component, it avoids the use of a pilot estimator; however this method suffers from the problem of an occasional non-robust solution. Recently, Basak et al. \cite{Basak/etc:2020} have proposed an iterative algorithm for finding the optimal value of the tuning parameter in case of both IID and non-homogeneous data which has been called the `Iterated Warwick-Jones' (IWJ) algorithm. This method appears to be able to remove the dependence on the pilot estimator which was inherent in the original Warwick-Jones approach.

 In this paper, we will follow the IWJ method for finding the optimal value of $\alpha$. For the sake of completeness, we state the algorithm briefly, as adapted to our case. Suppose that the observed dataset is obtained from a contamination model $g_{i}=(1-\epsilon)f_{i}(.,\boldsymbol{\theta}^{*})+ \epsilon \delta_{y_i}$ for each $i$, where $g_{i}$ and $f_{i}(.,\boldsymbol{\theta})$ are defined in Section \ref{SEC:2}, $\delta_{y_i}$s are some contaminating densities and $\boldsymbol{\theta}^{*}$ is the target parameter value. Let $\widehat{\boldsymbol{\theta}}_{\alpha,n}$ and $\boldsymbol{\theta}^{g}$ be the MDPDE and the corresponding best fitting parameters of $\boldsymbol{\theta}$ as defined in Section \ref{SEC:2}. Then the asymptotic summed MSE (AMSE) of the MDPDE $\widehat{\boldsymbol{\theta}}_{\alpha,n}$ with respect to the target $\boldsymbol{\theta}^{*}$ is given by 
\begin{equation}
	AMSE\left(\widehat{\boldsymbol{\theta}}_{\alpha,n}\right)= \left(\boldsymbol{\theta}^{g}-\boldsymbol{\theta}^{*}\right)^{T} \left(\boldsymbol{\theta}^{g}-\boldsymbol{\theta}^{*}\right)+ \frac{1}{n} trace \left[\boldsymbol{\Psi}_{\alpha}^{-1}\left(  \boldsymbol{\theta}^{g}\right)  \boldsymbol{\Omega}_{\alpha}\left(
	\boldsymbol{\theta}^{g}\right)  \boldsymbol{\Psi}_{\alpha
	}^{-1}\left(  \boldsymbol{\theta}^{g}\right)\right].
	\label{6.1}
\end{equation}
The expression on the right hand side of the above equation is a function of two unknown quantities, $\boldsymbol{\theta}^{g}$ and $\boldsymbol{\theta}^{*}$. We replace $\boldsymbol{\theta}^{g}$ by its consistent estimator $\widehat{\boldsymbol{\theta}}_{\alpha,n}$. However we do not have an obvious surrogate for $\boldsymbol{\theta}^{*}$ and therefore we have to use an appropriate robust pilot estimator $\widehat{\boldsymbol {\theta}}_{P}$ for it; the MDPDEs for $\alpha=0.5 ~\text{and}~ 1$ are popular choices for this purpose. Thus, an empirical estimate of the summed AMSE at tuning parameter $\alpha$ is given by 

\begin{equation}
	\widehat{AMSE}\left(\widehat{\boldsymbol{\theta}}_{\alpha,n}\right)= \left(\widehat{\boldsymbol{\theta}}_{\alpha,n}-\widehat{\boldsymbol{\theta}}_{P}\right)^{T} \left(\widehat{\boldsymbol{\theta}}_{\alpha,n}-\widehat{\boldsymbol{\theta}}_{P}\right)+ \frac{1}{n} trace \left[\boldsymbol{\Psi}_{n}^{-1}\left(  \widehat{\boldsymbol{\theta}
	}_{\alpha,n}\right)  \boldsymbol{\Omega}_{n}\left(
	\widehat{\boldsymbol{\theta}}_{\alpha,n}\right)  \boldsymbol{\Psi}_{n}^{-1}\left(  \widehat{\boldsymbol{\theta}}_{\alpha,n}\right)\right].
	\label{6.2}
\end{equation}
As the estimated summed $\widehat{AMSE}\left(\widehat{\boldsymbol{\theta}}_{\alpha,n}\right)$ is a function of $\alpha$, we can minimize this measure over $\alpha \in [0,1]$ to get the optimal tuning parameter for a given dataset. The algorithm proceeds as follows:\\
\begin{enumerate}[label=(\arabic*)]
	\item Start with a suitable robust pilot estimator $\widehat{\boldsymbol{\theta}}_{P}$ of $\boldsymbol{\theta}$.
	
	\item Calculate $\widehat{AMSE}(\widehat{\boldsymbol{\theta}}_{\alpha,n})$ for a large number of equispaced $\alpha \in [0,1]$.
	
	\item Find the value of $\alpha$ that minimizes estimated $\widehat{AMSE}(\widehat{\boldsymbol{\theta}}_{\alpha,n})$ in (\ref{6.2}) over the grid of $\alpha$ values considered in the previous step. Denote this by $\alpha_1$ and let the corresponding estimate of $\boldsymbol{\theta}$ be denoted as $\widehat{\boldsymbol{\theta}}_{1,\alpha_{1}}$.
	
	\item Choose $\widehat{\boldsymbol{\theta}}_{1,\alpha_{1}}$ as the updated pilot estimate of $\boldsymbol{\theta}$ for the next step.
	
	\item Repeat steps (3) and (4) until there is no further
	change in the estimate of $\boldsymbol{\theta}$ (or, equivalently, the estimate of the tuning parameter $\alpha$).
\end{enumerate}

The main advantage of this method is that the final choice of optimal $\alpha$ does not depend on the original pilot estimator $\widehat{\boldsymbol{\theta}}_{P}$. However, we still have to start the process at a suitable robust starting value, and here we use the MDPDE with $\alpha = 0.5$ as recommended by Ghosh and Basu (2015) \cite{Ghosh/Basu:2015}.

Notice that the above derivation of the optimal tuning parameter $\alpha$ uses the minimization of the empirical summed AMSE \textcolor{black}{which has been well-studied in the literature for providing} an optimal solution to the problem of estimating the parameter vector. \textcolor{black}{However, there is no general approach for the choice of an optimum $\alpha$ in robust testing of hypothesis. In our empirical investigations we have observed that, at least for the two datasets considered here, a better approach for finding the optimum value of $\alpha$ providing stable test results under data contamination would be to minimize an estimate of the asymptotic MSE of only the parameter(s) associated with the underlying hypothesis (rather than all the parameters taken together). This optimum $\alpha$ can then be used to estimate the \underline{entire} model (involving all the parameters) and use the resulting estimate to perform the targeted test of hypothesis. In other words, we are still proposing to use the same $\alpha$ to estimate the \underline{entire} model but that particular $\alpha$ value is chosen by minimising the MSE of the particular parameter(s) involved in our target hypothesis for getting better (stable) results in its testing. Although this appears to be intuitively expected (as the power of these Wald-types tests depend only on the variance of the parameters involved in the hypothesis), it needs further detailed investigation to understand the methodological implications, which we hope to take up in our future research.}



\textcolor{black}{For the AIS data, while testing for BMI,} we get the optimal value of the tuning parameter $\alpha$ to be $0.99$ \textcolor{black}{by minimizing the MSE of the estimated coefficient of the BMI variable for full data case}. Similarly, if we consider the covariate LBM in the AIS data example, the value of the corresponding optimal $\alpha$ becomes $0.93$. A similar exercise with the two covariates in case of Crime data leads to a common optimum $\alpha$ value of $0.82$ in each case. \textcolor{black}{The estimates along with the standard error for these optimal $\alpha$ values are given in Table \ref{TAB:AIS_Crime_opt_alp}.}

\begin{table}[h]
	\caption{\color{black} Parameter estimates along with standard error (SE) and p-values for both datasets at different optimal $\alpha$. 
		Covariate 1 stands for \textit{BMI} in case of the AIS data and \textit{Poverty} for the Crime data. 
		Covariate 2 indicates \textit{LBM} for the AIS data and \textit{Single} for the Crime data.}
	\color{black}
	\centering
		\begin{tabular}{c | c c| c }
			\hline
			Parameter & \multicolumn{3}{c}{ MDPDE at different $\alpha$}\\
			&  0.93 & 0.99 & 0.82 \\
			\hline
			&\multicolumn{2}{c|}{ AIS data} & Crime data\\
			\hline
			Intercept		&	$-46.2577$ & $-46.3154$	& $-18.0913$
			\\
			SE	 & (14.5464)	&	(14.5231)& (1.3928)\\
			p-value		&	(0.0012) &(0.0014) &  (1.98E-22)\\\hline
			
			Covariate 1		& 1.6484 &	1.6332& 0.4696	\\
			SE		& (0.8802) &	(0.8869)	& (0.0`543) 	\\
			p-value	& (0.0534) &	(0.0592)& (3.79E-05) \\ \hline
			
			Covariate 2		& 0.5086 &	0.5219	& 1.5112	\\
			SE		& (0.1913) &	(0.1946)& (0.1043) \\
			p-value	& (0.0128)	&	(0.0097)&(3.81E-05)\\ \hline
			
			$\sigma$	& 61.9796&	61.1098	 & 3.7089	\\
			SE	 &(4.7001)	&	(4.7862) & (0.5808) \\ \hline
			
			$\gamma$		& 10.6252&	10.7048& 7.9314	\\
			SE	& (5.4519)	&	(5.5635) & (4.6014)\\
			p-value	 &(0.0597)	&	(0.0631) & (0.0394)	 	\\ 
			 
			\hline
	\end{tabular}
	\label{TAB:AIS_Crime_opt_alp}
\end{table}

\section{Concluding remarks}

In this paper we have used the minimum density power divergence estimation method to fit linear regression models under skew errors. We have demonstrated, through influence function analysis, simulation studies and real data examples how the MDPDE can provide good fits for the regression model with skew normal errors. In particular this exploration reveals how procedures with large $\alpha$ can withstand the effect of model specific outliers unlike the likelihood based solutions (or other procedures with small values of $\alpha$) where the decision may be driven by a handful of outliers even in large data sets. So, in future, this method may be extended for the linear regression set up under any other positively skewed distributions, such as extreme valued, gamma, Gompertz etc. Further, the estimation procedure based on MDPDE may be applied to the Cox proportional hazards regression model (specially the frailty model) based on right censored survival data with the frailty distribution being skew normal.

\section*{Declarations}

\noindent
\textbf{Funding:} There in no funding to report for this work.

\noindent
\textbf{Conflicts of interests:} Authors declare no conflict of interest.

\end{document}